\wlog{File: fontch.tex
        [v2.2 (2010/04/12) font change package for plain TeX]}
%
\font\sixi=cmmi6
\font\sixsy=cmsy6
\skewchar\sixi='177
\skewchar\sixsy='60
\font\eighti=cmmi8
\font\eightsy=cmsy8
\skewchar\eighti='177
\skewchar\eightsy='60
\font\ninei=cmmi9
\font\ninesy=cmsy9
\skewchar\ninei='177
\skewchar\ninesy='60
\font\tenmib=cmmib10
\font\tenbsy=cmbsy10
\skewchar\tenmib='177
\skewchar\tenbsy='60
\font\twelvei=cmmi12
\font\twelvesy=cmsy10 scaled 1200
\font\twelveex=cmex10 scaled 1200
\font\twelvemib=cmmib10 scaled 1200
\font\twelvebsy=cmbsy10 scaled 1200
\skewchar\twelvei='177
\skewchar\twelvemib='177
\skewchar\twelvesy='60
\skewchar\twelvebsy='60
\font\fourteeni=cmmi12 scaled 1167
\font\fourteensy=cmsy10 scaled 1400
\font\fourteenex=cmex10 scaled 1400
\font\fourteenmib=cmmib10 scaled 1400
\font\fourteenbsy=cmbsy10 scaled 1400
\skewchar\fourteeni='177
\skewchar\fourteenmib='177
\skewchar\fourteensy='60
\skewchar\fourteenbsy='60
\font\twentyi=cmmi12 scaled 1667
\font\twentysy=cmsy10 scaled 2000
\font\twentyex=cmex10 scaled 2000
\font\twentymib=cmmib10 scaled 2000
\font\twentybsy=cmbsy10 scaled 2000
\skewchar\twentyi='177
\skewchar\twentymib='177
\skewchar\twentysy='60
\skewchar\twentybsy='60
\font\twentyfouri=cmmi12 scaled 2000
\font\twentyfoursy=cmsy10 scaled 2400
\font\twentyfourex=cmex10 scaled 2400
\font\twentyfourmib=cmmib10 scaled 2400
\font\twentyfourbsy=cmsy10 scaled 2400
\skewchar\twentyfouri='177
\skewchar\twentyfourmib='177
\skewchar\twentyfoursy='60
\skewchar\twentyfourbsy='60
\newdimen\bigsize
\newdimen\Bigsize
\newdimen\biggsize
\newdimen\Biggsize
\catcode`\@=11
\def\big#1{{\hbox{$\left#1\vbox to \bigsize{}\right.\n@space$}}}
\def\Big#1{{\hbox{$\left#1\vbox to \Bigsize{}\right.\n@space$}}}
\def\bigg#1{{\hbox{$\left#1\vbox to \biggsize{}\right.\n@space$}}}
\def\Bigg#1{{\hbox{$\left#1\vbox to \Biggsize{}\right.\n@space$}}}
\ifx\AMSfont\@undefined%
{}
\else
\let\ifAMS=\relax
\fi
\ifx\DStroke\@undefined%
{}
\else
\let\ifDS=\relax
\fi
\ifx\LMTone\@undefined%
{}
\else
\let\ifLMT=\relax
\fi
\ifx\LMTSone\@undefined%
{}
\else
\let\ifLMTS=\relax
\fi
\catcode`\@=12
\ifx\ifAMS\relax%
\immediate\write16{*** AMS option activated. AMS fonts are used ***}
\font\sevenex=cmex7
\font\eightex=cmex8
\font\nineex=cmex9
\font\fiveex=cmex7 scaled 714
\font\sixex=cmex7 scaled 857
\font\fivemib=cmmib5
\font\sixmib=cmmib6
\font\sevenmib=cmmib7
\font\eightmib=cmmib8
\font\ninemib=cmmib9
\font\fivebsy=cmbsy5
\font\sixbsy=cmbsy6
\font\sevenbsy=cmbsy7
\font\eightbsy=cmbsy8
\font\ninebsy=cmbsy9
\else
\font\fiveex=cmex10 scaled 500
\font\sixex=cmex10 scaled 600
\font\sevenex=cmex10 scaled 700
\font\eightex=cmex10 scaled 800
\font\nineex=cmex10 scaled 900
\font\fivemib=cmmib10 scaled 500
\font\sixmib=cmmib10 scaled 600
\font\sevenmib=cmmib10 scaled 700
\font\eightmib=cmmib10 scaled 800
\font\ninemib=cmmib10 scaled 900
\font\fivebsy=cmbsy10 scaled 500
\font\sixbsy=cmbsy10 scaled 600
\font\sevenbsy=cmbsy10 scaled 700
\font\eightbsy=cmbsy10 scaled 800
\font\ninebsy=cmbsy10 scaled 900
\fi
\skewchar\fivemib='177
\skewchar\sixmib='177
\skewchar\sevenmib='177
\skewchar\eightmib='177
\skewchar\ninemib='177
\skewchar\fivebsy='60
\skewchar\sixbsy='60
\skewchar\sevenbsy='60
\skewchar\eightbsy='60
\skewchar\ninebsy='60
\ifx\ifLMT\relax%
\immediate\write16{*** T1 option activated. cork-LM fonts are used ***}
\font\tenrm=ec-lmr10
\font\tenbf=ec-lmbx10
\font\tensl=ec-lmro10
\font\tentt=ec-lmtt10
\font\tenit=ec-lmri10
\font\tenss=ec-lmss10
\font\tencsc=ec-lmcsc10
\font\eightrm=ec-lmr8
\font\eightbf=ec-lmbx8
\font\eightsl=ec-lmro8
\font\eighttt=ec-lmtt8
\font\eightit=ec-lmri8
\font\eightss=ec-lmss8
\font\eightcsc=ec-lmcsc10 at 8pt
\font\fiverm=ec-lmr5
\font\fivebf=ec-lmbx5
\font\fivess=ec-lmss8 scaled 625
\font\sixrm=ec-lmr6
\font\sixbf=ec-lmbx6
\font\sixss=ec-lmss8 scaled 750
\font\sevenrm=ec-lmr7
\font\sevenbf=ec-lmbx7
\font\sevenss=ec-lmss8 scaled 875
\font\ninerm=ec-lmr9
\font\ninebf=ec-lmbx9
\font\niness=ec-lmss9
\font\twelverm=ec-lmr12
\font\twelvebf=ec-lmbx12
\font\twelvesl=ec-lmro12
\font\twelvett=ec-lmtt12
\font\twelveit=ec-lmri12
\font\twelvess=ec-lmss12
\font\twelvecsc=ec-lmcsc10 scaled 1200
\font\fourteenrm=ec-lmr12 scaled 1167
\font\fourteenbf=ec-lmbx12 scaled 1167
\font\fourteensl=ec-lmro12 scaled 1167
\font\fourteentt=ec-lmtt12 scaled 1167
\font\fourteenit=ec-lmri12 scaled 1167
\font\fourteenss=ec-lmss12 scaled 1167
\font\fourteencsc=ec-lmcsc10 scaled 1400
\font\twentyrm=ec-lmr12 scaled 1667
\font\twentybf=ec-lmbx12 scaled 1667
\font\twentysl=ec-lmro12 scaled 1667
\font\twentytt=ec-lmtt12 scaled 1667
\font\twentyit=ec-lmri12 scaled 1667
\font\twentyss=ec-lmss12 scaled 1667
\font\twentycsc=ec-lmcsc10 scaled 2000
\font\twentyfourrm=ec-lmr12 scaled 2000
\font\twentyfourbf=ec-lmbx12 scaled 2000
\font\twentyfoursl=ec-lmro12 scaled 2000
\font\twentyfourtt=ec-lmtt12 scaled 2000
\font\twentyfourit=ec-lmri12 scaled 2000
\font\twentyfourss=ec-lmss12 scaled 2000
\font\twentyfourcsc=ec-lmcsc10 scaled 2400
\chardef\i="19 \chardef\j="1A
\chardef\ae="E6 \chardef\AE="C6
\chardef\oe="F7 \chardef\OE="D7
\chardef\o="F8 \chardef\O="D8
\chardef\ss="FF \chardef\SS="DF
\chardef\aa="E5 \chardef\AA="C5
\chardef\l="AA \chardef\L="8A
\mathchardef\Gamma="0100
\mathchardef\Delta="0101
\mathchardef\Theta="0102
\mathchardef\Lambda="0103
\mathchardef\Xi="0104
\mathchardef\Pi="0105
\mathchardef\Sigma="0106
\mathchardef\Upsilon="0107
\mathchardef\Phi="0108
\mathchardef\Psi="0109
\mathchardef\Omega="010A
\def\hbar{{\mathchar"7009\mkern-9muh}}

\chardef\lh="A9 \chardef\Lh="89
\chardef\dh="A4 \chardef\th="B4
\chardef\dbar="9E
\chardef\Edh="D0 \chardef\edh="F0
\chardef\ij="BC \chardef\IJ="9C
\chardef\nj="AD \chardef\NJ="8D
\chardef\thorn="FE \chardef\Thorn="DE
\chardef\smallzero="18
\chardef\lguille="E \chardef\rguille="F
\chardef\vispace="20

\def\`#1{{\accent0 #1}}
\def\'#1{{\accent1 #1}}
\def\^#1{{\accent2 #1}}
\def\~#1{{\accent3 #1}}
\def\"#1{{\accent4 #1}}


\def\=#1{{\accent9 #1}}
\def\.#1{{\accent10 #1}}

\def\c#1{\setbox0=\hbox{#1}\ifdim\ht0=1ex \accent"B #1%
     \else{\ooalign{\unhbox0\crcr\hidewidth\char"B\hidewidth}}\fi}
\def\k#1{\setbox0=\hbox{#1}\ifdim\ht0=1ex \accent"C #1%
     \else{\ooalign{\unhbox0\crcr\hidewidth\char"C\hidewidth}}\fi}

%

\def\bar{\mathaccent"7009 }

\def\hat{\mathaccent"7002 }


%
\catcode`\^^80=11 \uccode`\^^80=`\^^80 \lccode`\^^80=`\^^a0
\catcode`\^^81=11 \uccode`\^^81=`\^^81 \lccode`\^^81=`\^^a1
\catcode`\^^82=11 \uccode`\^^82=`\^^82 \lccode`\^^82=`\^^a2
\catcode`\^^83=11 \uccode`\^^83=`\^^83 \lccode`\^^83=`\^^a3
\catcode`\^^84=11 \uccode`\^^84=`\^^84 \lccode`\^^84=`\^^a4
\catcode`\^^85=11 \uccode`\^^85=`\^^85 \lccode`\^^85=`\^^a5
\catcode`\^^86=11 \uccode`\^^86=`\^^86 \lccode`\^^86=`\^^a6
\catcode`\^^87=11 \uccode`\^^87=`\^^87 \lccode`\^^87=`\^^a7
\catcode`\^^88=11 \uccode`\^^88=`\^^88 \lccode`\^^88=`\^^a8
\catcode`\^^89=11 \uccode`\^^89=`\^^89 \lccode`\^^89=`\^^a9
\catcode`\^^8a=11 \uccode`\^^8a=`\^^8a \lccode`\^^8a=`\^^aa
\catcode`\^^8b=11 \uccode`\^^8b=`\^^8b \lccode`\^^8b=`\^^ab
\catcode`\^^8c=11 \uccode`\^^8c=`\^^8c \lccode`\^^8c=`\^^ac
\catcode`\^^8d=11 \uccode`\^^8d=`\^^8d \lccode`\^^8d=`\^^ad
\catcode`\^^8e=11 \uccode`\^^8e=`\^^8e \lccode`\^^8e=`\^^ae
\catcode`\^^8f=11 \uccode`\^^8f=`\^^8f \lccode`\^^8f=`\^^af
\catcode`\^^90=11 \uccode`\^^90=`\^^90 \lccode`\^^90=`\^^b0
\catcode`\^^91=11 \uccode`\^^91=`\^^91 \lccode`\^^91=`\^^b1
\catcode`\^^92=11 \uccode`\^^92=`\^^92 \lccode`\^^92=`\^^b2
\catcode`\^^93=11 \uccode`\^^93=`\^^93 \lccode`\^^93=`\^^b3
\catcode`\^^94=11 \uccode`\^^94=`\^^94 \lccode`\^^94=`\^^b4
\catcode`\^^95=11 \uccode`\^^95=`\^^95 \lccode`\^^95=`\^^b5
\catcode`\^^96=11 \uccode`\^^96=`\^^96 \lccode`\^^96=`\^^b6
\catcode`\^^97=11 \uccode`\^^97=`\^^97 \lccode`\^^97=`\^^b7
\catcode`\^^98=11 \uccode`\^^98=`\^^98 \lccode`\^^98=`\^^b8
\catcode`\^^99=11 \uccode`\^^99=`\^^99 \lccode`\^^99=`\^^b9
\catcode`\^^9a=11 \uccode`\^^9a=`\^^9a \lccode`\^^9a=`\^^ba
\catcode`\^^9b=11 \uccode`\^^9b=`\^^9b \lccode`\^^9b=`\^^bb
\catcode`\^^9c=11 \uccode`\^^9c=`\^^9c \lccode`\^^9c=`\^^bc
\catcode`\^^c0=11 \uccode`\^^c0=`\^^c0 \lccode`\^^c0=`\^^e0
\catcode`\^^c1=11 \uccode`\^^c1=`\^^c1 \lccode`\^^c1=`\^^e1
\catcode`\^^c2=11 \uccode`\^^c2=`\^^c2 \lccode`\^^c2=`\^^e2
\catcode`\^^c3=11 \uccode`\^^c3=`\^^c3 \lccode`\^^c3=`\^^e3
\catcode`\^^c4=11 \uccode`\^^c4=`\^^c4 \lccode`\^^c4=`\^^e4
\catcode`\^^c5=11 \uccode`\^^c5=`\^^c5 \lccode`\^^c5=`\^^e5
\catcode`\^^c6=11 \uccode`\^^c6=`\^^c6 \lccode`\^^c6=`\^^e6
\catcode`\^^c7=11 \uccode`\^^c7=`\^^c7 \lccode`\^^c7=`\^^e7
\catcode`\^^c8=11 \uccode`\^^c8=`\^^c8 \lccode`\^^c8=`\^^e8
\catcode`\^^c9=11 \uccode`\^^c9=`\^^c9 \lccode`\^^c9=`\^^e9
\catcode`\^^ca=11 \uccode`\^^ca=`\^^ca \lccode`\^^ca=`\^^ea
\catcode`\^^cb=11 \uccode`\^^cb=`\^^cb \lccode`\^^cb=`\^^eb
\catcode`\^^cc=11 \uccode`\^^cc=`\^^cc \lccode`\^^cc=`\^^ec
\catcode`\^^cd=11 \uccode`\^^cd=`\^^cd \lccode`\^^cd=`\^^ed
\catcode`\^^ce=11 \uccode`\^^ce=`\^^ce \lccode`\^^ce=`\^^ee
\catcode`\^^cf=11 \uccode`\^^cf=`\^^cf \lccode`\^^cf=`\^^ef
\catcode`\^^d0=11 \uccode`\^^d0=`\^^d0 \lccode`\^^d0=`\^^f0
\catcode`\^^d1=11 \uccode`\^^d1=`\^^d1 \lccode`\^^d1=`\^^f1
\catcode`\^^d2=11 \uccode`\^^d2=`\^^d2 \lccode`\^^d2=`\^^f2
\catcode`\^^d3=11 \uccode`\^^d3=`\^^d3 \lccode`\^^d3=`\^^f3
\catcode`\^^d4=11 \uccode`\^^d4=`\^^d4 \lccode`\^^d4=`\^^f4
\catcode`\^^d5=11 \uccode`\^^d5=`\^^d5 \lccode`\^^d5=`\^^f5
\catcode`\^^d6=11 \uccode`\^^d6=`\^^d6 \lccode`\^^d6=`\^^f6
\catcode`\^^d7=11 \uccode`\^^d7=`\^^d7 \lccode`\^^d7=`\^^f7
\catcode`\^^d8=11 \uccode`\^^d8=`\^^d8 \lccode`\^^d8=`\^^f8
\catcode`\^^d9=11 \uccode`\^^d9=`\^^d9 \lccode`\^^d9=`\^^f9
\catcode`\^^da=11 \uccode`\^^da=`\^^da \lccode`\^^da=`\^^fa
\catcode`\^^db=11 \uccode`\^^db=`\^^db \lccode`\^^db=`\^^fb
\catcode`\^^dc=11 \uccode`\^^dc=`\^^dc \lccode`\^^dc=`\^^fc
\catcode`\^^dd=11 \uccode`\^^dd=`\^^dd \lccode`\^^dd=`\^^fd
\catcode`\^^de=11 \uccode`\^^de=`\^^de \lccode`\^^de=`\^^fe
\catcode`\^^df=11 \uccode`\^^df=`\^^df \lccode`\^^df=`\^^ff
\catcode`\^^a0=11 \uccode`\^^a0=`\^^80 \lccode`\^^a0=`\^^a0
\catcode`\^^a1=11 \uccode`\^^a1=`\^^81 \lccode`\^^a1=`\^^a1
\catcode`\^^a2=11 \uccode`\^^a2=`\^^82 \lccode`\^^a2=`\^^a2
\catcode`\^^a3=11 \uccode`\^^a3=`\^^83 \lccode`\^^a3=`\^^a3
\catcode`\^^a4=11 \uccode`\^^a4=`\^^84 \lccode`\^^a4=`\^^a4
\catcode`\^^a5=11 \uccode`\^^a5=`\^^85 \lccode`\^^a5=`\^^a5
\catcode`\^^a6=11 \uccode`\^^a6=`\^^86 \lccode`\^^a6=`\^^a6
\catcode`\^^a7=11 \uccode`\^^a7=`\^^87 \lccode`\^^a7=`\^^a7
\catcode`\^^a8=11 \uccode`\^^a8=`\^^88 \lccode`\^^a8=`\^^a8
\catcode`\^^a9=11 \uccode`\^^a9=`\^^89 \lccode`\^^a9=`\^^a9
\catcode`\^^aa=11 \uccode`\^^aa=`\^^8a \lccode`\^^aa=`\^^aa
\catcode`\^^ab=11 \uccode`\^^ab=`\^^8b \lccode`\^^ab=`\^^ab
\catcode`\^^ac=11 \uccode`\^^ac=`\^^8c \lccode`\^^ac=`\^^ac
\catcode`\^^ad=11 \uccode`\^^ad=`\^^8d \lccode`\^^ad=`\^^ad
\catcode`\^^ae=11 \uccode`\^^ae=`\^^8e \lccode`\^^ae=`\^^ae
\catcode`\^^af=11 \uccode`\^^af=`\^^8f \lccode`\^^af=`\^^af
\catcode`\^^b0=11 \uccode`\^^b0=`\^^90 \lccode`\^^b0=`\^^b0
\catcode`\^^b1=11 \uccode`\^^b1=`\^^91 \lccode`\^^b1=`\^^b1
\catcode`\^^b2=11 \uccode`\^^b2=`\^^92 \lccode`\^^b2=`\^^b2
\catcode`\^^b3=11 \uccode`\^^b3=`\^^93 \lccode`\^^b3=`\^^b3
\catcode`\^^b4=11 \uccode`\^^b4=`\^^94 \lccode`\^^b4=`\^^b4
\catcode`\^^b5=11 \uccode`\^^b5=`\^^95 \lccode`\^^b5=`\^^b5
\catcode`\^^b6=11 \uccode`\^^b6=`\^^96 \lccode`\^^b6=`\^^b6
\catcode`\^^b7=11 \uccode`\^^b7=`\^^97 \lccode`\^^b7=`\^^b7
\catcode`\^^b8=11 \uccode`\^^b8=`\^^98 \lccode`\^^b8=`\^^b8
\catcode`\^^b9=11 \uccode`\^^b9=`\^^99 \lccode`\^^b9=`\^^b9
\catcode`\^^ba=11 \uccode`\^^ba=`\^^9a \lccode`\^^ba=`\^^ba
\catcode`\^^bb=11 \uccode`\^^bb=`\^^9b \lccode`\^^bb=`\^^bb
\catcode`\^^e0=11 \uccode`\^^e0=`\^^c0 \lccode`\^^e0=`\^^e0
\catcode`\^^e1=11 \uccode`\^^e1=`\^^c1 \lccode`\^^e1=`\^^e1
\catcode`\^^e2=11 \uccode`\^^e2=`\^^c2 \lccode`\^^e2=`\^^e2
\catcode`\^^e3=11 \uccode`\^^e3=`\^^c3 \lccode`\^^e3=`\^^e3
\catcode`\^^e4=11 \uccode`\^^e4=`\^^c4 \lccode`\^^e4=`\^^e4
\catcode`\^^e5=11 \uccode`\^^e5=`\^^c5 \lccode`\^^e5=`\^^e5
\catcode`\^^e6=11 \uccode`\^^e6=`\^^c6 \lccode`\^^e6=`\^^e6
\catcode`\^^e7=11 \uccode`\^^e7=`\^^c7 \lccode`\^^e7=`\^^e7
\catcode`\^^e8=11 \uccode`\^^e8=`\^^c8 \lccode`\^^e8=`\^^e8
\catcode`\^^e9=11 \uccode`\^^e9=`\^^c9 \lccode`\^^e9=`\^^e9
\catcode`\^^ea=11 \uccode`\^^ea=`\^^ca \lccode`\^^ea=`\^^ea
\catcode`\^^eb=11 \uccode`\^^eb=`\^^cb \lccode`\^^eb=`\^^eb
\catcode`\^^ec=11 \uccode`\^^ec=`\^^cc \lccode`\^^ec=`\^^ec
\catcode`\^^ed=11 \uccode`\^^ed=`\^^cd \lccode`\^^ed=`\^^ed
\catcode`\^^ee=11 \uccode`\^^ee=`\^^ce \lccode`\^^ee=`\^^ee
\catcode`\^^ef=11 \uccode`\^^ef=`\^^cf \lccode`\^^ef=`\^^ef
\catcode`\^^f0=11 \uccode`\^^f0=`\^^d0 \lccode`\^^f0=`\^^f0
\catcode`\^^f1=11 \uccode`\^^f1=`\^^d1 \lccode`\^^f1=`\^^f1
\catcode`\^^f2=11 \uccode`\^^f2=`\^^d2 \lccode`\^^f2=`\^^f2
\catcode`\^^f3=11 \uccode`\^^f3=`\^^d3 \lccode`\^^f3=`\^^f3
\catcode`\^^f4=11 \uccode`\^^f4=`\^^d4 \lccode`\^^f4=`\^^f4
\catcode`\^^f5=11 \uccode`\^^f5=`\^^d5 \lccode`\^^f5=`\^^f5
\catcode`\^^f6=11 \uccode`\^^f6=`\^^d6 \lccode`\^^f6=`\^^f6
\catcode`\^^f7=11 \uccode`\^^f7=`\^^d7 \lccode`\^^f7=`\^^f7
\catcode`\^^f8=11 \uccode`\^^f8=`\^^d8 \lccode`\^^f8=`\^^f8
\catcode`\^^f9=11 \uccode`\^^f9=`\^^d9 \lccode`\^^f9=`\^^f9
\catcode`\^^fa=11 \uccode`\^^fa=`\^^da \lccode`\^^fa=`\^^fa
\catcode`\^^fb=11 \uccode`\^^fb=`\^^db \lccode`\^^fb=`\^^fb
\catcode`\^^fc=11 \uccode`\^^fc=`\^^dc \lccode`\^^fc=`\^^fc
\catcode`\^^fd=11 \uccode`\^^fd=`\^^dd \lccode`\^^fd=`\^^fd
\catcode`\^^fe=11 \uccode`\^^fe=`\^^de \lccode`\^^fe=`\^^fe
\catcode`\^^ff=11 \uccode`\^^ff=`\^^df \lccode`\^^ff=`\^^ff
\else
\font\eightrm=cmr8
\font\eightbf=cmbx8
\font\eightsl=cmsl8
\font\eighttt=cmtt8
\font\eightit=cmti8
\font\eightss=cmss8
\font\fivess=cmss8 scaled 625
\font\sixrm=cmr6
\font\sixbf=cmbx6
\font\sixss=cmss8 scaled 750
\font\sevenss=cmss8 scaled 875
\font\ninerm=cmr9
\font\ninebf=cmbx9
\font\niness=cmss9
\font\tenss=cmss10
\font\tencsc=cmcsc10
\font\twelverm=cmr12
\font\twelvebf=cmbx12
\font\twelvesl=cmsl12
\font\twelvett=cmtt12
\font\twelveit=cmti12
\font\twelvess=cmss12
\font\twelvecsc=cmcsc10 scaled 1200
\font\fourteenrm=cmr12 scaled 1167
\font\fourteenbf=cmbx12 scaled 1167
\font\fourteensl=cmsl12 scaled 1167
\font\fourteentt=cmtt12 scaled 1167
\font\fourteenit=cmti12 scaled 1167
\font\fourteenss=cmss12 scaled 1167
\font\fourteencsc=cmcsc10 scaled 1400
\font\twentyrm=cmr12 scaled 1667
\font\twentybf=cmbx12 scaled 1667
\font\twentysl=cmsl12 scaled 1667
\font\twentytt=cmtt12 scaled 1667
\font\twentyit=cmti12 scaled 1667
\font\twentyss=cmss12 scaled 1667
\font\twentycsc=cmcsc10 scaled 2000
\font\twentyfourrm=cmr12 scaled 2000
\font\twentyfourbf=cmbx12 scaled 2000
\font\twentyfoursl=cmsl12 scaled 2000
\font\twentyfourtt=cmtt12 scaled 2000
\font\twentyfourit=cmti12 scaled 2000
\font\twentyfourss=cmss12 scaled 2000
\font\twentyfourcsc=cmcsc10 scaled 2400
\ifx\ifAMS\relax%
\font\eightcsc=cmcsc8
\else
\font\eightcsc=cmcsc10 scaled 800
\fi
\fi
\ifx\ifLMTS\relax%
\immediate\write16{*** TS1 option activated. ts1-LM fonts are used ***}
\input TS1mac.tex
\fi
\ifx\ifDS\relax%
\immediate\write16{*** Doublestroke option activated. DS fonts used ***}
\font\fiveds=dsrom8 scaled 625
\font\sixds=dsrom8 scaled 750
\font\sevends=dsrom8 scaled 875
\font\eightds=dsrom8
\font\nineds=dsrom10 scaled 900
\font\tends=dsrom10
\font\twelveds=dsrom12
\font\fourteends=dsrom12 scaled 1167
\font\twentyds=dsrom12 scaled 1667
\font\twentyfourds=dsrom12 scaled 2000
\fi
\newfam\mibfam
\newfam\bsyfam
\newfam\ssfam
\newfam\cscfam
\ifx\ifDS\relax%
\input DSmac.tex
\fi
\def\ds{}
\def\tenpoint{
\textfont0=\tenrm \scriptfont0=\sevenrm \scriptscriptfont0=\fiverm
\textfont1=\teni \scriptfont1=\seveni \scriptscriptfont1=\fivei
\def\mit{\fam1 } \def\oldstyle{\fam1 \teni}
\textfont2=\tensy \scriptfont2=\sevensy \scriptscriptfont2=\fivesy
\def\cal{\fam2 }
\textfont3=\tenex \scriptfont3=\sevenex \scriptscriptfont3=\fiveex
\textfont\itfam=\tenit
\textfont\slfam=\tensl
\textfont\ttfam=\tentt
\textfont\ssfam=\tenss
\scriptfont\ssfam=\sevenss \scriptscriptfont\ssfam=\fivess
\textfont\bffam=\tenbf
\scriptfont\bffam=\sevenbf \scriptscriptfont\bffam=\fivebf
\textfont\cscfam=\tencsc
\textfont\mibfam=\tenmib
\scriptfont\mibfam=\sevenmib \scriptscriptfont\mibfam=\fivemib
\textfont\bsyfam=\tenbsy
\scriptfont\bsyfam=\sevenbsy \scriptscriptfont\bsyfam=\fivebsy
\ifx\ifDS\relax%
\textfont\dsfam=\tends
\scriptfont\dsfam=\sevends \scriptscriptfont\dsfam=\fiveds
\def\ds{\fam\dsfam\tends}
\else
{}
\fi
\ifx\ifLMTS\relax%
\def\rm{\fam0 \tenrm \let\tcfont=\tencrm}
\def\it{\fam\itfam\tenit \let\tcfont=\tencit}
\def\sl{\fam\slfam\tensl \let\tcfont=\tencsl}
\def\tt{\fam\ttfam\tentt \let\tcfont=\tenctt}
\def\sf{\fam\ssfam\tenss \let\tcfont=\tencss}
\def\bf{\fam\bffam\tenbf \let\tcfont=\tencbf}
\else
\def\rm{\fam0 \tenrm}
\def\it{\fam\itfam\tenit}
\def\sl{\fam\slfam\tensl}
\def\tt{\fam\ttfam\tentt}
\def\sf{\fam\ssfam\tenss}
\def\bf{\fam\bffam\tenbf}
\fi
\def\sc{\fam\cscfam\tencsc}
\def\mb{\fam\mibfam} \def\boldstyle{\fam\mibfam\tenmib}
\def\bcal{\fam\bsyfam\tenbsy}
\def\unsetmathbold{
\textfont1=\teni \scriptfont1=\seveni \scriptscriptfont1=\fivei
\textfont2=\tensy \scriptfont2=\sevensy \scriptscriptfont2=\fivesy
\def\oldstyle{\fam1 \teni} }
\def\setmathbold{
\textfont1=\tenmib \scriptfont1=\sevenmib \scriptscriptfont1=\fivemib
\textfont2=\tenbsy \scriptfont2=\sevenbsy \scriptscriptfont2=\fivebsy
\def\oldstyle{\fam1 \tenmib} }
\normalbaselineskip=12pt minus 1pt
\setbox\strutbox=\hbox{\vrule height8.5pt depth3.5pt width0pt}
\bigsize=8.5pt
\Bigsize=11.5pt
\biggsize=14.5pt
\Biggsize=17.5pt
\normalbaselines\rm}
\def\twelvepoint{
\textfont0=\twelverm \scriptfont0=\eightrm \scriptscriptfont0=\sixrm
\textfont1=\twelvei \scriptfont1=\eighti \scriptscriptfont1=\sixi
\def\mit{\fam1 } \def\oldstyle{\fam1 \twelvei}
\textfont2=\twelvesy \scriptfont2=\eightsy \scriptscriptfont2=\sixsy
\def\cal{\fam2 }
\textfont3=\twelveex \scriptfont3=\eightex \scriptscriptfont3=\sixex
\textfont\itfam=\twelveit
\textfont\slfam=\twelvesl
\textfont\ttfam=\twelvett
\textfont\ssfam=\twelvess
\scriptfont\ssfam=\eightss \scriptscriptfont\ssfam=\sixss
\textfont\bffam=\twelvebf
\scriptfont\bffam=\eightbf \scriptscriptfont\bffam=\sixbf
\textfont\cscfam=\twelvecsc
\textfont\mibfam=\twelvemib
\scriptfont\mibfam=\eightmib \scriptscriptfont\mibfam=\sixmib
\textfont\bsyfam=\twelvebsy
\scriptfont\bsyfam=\eightbsy \scriptscriptfont\bsyfam=\sixbsy
\ifx\ifDS\relax%
\textfont\dsfam=\twelveds
\scriptfont\dsfam=\eightds \scriptscriptfont\dsfam=\sixds
\def\ds{\fam\dsfam\twelveds}
\else
{}
\fi
\ifx\ifLMTS\relax%
\def\rm{\fam0 \twelverm \let\tcfont=\twelvecrm}
\def\it{\fam\itfam\twelveit \let\tcfont=\twelvecit}
\def\sl{\fam\slfam\twelvesl \let\tcfont=\twelvecsl}
\def\tt{\fam\ttfam\twelvett \let\tcfont=\twelvectt}
\def\sf{\fam\ssfam\twelvess \let\tcfont=\twelvecss}
\def\bf{\fam\bffam\twelvebf \let\tcfont=\twelvecss}
\else
\def\rm{\fam0 \twelverm}
\def\it{\fam\itfam\twelveit}
\def\sl{\fam\slfam\twelvesl}
\def\tt{\fam\ttfam\twelvett}
\def\sf{\fam\ssfam\twelvess}
\def\bf{\fam\bffam\twelvebf}
\fi
\def\sc{\fam\cscfam\twelvecsc}
\def\mb{\fam\mibfam} \def\boldstyle{\fam\mibfam\twelvemib}
\def\bcal{\fam\bsyfam}
\def\unsetmathbold{
\textfont1=\twelvei \scriptfont1=\eighti \scriptscriptfont1=\sixi
\textfont2=\twelvesy \scriptfont2=\eightsy \scriptscriptfont2=\sixsy
\def\oldstyle{\fam1 \twelvei} }
\def\setmathbold{
\textfont1=\twelvemib \scriptfont1=\eightmib \scriptscriptfont1=\sixmib
\textfont2=\twelvebsy \scriptfont2=\eightbsy \scriptscriptfont2=\sixbsy
\def\oldstyle{\fam1 \twelvemib} }
\normalbaselineskip=14pt minus 1pt
\setbox\strutbox=\hbox{\vrule height10pt depth4pt width0pt}
\bigsize=10.2pt
\Bigsize=13.8pt
\biggsize=17.4pt
\Biggsize=21pt
\normalbaselines\rm}
\def\eightpoint{
\textfont0=\eightrm \scriptfont0=\sixrm \scriptscriptfont0=\fiverm
\textfont1=\eighti \scriptfont1=\sixi \scriptscriptfont1=\fivei
\def\mit{\fam1 } \def\oldstyle{\fam1 \eighti}
\textfont2=\eightsy \scriptfont2=\sixsy \scriptscriptfont2=\fivesy
\def\cal{\fam2 }
\textfont3=\eightex \scriptfont3=\sixex \scriptscriptfont3=\fiveex
\textfont\itfam=\eightit
\textfont\slfam=\eightsl
\textfont\ttfam=\eighttt
\textfont\ssfam=\eightss
\scriptfont\ssfam=\sixss \scriptscriptfont\ssfam=\fivess
\textfont\bffam=\eightbf
\scriptfont\bffam=\sixbf \scriptscriptfont\bffam=\fivebf
\textfont\cscfam=\eightcsc
\textfont\mibfam=\eightmib
\scriptfont\mibfam=\sixmib \scriptscriptfont\mibfam=\fivemib
\textfont\bsyfam=\eightbsy
\scriptfont\bsyfam=\sixbsy \scriptscriptfont\bsyfam=\fivebsy
\ifx\ifDS\relax%
\textfont\dsfam=\eightds
\scriptfont\dsfam=\sixds \scriptscriptfont\dsfam=\fiveds
\def\ds{\fam\dsfam\eightds}
\else
{}
\fi
\ifx\ifLMTS\relax%
\def\rm{\fam0 \eightrm \let\tcfont=\eightcrm}
\def\it{\fam\itfam\eightit \let\tcfont=\eightcit}
\def\sl{\fam\slfam\eightsl \let\tcfont=\eightcsl}
\def\tt{\fam\ttfam\eighttt \let\tcfont=\eightctt}
\def\sf{\fam\ssfam\eightss \let\tcfont=\eightcss}
\def\bf{\fam\bffam\eightbf \let\tcfont=\eightcbf}
\else
\def\rm{\fam0 \eightrm}
\def\it{\fam\itfam\eightit}
\def\sl{\fam\slfam\eightsl}
\def\tt{\fam\ttfam\eighttt}
\def\sf{\fam\ssfam\eightss}
\def\bf{\fam\bffam\eightbf}
\fi
\def\sc{\fam\cscfam\eightcsc}
\def\mb{\fam\mibfam} \def\boldstyle{\fam\mibfam\eightmib}
\def\bcal{\fam\bsyfam}
\def\unsetmathbold{
\textfont1=\eighti \scriptfont1=\sixi \scriptscriptfont1=\fivei
\textfont2=\eightsy \scriptfont2=\sixsy \scriptscriptfont2=\fivesy
\def\oldstyle{\fam1 \eighti} }
\def\setmathbold{
\textfont1=\eightmib \scriptfont1=\sixmib \scriptscriptfont1=\fivemib
\textfont2=\eightbsy \scriptfont2=\sixbsy \scriptscriptfont2=\fivebsy
\def\oldstyle{\fam1 \eightmib} }
\normalbaselineskip=9pt
\setbox\strutbox=\hbox{\vrule height7pt depth2pt width0pt}
\bigsize=6.8pt
\Bigsize=9.2pt
\biggsize=11.6pt
\Biggsize=14pt
\normalbaselines\rm}
\def\fourteenpoint{
\textfont0=\fourteenrm \scriptfont0=\ninerm \scriptscriptfont0=\sevenrm
\textfont1=\fourteeni \scriptfont1=\ninei \scriptscriptfont1=\seveni
\def\mit{\fam1 } \def\oldstyle{\fam1 \fourteeni}
\textfont2=\fourteensy \scriptfont2=\ninesy \scriptscriptfont2=\sevensy
\def\cal{\fam2 }
\textfont3=\fourteenex \scriptfont3=\nineex \scriptscriptfont3=\sevenex
\textfont\itfam=\fourteenit
\textfont\slfam=\fourteensl
\textfont\ttfam=\fourteentt
\textfont\ssfam=\fourteenss
\scriptfont\ssfam=\niness \scriptscriptfont\ssfam=\sevenss
\textfont\bffam=\fourteenbf
\scriptfont\bffam=\ninebf \scriptscriptfont\bffam=\sevenbf
\textfont\cscfam=\fourteencsc
\textfont\mibfam=\fourteenmib
\scriptfont\mibfam=\ninemib \scriptscriptfont\mibfam=\sevenmib
\textfont\bsyfam=\fourteenbsy
\scriptfont\bsyfam=\ninebsy \scriptscriptfont\bsyfam=\sevenbsy
\ifx\ifDS\relax%
\textfont\dsfam=\fourteends
\scriptfont\dsfam=\nineds \scriptscriptfont\dsfam=\sevends
\def\ds{\fam\dsfam\fourteends}
\else
{}
\fi
\ifx\ifLMTS\relax%
\def\rm{\fam0 \fourteenrm \let\tcfont=\fourteencrm}
\def\it{\fam\itfam\fourteenit \let\tcfont=\fourteencit}
\def\sl{\fam\slfam\fourteensl \let\tcfont=\fourteencsl}
\def\tt{\fam\ttfam\fourteentt \let\tcfont=\fourteenctt}
\def\bf{\fam\bffam\fourteenbf \let\tcfont=\fourteencbf}
\def\sf{\fam\ssfam\fourteenss \let\tcfont=\fourteencss}
\else
\def\rm{\fam0 \fourteenrm}
\def\it{\fam\itfam\fourteenit}
\def\sl{\fam\slfam\fourteensl}
\def\tt{\fam\ttfam\fourteentt}
\def\bf{\fam\bffam\fourteenbf}
\def\sf{\fam\ssfam\fourteenss}
\fi
\def\sc{\fam\cscfam\fourteencsc}
\def\mb{\fam\mibfam} \def\boldstyle{\fam\mibfam\fourteenmib}
\def\bcal{\fam\bsyfam}
\def\unsetmathbold{
\textfont1=\fourteeni \scriptfont1=\ninei \scriptscriptfont1=\seveni
\textfont2=\fourteensy \scriptfont2=\ninesy \scriptscriptfont2=\sevensy
\def\oldstyle{\fam1 \fourteeni} }
\def\setmathbold{
\textfont1=\fourteenmib \scriptfont1=\ninemib \scriptscriptfont1=\sevenmib
\textfont2=\fourteenbsy \scriptfont2=\ninebsy \scriptscriptfont2=\sevenbsy
\def\oldstyle{\fam1 \fourteenmib} }
\normalbaselineskip=17pt minus 1pt
\setbox\strutbox=\hbox{\vrule height12pt depth5pt width0pt}
\bigsize=11.9pt
\Bigsize=16.1pt
\biggsize=20.3pt
\Biggsize=24.5pt
\normalbaselines\rm}
\def\twentypoint{
\textfont0=\twentyrm \scriptfont0=\fourteenrm \scriptscriptfont0=\tenrm
\textfont1=\twentyi \scriptfont1=\fourteeni \scriptscriptfont1=\teni
\def\mit{\fam1 } \def\oldstyle{\fam1 \twentyi}
\textfont2=\twentysy \scriptfont2=\fourteensy \scriptscriptfont2=\tensy
\def\cal{\fam2 }
\textfont3=\twentyex \scriptfont3=\fourteenex \scriptscriptfont3=\tenex
\textfont\itfam=\twentyit
\textfont\slfam=\twentysl
\textfont\ttfam=\twentytt
\textfont\ssfam=\twentyss
\scriptfont\ssfam=\fourteenss \scriptscriptfont\ssfam=\tenss
\textfont\bffam=\twentybf
\scriptfont\bffam=\fourteenbf \scriptscriptfont\bffam=\tenbf
\textfont\cscfam=\twentycsc
\textfont\mibfam=\twentymib
\scriptfont\mibfam=\fourteenmib \scriptscriptfont\mibfam=\tenmib
\textfont\bsyfam=\twentybsy
\scriptfont\bsyfam=\fourteenbsy \scriptscriptfont\bsyfam=\tenbsy
\ifx\ifDS\relax%
\textfont\dsfam=\twentyds
\scriptfont\dsfam=\fourteends \scriptscriptfont\dsfam=\tends
\def\ds{\fam\dsfam\twentyds}
\else
{}
\fi
\ifx\ifLMTS\relax%
\def\rm{\fam0 \twentyrm \let\tcfont=\twentycrm}
\def\it{\fam\itfam\twentyit \let\tcfont=\twentycit}
\def\sl{\fam\slfam\twentysl \let\tcfont=\twentycsl}
\def\tt{\fam\ttfam\twentytt \let\tcfont=\twentyctt}
\def\sf{\fam\ssfam\twentyss \let\tcfont=\twentycss}
\def\bf{\fam\bffam\twentybf \let\tcfont=\twentycbf}
\else
\def\rm{\fam0 \twentyrm}
\def\it{\fam\itfam\twentyit}
\def\sl{\fam\slfam\twentysl}
\def\tt{\fam\ttfam\twentytt}
\def\sf{\fam\ssfam\twentyss}
\def\bf{\fam\bffam\twentybf}
\fi
\def\sc{\fam\cscfam\twentycsc}
\def\mb{\fam\mibfam} \def\boldstyle{\fam\mibfam\twentymib}
\def\unsetmathbold{
\textfont1=\twentyi \scriptfont1=\fourteeni \scriptscriptfont1=\teni
\textfont2=\twentysy \scriptfont2=\fourteensy \scriptscriptfont2=\tensy
\def\oldstyle{\fam1 \twentyi} }
\def\setmathbold{
\textfont1=\twentymib \scriptfont1=\fourteenmib \scriptscriptfont1=\tenmib
\textfont2=\twentybsy \scriptfont2=\fourteenbsy \scriptscriptfont2=\tenbsy
\def\oldstyle{\fam1 \twentymib} }
\normalbaselineskip=24pt minus 2pt
\setbox\strutbox=\hbox{\vrule height17pt depth7pt width0pt}
\bigsize=17pt
\Bigsize=23pt
\biggsize=29pt
\Biggsize=35pt
\normalbaselines\rm}
\def\twentyfourpoint{
\textfont0=\twentyfourrm \scriptfont0=\fourteenrm \scriptscriptfont0=\twelverm
\textfont1=\twentyfouri \scriptfont1=\fourteeni \scriptscriptfont1=\twelvei
\def\mit{\fam1 } \def\oldstyle{\fam1 \twentyfouri}
\textfont2=\twentyfoursy \scriptfont2=\fourteensy \scriptscriptfont2=\twelvesy
\def\cal{\fam2 }
\textfont3=\twentyfourex \scriptfont3=\fourteenex \scriptscriptfont3=\twelveex
\textfont\itfam=\twentyfourit
\textfont\slfam=\twentyfoursl
\textfont\ttfam=\twentyfourtt
\textfont\ssfam=\twentyfourss
\scriptfont\ssfam=\fourteenss \scriptscriptfont\ssfam=\twelvess
\textfont\bffam=\twentyfourbf
\scriptfont\bffam=\fourteenbf \scriptscriptfont\bffam=\twelvebf
\textfont\cscfam=\twentyfourcsc
\textfont\mibfam=\twentyfourmib
\scriptfont\mibfam=\fourteenmib \scriptscriptfont\mibfam=\twelvemib
\textfont\bsyfam=\twentyfourbsy
\scriptfont\bsyfam=\fourteenbsy \scriptscriptfont\bsyfam=\twelvebsy
\ifx\ifDS\relax%
\textfont\dsfam=\twentyfourds
\scriptfont\dsfam=\fourteends \scriptscriptfont\dsfam=\twelveds
\def\ds{\fam\dsfam\twentyfourds}
\else
{}
\fi
\ifx\ifLMTS\relax%
\def\rm{\fam0 \twentyfourrm \let\tcfont=\twentyfourcrm}
\def\it{\fam\itfam\twentyfourit \let\tcfont=\twentyfourcit}
\def\sl{\fam\itfam\twentyfoursl \let\tcfont=\twentyfourcsl}
\def\tt{\fam\ttfam\twentyfourtt \let\tcfont=\twentyfourctt}
\def\sf{\fam\ssfam\twentyfourss \let\tcfont=\twentyfourcss}
\def\bf{\fam\bffam\twentyfourbf \let\tcfont=\twentyfourcbf}
\else
\def\rm{\fam0 \twentyfourrm}
\def\it{\fam\itfam\twentyfourit}
\def\sl{\fam\itfam\twentyfoursl}
\def\tt{\fam\ttfam\twentyfourtt}
\def\sf{\fam\ssfam\twentyfourss}
\def\bf{\fam\bffam\twentyfourbf}
\fi
\def\sc{\fam\cscfam\twentyfourcsc}
\def\mb{\fam\mibfam} \def\boldstyle{\fam\mibfam\twentyfourmib}
\def\bcal{\fam\bsyfam}
\def\unsetmathbold{
\textfont1=\twentyfouri \scriptfont1=\fourteeni \scriptscriptfont1=\twelvei
\textfont2=\twentyfoursy \scriptfont2=\fourteensy \scriptscriptfont2=\twelvesy
\def\oldstyle{\fam1 \twentyfouri} }
\def\setmathbold{
\textfont1=\twentyfourmib \scriptfont1=\fourteenmib
\scriptscriptfont1=\twelvemib
\textfont2=\twentyfourbsy \scriptfont2=\fourteenbsy
\scriptscriptfont2=\twelvebsy
\def\oldstyle{\fam1 \twentyfourmib} }
\normalbaselineskip=28pt minus 2pt
\setbox\strutbox=\hbox{\vrule height20pt depth8pt width0pt}
\bigsize=20.4pt
\Bigsize=27.6pt
\biggsize=34.8pt
\Biggsize=42pt
\normalbaselines\rm}
\def\doublespace{
\normalbaselineskip=24pt minus 1pt
\setbox\strutbox=\hbox{\vrule height17pt depth 7pt width0pt}
\normalbaselines}
\tenpoint

%
%
%
\def\unredoffs{} \def\redoffs{\voffset=-.31truein\hoffset=-.48truein}
\def\speclscape{}
%
%
%
%
%
\newbox\leftpage \newdimen\fullhsize \newdimen\hstitle \newdimen\hsbody
\tolerance=1000\hfuzz=2pt
\catcode`\@=11 
\def\hypernoname{}
\ifx\pdfoutput\UNd@FiNeD\else\ifnum\pdfoutput>0
  \def\hyperdef#1#2#3#4{\ifvmode\leavevmode\fi\pdfdest name{#2.#3} xyz\relax #4}%
  \def\hyperref#1#2#3#4{\pdfstartlink attr{/Border[0 0 0]} goto name{#2.#3}\relax #4\pdfendlink}%
\fi\fi
\ifx\hyperdef\UNd@FiNeD\def\hyperdef#1#2#3#4{#4}\def\hyperref#1#2#3#4{#4}\fi
\def\bigans{b }
\def\answ{b }
%
\ifx\answ\bigans\message{(This will come out unreduced.}
  \magnification=1200\unredoffs\baselineskip=16pt plus 2pt minus 1pt
  \hsbody=\hsize \hstitle=\hsize 
\else\message{(This will be reduced.} \let\l@r=L
  \magnification=1000\baselineskip=16pt plus 2pt minus 1pt \vsize=7truein
  \redoffs \hstitle=8truein\hsbody=4.75truein\fullhsize=10truein\hsize=\hsbody
  \output={\ifnum\pageno=0 
        \shipout\vbox{\speclscape{\hsize\fullhsize\makeheadline}
          \hbox to \fullhsize{\hfill\pagebody\hfill}}\advancepageno
      \else
        \almostshipout{\leftline{\vbox{\pagebody\makefootline}}}\advancepageno
      \fi}
  \def\almostshipout#1{\if L\l@r \count1=1 \message{[\the\count0.\the\count1]}
      \global\setbox\leftpage=#1 \global\let\l@r=R
    \else \count1=2
      \shipout\vbox{\speclscape{\hsize\fullhsize\makeheadline}
        \hbox to\fullhsize{\box\leftpage\hfil#1}}  \global\let\l@r=L\fi}
\fi
%
\newcount\yearltd\yearltd=\year\advance\yearltd by -1900

\def\Title#1#2{\nopagenumbers\abstractfont\hsize=\hstitle\rightline{#1}%
  \vskip 1in\centerline{\titlefont #2}\abstractfont\vskip .5in\pageno=0}
%
%
%

\def\draftmode{\message{ DRAFTMODE }\def\draftdate{{\rm preliminary draft:
      \number\month/\number\day/\number\yearltd\ \ \hourmin}}%
  \headline={\hfil\draftdate}\writelabels\baselineskip=20pt plus 2pt minus 2pt
    {\count255=\time\divide\count255 by 60 \xdef\hourmin{\number\count255}
      \multiply\count255 by-60\advance\count255 by\time
      \xdef\hourmin{\hourmin:\ifnum\count255<10 0\fi\the\count255}}}
\def\nolabels{\def\wrlabeL##1{}\def\eqlabeL##1{}\def\reflabeL##1{}}
\def\writelabels{\def\wrlabeL##1{\leavevmode\vadjust{\rlap{\smash%
        {\line{{\escapechar=` \hfill\rlap{\sevenrm\hskip.03in\string##1}}}}}}}%
  \def\eqlabeL##1{{\escapechar-1\rlap{\sevenrm\hskip.05in\string##1}}}%
  \def\reflabeL##1{\noexpand\llap{\noexpand\sevenrm\string\string\string##1}}}
\nolabels
%
\global\newcount\secno \global\secno=0
\global\newcount\meqno \global\meqno=1
\def\s@csym{}
\def\newsec#1{\global\advance\secno by1%
    {\toks0{#1}\message{(\the\secno. \the\toks0)}}%
  \global\subsecno=0\eqnres@t\let\s@csym\secsym\xdef\secn@m{\the\secno}\noindent
  {\bf\hyperdef\hypernoname{section}{\the\secno}{\the\secno.} #1}%
  \writetoca{{\string\hyperref{}{section}{\the\secno}{\the\secno.}} {#1}}%
  \par\nobreak\medskip\nobreak}
\def\eqnres@t{\xdef\secsym{\the\secno.}\global\meqno=1\bigbreak\bigskip}
\def\sequentialequations{\def\eqnres@t{\bigbreak}}\xdef\secsym{}
\global\newcount\subsecno \global\subsecno=0
\def\subsec#1{\global\advance\subsecno by1%
    {\toks0{#1}\message{(\s@csym\the\subsecno. \the\toks0)}}%
  \ifnum\lastpenalty>9000\else\bigbreak\fi
  \noindent{\it\hyperdef\hypernoname{subsection}{\secn@m.\the\subsecno}%
    {\secn@m.\the\subsecno.} #1}\writetoca{\string\quad
    {\string\hyperref{}{subsection}{\secn@m.\the\subsecno}{\secn@m.\the\subsecno.}}
    {#1}}\par\nobreak\medskip\nobreak}
\def\appendix#1#2{\global\meqno=1\global\subsecno=0\xdef\secsym{\hbox{#1.}}%
  \bigbreak\bigskip\noindent{\bf Appendix \hyperdef\hypernoname{appendix}{#1}%
    {#1.} #2}{\toks0{(#1. #2)}\message{\the\toks0}}%
  \xdef\s@csym{#1.}\xdef\secn@m{#1}%
  \writetoca{\string\hyperref{}{appendix}{#1}{Appendix {#1.}} {#2}}%
  \par\nobreak\medskip\nobreak}
%
%
\def\checkm@de#1#2{\ifmmode{\def\f@rst##1{##1}\hyperdef\hypernoname{equation}%
    {#1}{#2}}\else\hyperref{}{equation}{#1}{#2}\fi}
\def\eqnn#1{\DefWarn#1\xdef #1{(\noexpand\relax\noexpand\checkm@de%
    {\s@csym\the\meqno}{\secsym\the\meqno})}%
  \wrlabeL#1\writedef{#1\leftbracket#1}\global\advance\meqno by1}
\def\f@rst#1{\c@t#1a\em@ark}\def\c@t#1#2\em@ark{#1}
\def\eqna#1{\DefWarn#1\wrlabeL{#1$\{\}$}%
  \xdef #1##1{(\noexpand\relax\noexpand\checkm@de%
    {\s@csym\the\meqno\noexpand\f@rst{##1}}{\hbox{$\secsym\the\meqno##1$}})}
  \writedef{#1\numbersign1\leftbracket#1{\numbersign1}}\global\advance\meqno by1}
\def\eqn#1#2{\DefWarn#1%
  \xdef #1{(\noexpand\hyperref{}{equation}{\s@csym\the\meqno}%
    {\secsym\the\meqno})}$$#2\eqno(\hyperdef\hypernoname{equation}%
    {\s@csym\the\meqno}{\secsym\the\meqno})\eqlabeL#1$$%
  \writedef{#1\leftbracket#1}\global\advance\meqno by1}
\def\xeqn{\expandafter\xe@n}\def\xe@n(#1){#1}
\def\xeqna#1{\expandafter\xe@n#1}
\def\eqns#1{(\e@ns #1{\hbox{}})}
\def\e@ns#1{\ifx\UNd@FiNeD#1\message{eqnlabel \string#1 is undefined.}%
    \xdef#1{(?.?)}\fi{\let\hyperref=\relax\xdef\next{#1}}%
  \ifx\next\em@rk\def\next{}\else%
    \ifx\next#1\xeqn#1\else\def\n@xt{#1}\ifx\n@xt\next#1\else\xeqna#1\fi
    \fi\let\next=\e@ns\fi\next}

\def\DefWarn#1{\ifx\UNd@FiNeD#1\else
    \immediate\write16{*** WARNING: the label \string#1 is already defined ***}\fi}
%
\newskip\footskip\footskip14pt plus 1pt minus 1pt 
\def\footnotefont{\ninepoint}\def\f@t#1{\footnotefont #1\@foot}
\def\f@@t{\baselineskip\footskip\bgroup\footnotefont\aftergroup\@foot\let\next}
\setbox\strutbox=\hbox{\vrule height9.5pt depth4.5pt width0pt}
\global\newcount\ftno \global\ftno=0
\def\foot{\global\advance\ftno by1\def\foot@rg{\hyperref{}{footnote}%
    {\the\ftno}{\the\ftno}\xdef\foot@rg{\noexpand\hyperdef\noexpand\hypernoname%
      {footnote}{\the\ftno}{\the\ftno}}}\footnote{$^{\foot@rg}$}}
%
\newwrite\ftfile
\def\footend{\def\foot{\global\advance\ftno by1\chardef\wfile=\ftfile
    \hyperref{}{footnote}{\the\ftno}{$^{\the\ftno}$}%
    \ifnum\ftno=1\immediate\openout\ftfile=\jobname.fts\fi%
    \immediate\write\ftfile{\noexpand\smallskip%
      \noexpand\item{\noexpand\hyperdef\noexpand\hypernoname{footnote}
                  {\the\ftno}{f\the\ftno}:\ }\pctsign}\findarg}%
  \def\footatend{\vfill\eject\immediate\closeout\ftfile{\parindent=20pt
      \centerline{\bf Footnotes}\nobreak\bigskip\input \jobname.fts }}}
\def\footatend{}
%
%
\global\newcount\refno \global\refno=1
\newwrite\rfile
\def\ref{[\hyperref{}{reference}{\the\refno}{\the\refno}]\nref}
\def\nref#1{\DefWarn#1%
  \xdef#1{[\noexpand\hyperref{}{reference}{\the\refno}{\the\refno}]}%
  \writedef{#1\leftbracket#1}%
  \ifnum\refno=1\immediate\openout\rfile=\jobname.refs\fi
  \chardef\wfile=\rfile\immediate\write\rfile{\noexpand\item{[\noexpand\hyperdef%
                  \noexpand\hypernoname{reference}{\the\refno}{\the\refno}]\ }%
    \reflabeL{#1\hskip.31in}\pctsign}\global\advance\refno by1\findarg}
\def\findarg#1#{\begingroup\obeylines\newlinechar=`\^^M\pass@rg}
{\obeylines\gdef\pass@rg#1{\writ@line\relax #1^^M\hbox{}^^M}%
  \gdef\writ@line#1^^M{\expandafter\toks0\expandafter{\striprel@x #1}%
      \edef\next{\the\toks0}\ifx\next\em@rk\let\next=\endgroup\else\ifx\next\empty%
        \else\immediate\write\wfile{\the\toks0}\fi\let\next=\writ@line\fi\next\relax}}
\def\striprel@x#1{} \def\em@rk{\hbox{}}
\def\lref{\begingroup\obeylines\lr@f}
\def\lr@f#1#2{\DefWarn#1\gdef#1{\let#1=\UNd@FiNeD\ref#1{#2}}\endgroup\unskip}

\def\addref#1{\immediate\write\rfile{\noexpand\item{}#1}} 
\def\listrefs{\footatend\vfill\supereject\immediate\closeout\rfile\writestoppt
  \baselineskip=\footskip\centerline{{\bf References}}\bigskip{\parindent=20pt%
    \frenchspacing\escapechar=` \input \jobname.refs\vfill\eject}\nonfrenchspacing}
\def\startrefs#1{\immediate\openout\rfile=\jobname.refs\refno=#1}
\def\xref{\expandafter\xr@f}\def\xr@f[#1]{#1}
\def\refs#1{\count255=1[\r@fs #1{\hbox{}}]}
\def\r@fs#1{\ifx\UNd@FiNeD#1\message{reflabel \string#1 is undefined.}%
    \nref#1{need to supply reference \string#1.}\fi%
  \vphantom{\hphantom{#1}}{\let\hyperref=\relax\xdef\next{#1}}%
  \ifx\next\em@rk\def\next{}%
  \else\ifx\next#1\ifodd\count255\relax\xref#1\count255=0\fi%
    \else#1\count255=1\fi\let\next=\r@fs\fi\next}
%

%
\newwrite\ffile\global\newcount\figno \global\figno=1
\def\fig{fig.~\hyperref{}{figure}{\the\figno}{\the\figno}\nfig}
\def\nfig#1{\DefWarn#1%
  \xdef#1{fig.~\noexpand\hyperref{}{figure}{\the\figno}{\the\figno}}%
  \writedef{#1\leftbracket fig.\noexpand~\xfig#1}%
  \ifnum\figno=1\immediate\openout\ffile=\jobname.figs\fi\chardef\wfile=\ffile%
  {\let\hyperref=\relax
    \immediate\write\ffile{\noexpand\medskip\noexpand\item{Fig.\ %
                  \noexpand\hyperdef\noexpand\hypernoname{figure}{\the\figno}{\the\figno}. }
      \reflabeL{#1\hskip.55in}\pctsign}}\global\advance\figno by1\findarg}
\def\listfigs{\vfill\eject\immediate\closeout\ffile{\parindent40pt
    \baselineskip14pt\centerline{{\bf Figure Captions}}\nobreak\medskip
    \escapechar=` \input \jobname.figs\vfill\eject}}
\def\xfig{\expandafter\xf@g}\def\xf@g fig.\penalty\@M\ {}
\def\figs#1{figs.~\f@gs #1{\hbox{}}}
\def\f@gs#1{{\let\hyperref=\relax\xdef\next{#1}}\ifx\next\em@rk\def\next{}\else
    \ifx\next#1\xfig #1\else#1\fi\let\next=\f@gs\fi\next}
\def\figin{\epsfcheck\figin}\def\figins{\epsfcheck\figins}
\def\epsfcheck{\ifx\epsfbox\UNd@FiNeD
    \message{(NO epsf.tex, FIGURES WILL BE IGNORED)}
    \gdef\figin##1{\vskip2in}\gdef\figins##1{\hskip.5in}
  \else\message{(FIGURES WILL BE INCLUDED)}%
    \gdef\figin##1{##1}\gdef\figins##1{##1}\fi}
\def\DefWarn#1{}
\def\figinsert{\goodbreak\midinsert}
\def\ifig#1#2#3{\DefWarn#1\xdef#1{fig.~\noexpand\hyperref{}{figure}%
    {\the\figno}{\the\figno}}\writedef{#1\leftbracket fig.\noexpand~\xfig#1}%
  \figinsert\figin{\centerline{#3}}\medskip\centerline{\vbox{\baselineskip12pt
      \advance\hsize by -1truein\noindent\wrlabeL{#1=#1}\footnotefont%
      {\bf Fig.~\hyperdef\hypernoname{figure}{\the\figno}{\the\figno}:} #2}}
  \bigskip\endinsert\global\advance\figno by1}
\newwrite\lfile
{\escapechar-1\xdef\pctsign{\string\%}\xdef\leftbracket{\string\{}
  \xdef\rightbracket{\string\}}\xdef\numbersign{\string\#}}
\def\writedefs{\immediate\openout\lfile=\jobname.defs \def\writedef##1{%
    {\let\hyperref=\relax\let\hyperdef=\relax\let\hypernoname=\relax
      \immediate\write\lfile{\string\def\string##1\rightbracket}}}}%
\def\writestop{\def\writestoppt{\immediate\write\lfile{\string\pageno
      \the\pageno\string\startrefs\leftbracket\the\refno\rightbracket
      \string\def\string\secsym\leftbracket\secsym\rightbracket
      \string\secno\the\secno\string\meqno\the\meqno}\immediate\closeout\lfile}}
\def\writestoppt{}\def\writedef#1{}
\def\seclab#1{\DefWarn#1%
  \xdef #1{\noexpand\hyperref{}{section}{\the\secno}{\the\secno}}%
  \writedef{#1\leftbracket#1}\wrlabeL{#1=#1}}
\def\subseclab#1{\DefWarn#1%
  \xdef #1{\noexpand\hyperref{}{subsection}{\secn@m.\the\subsecno}%
    {\secn@m.\the\subsecno}}\writedef{#1\leftbracket#1}\wrlabeL{#1=#1}}
\def\applab#1{\DefWarn#1%
  \xdef #1{\noexpand\hyperref{}{appendix}{\secn@m}{\secn@m}}%
  \writedef{#1\leftbracket#1}\wrlabeL{#1=#1}}
\newwrite\tfile \def\writetoca#1{}
\def\leaderfill{\leaders\hbox to 1em{\hss.\hss}\hfill}
\def\writetoc{\immediate\openout\tfile=\jobname.toc
  \def\writetoca##1{{\edef\next{\write\tfile{\noindent ##1
          \string\leaderfill {\string\hyperref{}{page}{\noexpand\number\pageno}%
            {\noexpand\number\pageno}} \par}}\next}}}
\newread\ch@ckfile
\def\listtoc{\immediate\closeout\tfile\immediate\openin\ch@ckfile=\jobname.toc
  \ifeof\ch@ckfile\message{no file \jobname.toc, no table of contents this pass}%
  \else\closein\ch@ckfile\centerline{\bf Contents}\nobreak\medskip%
    {\baselineskip=12pt\footnotefont\parskip=0pt\catcode`\@=11\input\jobname.toc
      \catcode`\@=12\bigbreak\bigskip}\fi}
\catcode`\@=12 
%
\edef\tfontsize{\ifx\answ\bigans scaled\magstep3\else scaled\magstep4\fi}
\font\titlerm=cmr10 \tfontsize \font\titlerms=cmr7 \tfontsize
\font\titlermss=cmr5 \tfontsize \font\titlei=cmmi10 \tfontsize
\font\titleis=cmmi7 \tfontsize \font\titleiss=cmmi5 \tfontsize
\font\titlesy=cmsy10 \tfontsize \font\titlesys=cmsy7 \tfontsize
\font\titlesyss=cmsy5 \tfontsize \font\titleit=cmti10 \tfontsize
\skewchar\titlei='177 \skewchar\titleis='177 \skewchar\titleiss='177
\skewchar\titlesy='60 \skewchar\titlesys='60 \skewchar\titlesyss='60
\def\titlefont{\def\rm{\fam0\titlerm}
  \textfont0=\titlerm \scriptfont0=\titlerms \scriptscriptfont0=\titlermss
  \textfont1=\titlei \scriptfont1=\titleis \scriptscriptfont1=\titleiss
  \textfont2=\titlesy \scriptfont2=\titlesys \scriptscriptfont2=\titlesyss
  \textfont\itfam=\titleit \def\it{\fam\itfam\titleit}\rm}
 \ifx\answ\bigans\else scaled\magstep1\fi
\ifx\answ\bigans\def\abstractfont{\tenpoint}\else
  \font\absit=cmti10 scaled \magstep1
  \font\abssl=cmsl10 scaled \magstep1
  \font\absrm=cmr10 scaled\magstep1 \font\absrms=cmr7 scaled\magstep1
  \font\absrmss=cmr5 scaled\magstep1 \font\absi=cmmi10 scaled\magstep1
  \font\absis=cmmi7 scaled\magstep1 \font\absiss=cmmi5 scaled\magstep1
  \font\abssy=cmsy10 scaled\magstep1 \font\abssys=cmsy7 scaled\magstep1
  \font\abssyss=cmsy5 scaled\magstep1 \font\absbf=cmbx10 scaled\magstep1
  \skewchar\absi='177 \skewchar\absis='177 \skewchar\absiss='177
  \skewchar\abssy='60 \skewchar\abssys='60 \skewchar\abssyss='60
  \def\abstractfont{\def\rm{\fam0\absrm}
    \textfont0=\absrm \scriptfont0=\absrms \scriptscriptfont0=\absrmss
    \textfont1=\absi \scriptfont1=\absis \scriptscriptfont1=\absiss
    \textfont2=\abssy \scriptfont2=\abssys \scriptscriptfont2=\abssyss
    \textfont\itfam=\absit \def\it{\fam\itfam\absit}\def\footnotefont{\tenpoint}%
    \textfont\slfam=\abssl \def\sl{\fam\slfam\abssl}%
    \textfont\bffam=\absbf \def\bf{\fam\bffam\absbf}\rm}\fi
\def\tenpoint{\def\rm{\fam0\tenrm}
  \textfont0=\tenrm \scriptfont0=\sevenrm \scriptscriptfont0=\fiverm
  \textfont1=\teni  \scriptfont1=\seveni  \scriptscriptfont1=\fivei
  \textfont2=\tensy \scriptfont2=\sevensy \scriptscriptfont2=\fivesy
  \textfont\itfam=\tenit \def\it{\fam\itfam\tenit}\def\footnotefont{\ninepoint}%
  \textfont\bffam=\tenbf \def\bf{\fam\bffam\tenbf}\def\sl{\fam\slfam\tensl}\rm}
\font\ninerm=cmr9 \font\sixrm=cmr6 \font\ninei=cmmi9 \font\sixi=cmmi6
\font\ninesy=cmsy9 \font\sixsy=cmsy6 \font\ninebf=cmbx9
\font\nineit=cmti9 \font\ninesl=cmsl9 \skewchar\ninei='177
\skewchar\sixi='177 \skewchar\ninesy='60 \skewchar\sixsy='60
\def\ninepoint{\def\rm{\fam0\ninerm}
  \textfont0=\ninerm \scriptfont0=\sixrm \scriptscriptfont0=\fiverm
  \textfont1=\ninei \scriptfont1=\sixi \scriptscriptfont1=\fivei
  \textfont2=\ninesy \scriptfont2=\sixsy \scriptscriptfont2=\fivesy
  \textfont\itfam=\ninei \def\it{\fam\itfam\nineit}\def\sl{\fam\slfam\ninesl}%
  \textfont\bffam=\ninebf \def\bf{\fam\bffam\ninebf}\rm}
%
%

\hyphenation{anom-aly anom-alies coun-ter-term coun-ter-terms}
\def\inv{^{\raise.15ex\hbox{${\scriptscriptstyle -}$}\kern-.05em 1}}

\def\Dsl{\,\raise.15ex\hbox{/}\mkern-13.5mu D} 
\def\dsl{\raise.15ex\hbox{/}\kern-.57em\partial}

 \def\Tr{{\rm Tr}}
\def\lspace{\ifx\answ\bigans{}\else\qquad\fi}
\def\lbspace{\ifx\answ\bigans{}\else\hskip-.2in\fi} 
\def\boxeqn#1{\vcenter{\vbox{\hrule\hbox{\vrule\kern3pt\vbox{\kern3pt
          \hbox{${\displaystyle #1}$}\kern3pt}\kern3pt\vrule}\hrule}}}
\def\mbox#1#2{\vcenter{\hrule \hbox{\vrule height#2in
      \kern#1in \vrule} \hrule}}  
%
 \def\CO{{\cal O}} 
   \def\CG{{\cal G}}
  \def\CI{{\cal I}} 
  \def\CD{{\cal D}}

\def\darr#1{\raise1.5ex\hbox{$\leftrightarrow$}\mkern-16.5mu #1}

\def\roughly#1{\raise.3ex\hbox{$#1$\kern-.75em\lower1ex\hbox{$\sim$}}}

\input epsf
\newif\ifdraft\draftfalse
\newif\ifinter\interfalse
\ifdraft\draftmode\else\interfalse\fi
\def\journal#1&#2(#3){\unskip, \sl #1\ \bf #2 \rm(19#3) }
\def\andjournal#1&#2(#3){\sl #1~\bf #2 \rm (19#3) }

\def\S{\hbox{$\bb S$}}

\def\frac#1#2{{#1\over#2}}

\def\inbar{\,\vrule height1.5ex width.4pt depth0pt}
\def\IC{\relax\hbox{$\inbar\kern-.3em{\rm C}$}}
\def\IR{\relax{\rm I\kern-.18em R}}
\def\IP{\relax{\rm I\kern-.18em P}}


%
\catcode`\@=11
\def\slash#1{\mathord{\mathpalette\c@ncel{#1}}}
\overfullrule=0pt

%
\def\draftnote#1{\ifdraft{\baselineskip2ex
    \vbox{\kern1em\hrule\hbox{\vrule\kern1em\vbox{\kern1ex
          \noindent \underbar{NOTE}: #1
          \vskip1ex}\kern1em\vrule}\hrule}}\fi}
\def\internote#1{\ifinter{\baselineskip2ex
    \vbox{\kern1em\hrule\hbox{\vrule\kern1em\vbox{\kern1ex
          \noindent \underbar{Internal Note}: #1
          \vskip1ex}\kern1em\vrule}\hrule}}\fi}


\newfam\frakfam
\font\teneufm=eufm10
\font\seveneufm=eufm7
\font\fiveeufm=eufm5
\textfont\frakfam=\teneufm
\scriptfont\frakfam=\seveneufm
\scriptscriptfont\frakfam=\fiveeufm
\def\frak{\fam\frakfam \teneufm}

\def\bb{
  \font\tenmsb=msbm10
  \font\sevenmsb=msbm7
  \font\fivemsb=msbm5
  \textfont1=\tenmsb
  \scriptfont1=\sevenmsb
  \scriptscriptfont1=\fivemsb
}

\newfam\bbfam
\font\tenmsb=msbm10
\font\sevenmsb=msbm7
\font\fivemsb=msbm5
\textfont\bbfam=\tenmsb
\scriptfont\bbfam=\sevenmsb
\scriptscriptfont\bbfam=\fivemsb
\def\bbC{{\fam\bbfam C}}

\def\bbT{{\fam\bbfam T}}
\def\bbZ{{\fam\bbfam Z}}
\def\bflambda{{\mb\mathchar"7115}}
\def\bfepsilon{{\mb\mathchar"710F}}

\def\mmod#1{ [\![  #1 ] \!] }

\def\Th{\theta}
\def\bbr#1{\mmod{#1}}
\def\Res{\mathop{\rm Res}}
\def\qed{\hfill\vrule height6pt width5pt depth0pt\medskip}
\def\proof{\par\ifdim\lastskip<\bigskipamount\removelastskip\penalty300\bigskip\fi\noindent{\it Proof.}\ }
\def\parhead#1{\bigbreak\noindent{\it #1}\ }
\def\parsub#1{\par\medbreak\noindent{\it #1}\ }
\def\CO{{\cal O}}
\def\CI{{\cal I}}
\def\CP{{\cal P}}
\def\CK{{\cal K}}
\def\CD{{\cal D}}

\def\tGamma{\widetilde\Gamma}
\def\tE{\widetilde\Theta}
\def\ord{\mathop{\rm ord}}
\def\emph#1{{\it #1\/}}
\def\binom#1#2{{#1\choose #2}}


\lref\RazamatYamazaki{
  S.~S.~Razamat and M.~Yamazaki,
  ``S-duality and the N=2 Lens Space Index,''
  JHEP {\bf 1310}, 048 (2013).
    [arXiv:1306.1543 [hep-th]].
}

\lref\MOY{
  K.~Maruyoshi, T.~Ota and M.~Yamazaki,
  ``Surface defects on lens space and transfer matrices,''
  [arXiv:2610.28165 [hep-th]].
}

\lref\RastelliRazamatReview{
  L.~Rastelli and S.~S.~Razamat,
  ``The superconformal index of theories of class S,''
  in {\it New Dualities of Supersymmetric Gauge Theories}, ed.\ J.~Teschner, Springer (2016), 261--305.
    [arXiv:1412.7131 [hep-th]].
}

\lref\AldayBFH{
  L.~F.~Alday, M.~Bullimore, M.~Fluder and L.~Hollands,
  ``Surface defects, the superconformal index and q-deformed Yang-Mills,''
  JHEP {\bf 1310}, 018 (2013).
    [arXiv:1303.4460 [hep-th]].
}

\lref\BullimoreFHR{
  M.~Bullimore, M.~Fluder, L.~Hollands and P.~Richmond,
  ``The superconformal index and an elliptic algebra of surface defects,''
  JHEP {\bf 1410}, 062 (2014).
    [arXiv:1401.3379 [hep-th]].
}

\lref\MaruyoshiYagi{
  K.~Maruyoshi and J.~Yagi,
  ``Surface defects as transfer matrices,''
  PTEP {\bf 2016}, 113B01 (2016).
    [arXiv:1606.01041 [hep-th]].
}

\lref\GaiottoRazamatSk{
  D.~Gaiotto and S.~S.~Razamat,
  ``${\cal N}=1$ theories of class ${\cal S}_k$,''
  JHEP {\bf 1507}, 073 (2015).
    [arXiv:1503.05159 [hep-th]].
}

\lref\KelsYamazakiLens{
  A.~P.~Kels and M.~Yamazaki,
  ``Elliptic hypergeometric sum/integral transformations and supersymmetric lens index,''
  SIGMA {\bf 14}, 013 (2018).
    [arXiv:1704.03159 [math-ph]].
}

\lref\Sklyanin{
  E.~K.~Sklyanin,
  ``Some algebraic structures connected with the Yang-Baxter equation,''
  Funct.\ Anal.\ Appl.\ {\bf 16}, 263 (1982).
}

\lref\Hasegawa{
  K.~Hasegawa,
  ``Ruijsenaars' commuting difference operators as commuting transfer matrices,''
  Commun.\ Math.\ Phys.\ {\bf 187}, 289 (1997).
    [arXiv:q-alg/9512029].
}

\lref\CherednikSklyanin{
  I.~V.~Cherednik,
  ``Some finite-dimensional representations of generalized Sklyanin algebras,''
  Funktsional.\ Anal.\ i Prilozhen.\ {\bf 19}, no.~1, 89--90 (1985).
}

\lref\NieriPasquetti{
  F.~Nieri and S.~Pasquetti,
  ``Factorisation and holomorphic blocks in 4d,''
  JHEP {\bf 1511}, 155 (2015).
    [arXiv:1507.00261 [hep-th]].
}

\lref\AharonyST{
  O.~Aharony, N.~Seiberg and Y.~Tachikawa,
  ``Reading between the lines of four-dimensional gauge theories,''
  JHEP {\bf 1308}, 115 (2013).
    [arXiv:1305.0318 [hep-th]].
}

\lref\FluderSong{
  M.~Fluder and J.~Song,
  ``Four-dimensional Lens Space Index from Two-dimensional Chiral Algebra,''
  JHEP {\bf 1807}, 073 (2018).
    [arXiv:1710.06029 [hep-th]].
}

\lref\AmaritiMarcassoli{
  A.~Amariti and A.~Marcassoli,
  ``Lens space index and global properties for 4d ${\cal N}=2$ models,''
  JHEP {\bf 2002}, 143 (2020).
    [arXiv:1911.13264 [hep-th]].
}

\lref\KelsStar{
  A.~P.~Kels,
  ``New solutions of the star-triangle relation with discrete and continuous spin variables,''
  J.\ Phys.\ A {\bf 48}, 435201 (2015).
    [arXiv:1504.07074 [math-ph]].
}

\lref\GahramanovKels{
  I.~Gahramanov and A.~P.~Kels,
  ``The star-triangle relation, lens partition function, and hypergeometric sum/integrals,''
  JHEP {\bf 1702}, 040 (2017).
    [arXiv:1610.09229 [math-ph]].
}

\lref\KelsYamazakiYBE{
  A.~P.~Kels and M.~Yamazaki,
  ``Lens elliptic gamma function solution of the Yang-Baxter equation at roots of unity,''
  J.\ Stat.\ Mech.\ {\bf 1802}, 023108 (2018).
    [arXiv:1709.07148 [math-ph]].
}

\lref\YamazakiQuivers{
  M.~Yamazaki,
  ``Quivers, YBE and 3-manifolds,''
  JHEP {\bf 1205}, 147 (2012).\hfil\break
  [arXiv:1203.5784 [hep-th]].
}

\lref\TerashimaYamazakiEmergent{
  Y.~Terashima and M.~Yamazaki,
  ``Emergent 3-manifolds from 4d Superconformal Indices,''
  Phys.\ Rev.\ Lett.\ {\bf 109}, 091602 (2012).
    [arXiv:1203.5792 [hep-th]].
}

\lref\YamazakiNewIntegrable{
  M.~Yamazaki,
  ``New Integrable Models from the Gauge/YBE Correspondence,''
  J.\ Statist.\ Phys.\ {\bf 154}, 895 (2014).
    [arXiv:1307.1128 [hep-th]].
}

\lref\YamazakiReview{
  M.~Yamazaki,
  ``Integrability As Duality: The Gauge/YBE Correspondence,''
  Phys.\ Rept.\ {\bf 859}, 1 (2020).
    [arXiv:1808.04374 [hep-th]].
}

\lref\RazamatEF{
  S.~S.~Razamat,
  ``On the N=2 superconformal index and eigenfunctions of the elliptic RS model,''
  Lett.\ Math.\ Phys.\ {\bf 104}, 673 (2014).
    [arXiv:1309.0278 [hep-th]].
}

\lref\RuijsenaarsSchneider{
  S.~N.~M.~Ruijsenaars and H.~Schneider,
  ``A New Class of Integrable Systems and Its Relation to Solitons,''
  Annals Phys.\ {\bf 170}, 370 (1986).
}

\lref\RuijsenaarsCMP{
  S.~N.~M.~Ruijsenaars,
  ``Complete Integrability of Relativistic Calogero-Moser Systems and Elliptic Function Identities,''
  Commun.\ Math.\ Phys.\ {\bf 110}, 191 (1987).
}

\lref\NoumiSano{
  M.~Noumi and A.~Sano,
  ``An infinite family of higher-order difference operators that commute with Ruijsenaars operators of type A,''
  Lett.\ Math.\ Phys.\ {\bf 111}, 91 (2021).
    [arXiv:2012.03135 [math-ph]].
}

\lref\SpiridonovRarefied{
  V.~P.~Spiridonov,
  ``Rarefied elliptic hypergeometric functions,''
  Adv.\ Math.\ {\bf 331}, 830 (2018).
    [arXiv:1609.00715 [math.CA]].
}

\lref\AldayRS{
  L.~F.~Alday, M.~Bullimore and M.~Fluder,
  ``On S-duality of the Superconformal Index on Lens Spaces and 2d TQFT,''
  JHEP {\bf 1305}, 122 (2013).
    [arXiv:1301.7486 [hep-th]].
}

\lref\BeniniNC{
  F.~Benini, T.~Nishioka and M.~Yamazaki,
  ``4d Index to 3d Index and 2d TQFT,''
  Phys.\ Rev.\ D {\bf 86}, 065015 (2012).
    [arXiv:1109.0283 [hep-th]].
}

\lref\GaddeKB{
  A.~Gadde, E.~Pomoni, L.~Rastelli and S.~S.~Razamat,
  ``S-duality and 2d Topological QFT,''
  JHEP {\bf 1003}, 032 (2010).
    [arXiv:0910.2225 [hep-th]].
}

\lref\GaddeUV{
  A.~Gadde, L.~Rastelli, S.~S.~Razamat and W.~Yan,
  ``Gauge Theories and Macdonald Polynomials,''
  Commun.\ Math.\ Phys.\  {\bf 319}, 147 (2013).
    [arXiv:1110.3740 [hep-th]].
}

\lref\GaiottoWE{
  D.~Gaiotto,
  ``N=2 dualities,''
  JHEP {\bf 1208}, 034 (2012).
    [arXiv:0904.2715 [hep-th]].
}

\lref\GaiottoHG{
  D.~Gaiotto, G.~W.~Moore and A.~Neitzke,
  ``Wall-crossing, Hitchin Systems, and the WKB Approximation,''
  Adv.\ Math.\ {\bf 234}, 239 (2013).
    [arXiv:0907.3987 [hep-th]].
}

\lref\GaiottoXA{
  D.~Gaiotto, L.~Rastelli and S.~S.~Razamat,
  ``Bootstrapping the superconformal index with surface defects,''
  JHEP {\bf 1301}, 022 (2013).
    [arXiv:1207.3577 [hep-th]].
}

\lref\KinneyEJ{
  J.~Kinney, J.~M.~Maldacena, S.~Minwalla and S.~Raju,
  ``An Index for 4 dimensional super conformal theories,''
  Commun.\ Math.\ Phys.\  {\bf 275}, 209 (2007).
    [hep-th/0510251].
}

\lref\RomelsbergerEG{
  C.~Romelsberger,
  ``Counting chiral primaries in N = 1, d=4 superconformal field theories,''
  Nucl.\ Phys.\ B {\bf 747}, 329 (2006).
    [hep-th/0510060].
}

\lref\WRLens{
  S.~S.~Razamat and B.~Willett,
  ``Global Properties of Supersymmetric Theories and the Lens Space,''
  Commun.\ Math.\ Phys.\ {\bf 334}, 661 (2015).
    [arXiv:1307.4381 [hep-th]].
}

\lref\SpiridonovZA{
  V.~P.~Spiridonov and G.~S.~Vartanov,
  ``Elliptic Hypergeometry of Supersymmetric Dualities,''
  Commun.\ Math.\ Phys.\  {\bf 304}, 797 (2011).
    [arXiv:0910.5944 [hep-th]].
}

\lref\DolanQI{
  F.~A.~Dolan and H.~Osborn,
  ``Applications of the Superconformal Index for Protected Operators and q-Hypergeometric Identities to N=1 Dual Theories,''
  Nucl.\ Phys.\ B {\bf 818}, 137 (2009).
    [arXiv:0801.4947 [hep-th]].
}

\lref\GaddeDDA{
  A.~Gadde and S.~Gukov,
  ``2d Index and Surface operators,''
  JHEP {\bf 1403}, 080 (2014).
    [arXiv:1305.0266 [hep-th]].
}

\lref\cherednik{
  I.~Cherednik,
  ``Double Affine Hecke Algebras,''
  London Mathematical Society Lecture Note Series, vol.~319,
  Cambridge University Press (2005).
}


\rightline{RIKEN-iTHEMS-Report-26}

\Title{
  \vbox{\baselineskip12pt \hbox{}}}
{\vbox{\centerline{Surface Defects in the ${\cal N}=2$ Lens Space Index}
    \vskip .1in
    \centerline{and Elliptic Integrability}}}

\centerline{Masahito Yamazaki$^{1,2,3,4,5}$}
\bigskip
{\baselineskip=13pt
  \centerline{$^{1}${\it Department of Physics, University of Tokyo, Tokyo 113-0033, Japan}}
  \vskip 3pt
  \centerline{$^{2}${\it Kavli Institute for the Physics and Mathematics of the Universe (Kavli IPMU),}}
  \centerline{\it UTIAS, University of Tokyo, Chiba 277-8583, Japan}
  \vskip 3pt
  \centerline{$^{3}${\it Trans-Scale Quantum Science Institute, University of Tokyo, Tokyo 113-0033, Japan}}
  \vskip 3pt
  \centerline{$^{4}${\it Center for Interdisciplinary Theoretical and Mathematical Sciences (iTHEMS),}}
  \centerline{\it RIKEN, Saitama 351-0198, Japan}
  \vskip 3pt
  \centerline{$^{5}${\it Center for Data-Driven Discovery (CD3),}}
  \centerline{\it Kavli IPMU, University of Tokyo, Chiba 277-8583, Japan}
  \par}

\vskip .4in
\centerline{\bf Abstract}

\noindent
We determine the surface-defect difference operators of the 4d ${\cal N}=2$ lens space index for class ${\cal S}$ theories of type $A_{N-1}$ on $(\S^3/{\bbZ}_r)\times\S^1$. For generic fugacities $(p,q,t)$ and arbitrary rank $N$, orbifold order $r$ and discrete holonomies, their coefficients are finite products of theta functions with nomes $p^r$ and $q^r$. After an explicit gauge transformation, Gaussian in the holonomies, each operator factorizes into two elliptic Ruijsenaars--Schneider-type operators with step $pq$, one with nome $p^r$ and one with nome $q^r$, acting on variables glued along a lattice. Using this factorization and a Wronski-type relation, we prove that all the operators commute. We also establish a sufficient criterion for self-adjointness with respect to the bilinear vector multiplet pairing. It applies to the basic mixed operator and, in suitable parameter chambers, to the pure $p$- and $q$-type operators; for general mixed operators the contour prescription remains open.
These results support the expectations from S-duality.

\vfill
\eject

\footline={\hss\tenrm\folio\hss}
\headline={\hyperdef\hypernoname{page}{\folio}{}\hfil}
\listtoc
\writetoc

\newsec{Introduction}

The supersymmetric index \refs{\KinneyEJ,\RomelsbergerEG,\DolanQI,\SpiridonovZA} of 4d ${\cal N}=2$ superconformal field theories
has been a useful probe of their strongly coupled dynamics.
For theories of class ${\cal S}$ \refs{\GaiottoWE,\GaiottoHG}, obtained by compactifying the 6d $(2,0)$ theory of type $A_{N-1}$
on a punctured Riemann surface ${\cal C}$, the index does not depend on the complex structure of ${\cal C}$
and hence defines a 2d topological QFT on ${\cal C}$ \GaddeKB\ (see \RastelliRazamatReview\ for a review).
S-duality then implies that the index can be computed in any pair-of-pants decomposition of ${\cal C}$.
As pointed out in \GaiottoXA, this constraint, combined with the analytic structure of the index,
determines the index completely.
The main tool is a set of difference operators, which compute the residues of the index in the fugacity of a
decoupled free hypermultiplet; physically, taking such a residue introduces a surface defect (see also \refs{\GaddeDDA,\AldayBFH,\BullimoreFHR}).
For the ordinary index these operators are the elliptic Ruijsenaars--Schneider (RS) operators
\refs{\RuijsenaarsSchneider,\RuijsenaarsCMP}, and their joint eigenfunctions, which reduce to Macdonald polynomials in the Macdonald limit \GaddeUV, diagonalize the index \refs{\GaiottoXA,\RazamatEF}.

The lens space index \BeniniNC\ is the twisted partition function on $\S^3/\bbZ_r\times \S^1$.
In addition to the fugacities, it depends on discrete $\bbZ_r$ holonomies for the gauge and flavor symmetries, and is therefore sensitive to the global structure of the gauge group \refs{\AharonyST,\WRLens,\AmaritiMarcassoli}.
It has also been studied in relation to 2d chiral algebras \FluderSong, and its building blocks give solutions of the Yang--Baxter equation \refs{\KelsStar,\GahramanovKels,\KelsYamazakiYBE}
in the context of the Gauge/YBE correspondence \refs{\YamazakiQuivers,\TerashimaYamazakiEmergent,\YamazakiNewIntegrable} (see \YamazakiReview\ for a review).
The analysis of \GaiottoXA\ was extended to the lens index in \RazamatYamazaki\ (see also \AldayRS),
where it was shown that the residues are computed by difference operators acting non-locally
on the lattice of discrete holonomies, i.e.\ by matrix-valued generalizations of the elliptic RS operators.
The explicit form of these operators was however obtained there only for the simplest pole, and otherwise in the limit $p\to0$.
S-duality also predicts that the operators commute with each other and are self-adjoint with respect to the vector multiplet measure,
but these properties were left unproven in \RazamatYamazaki.\foot{Footnote 4 of \RazamatYamazaki\ reads:
  ``We would like to urge the more mathematically-oriented readers to prove these properties explicitly.''}

In this paper we address these questions. Our results are summarized as follows.

We first determine the difference operator for all poles
$a^*=t^{1\over2}p^{U/N}q^{V/N}$, $U\equiv V$ mod $r$, for generic $(p,q,t)$ and arbitrary $N$, $r$ and holonomies (Theorem 3.1);
the other roots of the pole condition are related to these by the center of $SU(N)$.
The coefficients are given by a single formula as finite products of theta functions.
It contains the operators of \RazamatYamazaki\ as special cases, and reduces to the elliptic RS operators of \GaiottoXA\ for $r=1$.

We next prove a factorization theorem (Theorem 5.6).
After conjugation by an explicit function $\CG$ of the fugacities and the holonomies, Gaussian in the latter,
every operator becomes a product of two commuting operators of elliptic RS type with step $pq$:
one with nome $p^r$ acting on the variables $\xi_i=b_ip^{n_i}$,
and one with nome $q^r$ acting on $\eta_i=b_iq^{-n_i}$.
The lens system is therefore gauge-equivalent to two decoupled systems of elliptic RS type, glued along the lattice $\xi_i/\eta_i\in(pq)^{\bbZ}$.
The proof relates the lens index, up to Gaussian factors local to each puncture, to a ``lifted'' index built from the lens elliptic gamma function \refs{\BeniniNC,\KelsYamazakiLens} with the holonomy not reduced mod $r$ (see also \SpiridonovRarefied).

Using the factorization, we also prove that all the operators commute (Theorem 6.4).
The proof reduces to a Wronski-type relation among elliptic RS operators, which follows from an identity of Noumi and Sano \NoumiSano.

We moreover give a sufficient criterion for the operators to be self-adjoint with respect to the bilinear pairing defined by the vector multiplet measure,
under explicit analytic conditions on $(p,q)$ and the test functions (Theorem 7.3).
The proof combines a detailed-balance identity, which follows from the exchange symmetry of the bifundamental,
with a description of the poles of the kernel.
The criterion ensures that each term can be treated by a contour deformation without crossing poles.
For $r\ge 2$, for instance, the sufficient conditions for the operators with $(U,V)=(0,r)$ and $(r,0)$ exclude each other.
For the operators whose conditions hold in some parameter chamber, which include those with $(U,V)=(1,1)$, $(0,V)$ and $(U,0)$,
the identity extends by analytic continuation of the matrix elements along paths where this continuation exists; this gives an operator-dependent contour prescription.
For general mixed operators such a starting chamber need not exist: we give an example with $(U,V)=(2,2)$ in section 7.4.
Their self-adjointness requires an additional argument.

Surface defects on the lens space are also studied in the recent paper \MOY, for theories of type $A_1$, including the ${\cal N}=1$ theories of class ${\cal S}_k$ \GaiottoRazamatSk.
There the holonomy $\mu\in\bbZ_r$ of the $U(1)$ symmetry whose fugacity is tuned to the pole, set to zero in \RazamatYamazaki, is also turned on.
This leads to elementary defects with $\mu\neq0$, which generate all the other operators and are traces of L-operators of Sklyanin algebras \Sklyanin, as for $r=1$ \MaruyoshiYagi;
the $p$- and $q$-type algebras together form a lens space version of the elliptic modular double.
In the present paper we keep $\mu=0$ as in \RazamatYamazaki, but treat general $N$ and all poles.
The relation between the two papers is discussed in sections 4, 5 and 6.

This paper is organized as follows. In section 2 we review the lens index and set up our notation.
In section 3 we derive the residue formula. Section 4 discusses special cases and the comparison with \RazamatYamazaki.
Section 5 proves the factorization theorem, section 6 proves commutativity, and section 7 establishes the self-adjointness criterion and discusses its limitations.
We conclude in section 8 with some open questions. Appendix A contains the proof of Proposition 5.8.

\newsec{Lens Space Index}

We briefly review the 4d ${\cal N}=2$ lens index~\BeniniNC, in the conventions of \RazamatYamazaki.
The lens space $L(r,1)=\S^3/\bbZ_r$ is defined by the identification
$(z_1,z_2)\sim (e^{2\pi i/r}z_1,e^{-2\pi i/r}z_2)$ on $|z_1|^2+|z_2|^2=1$, which acts on the Hopf fiber.
The lens index is
\eqn\Idef{
  {\cal I}(p,q,t;a):={\Tr} \left[
    (-1)^F\,\left(\frac{t}{pq}\right)^{{\sf r}}\,
    p^{j_{2}+j_1}\,
    q^{j_{2}-j_1}\,
    t^{{\sf R}}\,
    \prod_i a_i^{f_i}
    \right]\,,
}
where the trace is over the Hilbert space on $\S^3/\bbZ_r$. Here $j_1, j_2$ are the Cartans of $SU(2)_1\times SU(2)_2$,
${\sf R}$ is the Cartan of $SU(2)_R$, ${\sf r}$ is the $U(1)_r$ charge, and $f_i$ are flavor charges.
We assume $0<|p|,|q|,|t|<1$ and $|t|>|pq|$.
We fix branches of $\log p$, $\log q$ and $\log t$ once and for all, and define all fractional powers such as $p^{1/N}$, $t^{1\over2}$ and $(pq/t)^{1/r}$ through them.
Unless stated otherwise, $p$ and $q$ are generic; in particular, we assume $p^a q^b\ne1$ for every $(a,b)\in\bbZ^2\setminus\{(0,0)\}$, and take the other fugacities away from poles of the displayed expressions.
This excludes collisions of the gamma-function poles and resonances of the difference steps with the theta-function periods.
The algebraic identities below are identities of meromorphic coefficients and extend wherever these coefficients are finite.
Residues at colliding poles, and contour integrals across singularities, require separate prescriptions.
Since $\pi_1(\S^3/\bbZ_r)=\bbZ_r$, we can turn on discrete holonomies for the gauge and flavor symmetries.
For $SU(N)$ the holonomy is ${\rm diag}(e^{2\pi i m_1/r},\dots,e^{2\pi i m_N/r})$ with $m_i\in\bbZ_r$ and $\sum_i m_i\equiv0$ mod $r$.
We indicate the holonomies of an index by subscripts, as in $\CI_{\bf m}({\bf z},\dots)$ below.

\subsec{\it Notation}

For an integer $k$ we denote by $\bbr{k}\in\{0,\dots,r-1\}$ its residue mod $r$, and we set
\eqn\fdef{
  f(k):=\bbr{k}\,(r-\bbr{k})\,.
}
We use $\Th(x;P):=(x;P)_\infty(P/x;P)_\infty$, the constant $\kappa:=(p^r;p^r)_\infty(q^r;q^r)_\infty$, and the elliptic gamma function
\eqn\ellgamma{
  \Gamma(x;P,Q):=\prod_{i,j\ge0}
  \frac{(1-x^{-1}P^{i+1}Q^{j+1})}{(1-xP^iQ^j)}\,,
}
which satisfies
\eqn\Gammaqp{
  \Gamma(xQ;P,Q)=\Th(x;P)\,\Gamma(x;P,Q)\,,\qquad \Gamma(xP;P,Q)=\Th(x;Q)\,\Gamma(x;P,Q)\,.
}
The basic building block of the lens index is the contribution of a chiral multiplet
with holonomy charge $k\in\bbZ$~\refs{\BeniniNC,\RazamatYamazaki},
\eqn\Gammak{
  \Gamma_k(x):=\Gamma(xp^{\bbr{k}};pq,p^r)\,\Gamma(xq^{r-\bbr{k}};pq,q^r)
  =\prod_{i,j\ge0}\frac{(1-x^{-1}p^{i+1}q^{j+1})^{[i-j\equiv -k]}}{(1-x\,p^iq^j)^{[i-j\equiv k]}}\,,
}
where $[\,\cdot\,]$ equals $1$ if the condition in the bracket holds and $0$ otherwise (here a congruence mod $r$; we use the same notation for equalities and inequalities below).
This is the lens elliptic gamma function \refs{\BeniniNC,\KelsYamazakiLens}. It satisfies
$\Gamma_k(x)\Gamma_{-k}(pq/x)=1$. It has simple poles at $x=p^{-u}q^{-v}$ with $u,v\ge0$ and $u-v\equiv k$,
and simple zeros at $x=p^{u}q^{v}$ with $u,v\ge1$ and $u-v\equiv -k$.

\subsec{\it The shift lemma}

The main technical tool is the following quasi-periodicity of $\Gamma_k$.

\proclaim Lemma 2.1 (Shift lemma). For $u,v\ge0$ set $X:=p^uq^v$ and $w:=u-v$. Then
$$
  \Gamma_{k-w}(Xx)=\Gamma_k(x)\,\Theta_{u,v}(k;x)\,,
$$
where
\eqn\shift{
  \Theta_{u,v}(k;x):=\prod_{\alpha=0}^{u-1}\Th\big(x\,p^{\alpha}q^{\bbr{\alpha-k}};q^r\big)
  \prod_{\beta=0}^{v-1}\Th\big(x\,p^{u+\bbr{k-u+\beta}}q^{\beta};p^r\big)\,.
}
Moreover,
\eqn\limit{\lim_{x\to X^{-1}}(1-xX)\,\Gamma_w(x)=\big(\kappa\,\Theta_{u,v}(w;X^{-1})\big)^{-1}\,.
}

\proof From the product formula \Gammak\ one finds the two elementary relations
$\Gamma_{k-1}(px)=\Th(xq^{\bbr{-k}};q^r)\,\Gamma_k(x)$ and $\Gamma_{k+1}(qx)=\Th(xp^{\bbr{k}};p^r)\,\Gamma_k(x)$.
Starting from $\Gamma_k(x)$, apply the first relation $u$ times (with $x\to px\to\cdots\to p^ux$) and then the second $v$ times (with $p^ux\to p^uqx\to\cdots\to p^uq^vx$).
The $\alpha$-th and $\beta$-th steps produce the corresponding factors of \shift.
For the residue, write $\Gamma_w(x)=\Gamma_{0}(Xx)/\Theta_{u,v}(w;x)$, which follows from \shift\ with $k=w$.
Then use $\lim_{y\to1}(1-y)\Gamma_0(y)=1/\kappa$.
This holds because $\Gamma_0(y)=\Gamma(y;pq,p^r)\Gamma(yq^r;pq,q^r)$ and, as $y\to1$, only the factor $1/(1-y)$ of the first gamma function is singular. The remaining products give $1/((p^r;p^r)_\infty(q^r;q^r)_\infty)$.
\qed

\subsec{\it Building blocks}

The lens index of the bifundamental hypermultiplet of $SU(N)_{\bf z}\times SU(N)_{\bf b}\times U(1)_a$,
with holonomies ${\bf m}$ and ${\bf n}$ for the two $SU(N)$'s, is
\eqn\IH{
  I_H^{({\bf m},{\bf n})}(a,{\bf b},{\bf z}):=\left(\frac{pq}{t}\right)^{\frac1{2r}\sum_{i,j}f(m_i+n_j)}
  \prod_{i,j=1}^N\Gamma_{m_i+n_j}(t^{\frac12}z_ib_ja)\,\Gamma_{-m_i-n_j}(t^{\frac12}z_i^{-1}b_j^{-1}a^{-1})\,.
}
The prefactor is the zero-point contribution. The index of the $SU(N)$ vector multiplet is
\eqn\IV{
  I_V^{({\bf m})}({\bf z}):=\left(\frac{pq}{t}\right)^{-\frac1r\sum_{i<j}f(m_i-m_j)}
  \left(\frac{\kappa}{\Gamma_0(t)}\right)^{N-1}\prod_{i\neq j}\frac{1}{\Gamma_{m_i-m_j}(z_i/z_j)\,\Gamma_{m_i-m_j}(tz_i/z_j)}\,,
}
The Haar measure of the unbroken gauge group is contained in the factors $1/\Gamma_0(z_i/z_j)$ with $m_i\equiv m_j$.
Gauging the diagonal $SU(N)$ of the trinion and of a theory $T[{\cal C}]$ with index $\CI_{\bf m}({\bf z},\dots)$ gives
\eqn\glue{
  \CI'_{{\bf n}}(a,{\bf b},\dots):=\sum_{{\bf m}}\frac{1}{N!}\oint\prod_{i=1}^{N-1}\frac{dz_i}{2\pi i z_i}\;
  I_H^{({\bf m},{\bf n})}(a,{\bf b},{\bf z})\,I_V^{({\bf m})}({\bf z})\,\CI_{-{\bf m}}({\bf z}^{-1},\dots)\,,
}
where the sum runs over ordered $N$-tuples ${\bf m}\in(\bbZ_r)^N$ with $\sum_im_i\equiv0$, and the contours are $|z_i|=1$.
This agrees with the conventions of \RazamatYamazaki. There the sum runs over unordered holonomies, with the combinatorial factor $1/\prod_iN_i!$.
We write $\bbT:=\{{\bf b}\in(\bbC^*)^N:\ |b_i|=1,\ \prod_ib_i=1\}$ for the unit torus of $SU(N)$.
We will use the same gluing measure to define the pairing
\eqn\pairing{
  \eqalign{
    \langle f,g\rangle&:=\sum_{{\bf n}}\oint_{\bbT}[d{\bf b}]\;\Delta_{\bf n}({\bf b})\,f_{\bf n}({\bf b})\,g_{-{\bf n}}({\bf b}^{-1})\,,\cr
    \Delta_{\bf n}({\bf b})&:=I_V^{({\bf n})}({\bf b})\,,\qquad [d{\bf b}]:=\frac1{N!}\prod_{i=1}^{N-1}\frac{db_i}{2\pi i b_i}\,,
  }}
where ${\bf n}$ runs over the same set as ${\bf m}$ in \glue.
This pairing is bilinear, with no complex conjugation.
Throughout the paper, self-adjointness means symmetry with respect to this pairing, $\langle\CO f,g\rangle=\langle f,\CO g\rangle$.

\newsec{Residues and Difference Operators}

\subsec{\it Poles and pinching configurations}

We initially take $|b_i|=1$ and $|t|^{1/2}<|a|<1$, so that the unit tori in \glue\ separate the two trinion pole families.
As explained in \refs{\GaiottoXA,\RazamatYamazaki}, poles in $a$ arise when the continued $z$-contours are pinched between poles of the trinion factor \IH.
To reach the poles below, we analytically continue in $a$, deforming the integration cycle so that these families remain separated until the pinch.
The residue formula refers to this continuation of \glue, rather than to an integral over fixed unit tori for all $a$.

In the initial domain the relevant poles inside the contours come from the factors
$$
  \Gamma_{-m_i-n_{\sigma(i)}}\big(t^{1\over2}z_i^{-1}b_{\sigma(i)}^{-1}a^{-1}\big)
$$
for some permutation $\sigma\in{\frak S}_N$. By \Gammak\ they are
\eqn\zpoles{
  z_i=\frac{t^{\frac12}\,p^{u_i}q^{v_i}}{b_{\sigma(i)}\,a}\,,\qquad u_i,v_i\ge0\,,\qquad u_i-v_i\equiv-(m_i+n_{\sigma(i)})\ \ {\rm mod}\ r\,.
}
This includes both families $z^{(q)}$ and $z^{(p)}$ of \RazamatYamazaki.
The contour is pinched when, in addition, $z_N=1/\prod_{i<N}z_i$ sits on the pole of the same type in its own row.
Using $\prod_ib_i=1$, this happens at
\eqn\apoles{
  a^N=t^{\frac N2}p^Uq^V\,,\qquad U:=\sum_{i=1}^Nu_i\,,\quad V:=\sum_{i=1}^Nv_i\,.
}
Summing the congruences in \zpoles\ and using $\sum_im_i\equiv\sum_in_i\equiv0$ gives $U\equiv V$ mod $r$.
For each allowed $(U,V)$, \apoles\ has $N$ roots. We choose the representative
\eqn\astar{
  a^*=t^{\frac12}\,p^{U/N}q^{V/N}\,,\qquad U,V\ge0\,,\qquad U\equiv V\ \ {\rm mod}\ r\,.
}
The full set is $a=\zeta a^*$ with $\zeta^N=1$, reproducing the pole condition of \RazamatYamazaki.
The other roots are related to this representative by the center action on the flavor fugacities:
$$
  \CI'_{\bf n}(\zeta a,{\bf b},\dots)=\CI'_{\bf n}(a,\zeta{\bf b},\dots)\,,\qquad \zeta^N=1\,.
$$
Indeed, the trinion depends on $a$ and ${\bf b}$ only through $ab_j$ and their inverses, and $\prod_j(\zeta b_j)=1$.
We therefore work with $a^*$ below and recover the remaining roots by this covariance.
Since the integrand is invariant under simultaneous permutations of $(z_i,m_i)$, the $N!$ choices of $\sigma$
cancel the factor $1/N!$ in \glue, and we may set $\sigma={\rm id}$.

We call a pair $({\bf u},{\bf v})\in\bbZ_{\ge0}^N\times\bbZ_{\ge0}^N$ with $\sum_iu_i=U$ and $\sum_iv_i=V$ a {\it pinching configuration} for the pole $(U,V)$.
Given ${\bf n}$, a pinching configuration determines the gauge holonomy through $m_i\equiv-n'_i$, where
$$
  n'_i:=n_i+w_i\,,\qquad w_i:=u_i-v_i\,.
$$
We write ${\bf u}:=(u_i)$, ${\bf v}:=(v_i)$ and ${\bf w}:=(w_i)={\bf u}-{\bf v}$ for the corresponding vectors in $\bbZ^N$, so that ${\bf n}'={\bf n}+{\bf w}$; in general, bold symbols denote vectors and the corresponding plain symbols with an index their components.
We also set
\eqn\wdef{
  X_i:=p^{u_i}q^{v_i}\,,\qquad {\bar X}:=\Big(\prod_{i=1}^NX_i\Big)^{1/N}=p^{U/N}q^{V/N}\,,\qquad b'_i:=\frac{{\bar X}\,b_i}{X_i}\qquad \Big(\prod_ib'_i=1\Big)\,,
}
so that at the pinching point \zpoles\ with $a=a^*$ one has $z_i={b'_i}^{-1}$. The last factor in \glue\ then becomes $\CI_{-{\bf m}}({\bf z}^{-1},\dots)=\CI_{{\bf n}'}({\bf b}',\dots)$.

\subsec{\it The residue formula}

\proclaim Theorem 3.1. Assume that the input index $\CI$ is Weyl invariant and holomorphic in a neighborhood of each relevant pinching point, and that the other flavor parameters are generic so that no additional pole families pinch simultaneously.
Continue \glue\ as described in section 3.1 and define $\Res_{a\to a^*}F:=\lim_{a\to a^*}(1-a^*/a)F(a)$. Then
\eqn\resformula{
  \eqalign{
    \Res_{a\to a^*}\CI'_{{\bf n}}(a,{\bf b},\dots)
    &=\frac{\Gamma_0(t)}{N\kappa}\,\big(\CO_{U,V}\,\CI\big)_{{\bf n}}({\bf b},\dots)\,,\cr
    \big(\CO_{U,V}\,\CI\big)_{{\bf n}}({\bf b})&:=\sum_{({\bf u},{\bf v})}C_{{\bf u},{\bf v}}({\bf b};{\bf n})\;\CI_{{\bf n}'}({\bf b}')\,,
  }}
where the second line defines the difference operator $\CO_{U,V}$, the sum runs over all pinching configurations for $(U,V)$, ${\bf b}'$ and ${\bf n}'$ are determined by $({\bf u},{\bf v})$ as in \wdef, and
\eqn\mainformula{
  C_{{\bf u},{\bf v}}({\bf b};{\bf n}):=\left(\frac{pq}{t}\right)^{Z}\prod_{i,j=1}^{N}
  \frac{\Theta_{u_j,v_j}\big(n'_j-n'_i;\;t\,b'_j/b'_i\big)}{\Theta_{u_j,v_j}\big(n'_i-n_j;\;b'_i/({\bar X}\,b_j)\big)}\,,
}
\eqn\Zdef{
  Z:=\frac{1}{2r}\sum_{i,j=1}^{N}f(n_j-n'_i)-\frac1r\sum_{i<j}f(n'_i-n'_j)\,.
}

\proof
Take the residues at $z_i=z_i^*$ for $i<N$, given by \zpoles, and then the residue in $a$ coming from row $N$.
By Lemma 2.1 the residue of $\Gamma_{w_i}(t^{1\over2}/(z_ib_ia))$ at $z_i=z_i^*$, taken with the measure $dz_i/(2\pi i z_i)$,
equals $1/(\kappa \Theta_{u_i,v_i}(w_i;X_i^{-1}))$.
For row $N$, after the other residues are taken, $t^{1\over2}/(z_Nb_Na)=t^{N/2}\prod_{i<N}X_i/a^N$, and hence
$1-X_N\,t^{1\over2}/(z_Nb_Na)=1-(a^*/a)^N\simeq N(1-a^*/a)$. This produces the same factor, times $1/N$.
The remaining factors are combined as follows.
\item{$\bullet$} The diagonal hypermultiplet factor is
$$
  \Gamma_{m_i+n_i}(t^{1\over2}z_i^*b_ia^*)=\Gamma_{-w_i}(tX_i)=\Gamma_0(t)\,\Theta_{u_i,v_i}(0;t)\,.
$$
Together with the residue factor $1/(\kappa \Theta_{u_i,v_i}(w_i;X_i^{-1}))$, it gives the $i=j$ factor of \mainformula\ times $\Gamma_0(t)/\kappa$.
\item{$\bullet$} For $i\neq j$, the hypermultiplet gives
$\Gamma_{n_j-n'_i}(tX_ib_j/b_i)\,\Gamma_{n'_i-n_j}(b_i/(X_ib_j))$,
and the vector multiplet gives $1/(\Gamma_{n'_j-n'_i}(tb'_j/b'_i)\,\Gamma_{n'_j-n'_i}(b'_j/b'_i))$, because $z^*_i/z^*_j=b'_j/b'_i$.
Since $tX_ib_j/b_i=X_j\cdot(tb'_j/b'_i)$ and $b'_j/b'_i=X_i\cdot b_j/(X_jb_i)$, Lemma 2.1 gives
$$
  \eqalign{
    \frac{\Gamma_{n_j-n'_i}(tX_ib_j/b_i)}{\Gamma_{n'_j-n'_i}(tb'_j/b'_i)}&=\Theta_{u_j,v_j}(n'_j-n'_i;tb'_j/b'_i)\,,\cr
    \frac{\Gamma_{n'_j-n_i}(b_j/(X_jb_i))}{\Gamma_{n'_j-n'_i}(b'_j/b'_i)}&=\frac{1}{\Theta_{u_i,v_i}(n'_j-n_i;b_j/(X_jb_i))}\,.
  }
$$
Relabeling $i\leftrightarrow j$ in the second relation gives the off-diagonal factors of \mainformula.
\item{$\bullet$} The zero-point factors of \IH\ and \IV\ at ${\bf m}=-{\bf n}'$ combine to $(pq/t)^Z$, since $f$ is even.
\item{$\bullet$} The constants combine to $\kappa^{-N}\Gamma_0(t)^N(\kappa/\Gamma_0(t))^{N-1}/N=\Gamma_0(t)/(N\kappa)$.
\item{$\bullet$} Finally, the last factor in \glue\ gives $\CI_{{\bf n}'}({\bf b}')$, as noted below \wdef.

\noindent For generic ${\bf b}$ different configurations give different arguments, so no cancellation occurs between them.
\qed

\noindent For the other roots $\zeta a^*$ of \apoles, $\zeta^N=1$, the center covariance of section 3.1 shows that the same normalized residue is given by the right-hand side of \resformula\ evaluated at $\zeta{\bf b}$.
The coefficients are unchanged, since they depend only on ratios of the $b_i$, while the shifted arguments become $\zeta{\bf b}'$; the same regularity assumptions apply at those pinching points.

\noindent We note some simple properties of the coefficients. First, $b'_i/({\bar X}b_j)=(b_i/b_j)/X_i$ and $tb'_j/b'_i=t(b_j/b_i)X_i/X_j$.
Since $\Theta_{0,0}=1$, only indices $j$ with $(u_j,v_j)\neq(0,0)$ contribute to the product.
Since $\Theta_{u_j,v_j}$ is a product of $u_j+v_j$ theta functions and $\sum_j(u_j+v_j)=U+V$,
each coefficient is, up to the factor $(pq/t)^Z$, a ratio of two products of $N(U+V)$ theta functions each, with nomes $p^r$ and $q^r$.
The trivial pole $U=V=0$ gives $\CO_{0,0}=1$.
Note also that all coefficients depend on the holonomies only through differences.
Hence, for $s\in\bbZ$ with $Ns\equiv0$ mod $r$, the common shift $(S_sf)_{\bf n}=f_{{\bf n}+(s,\dots,s)}$
commutes with every $\CO_{U,V}$. This is a $\bbZ_{\gcd(N,r)}$ symmetry of the whole system.

As discussed in \refs{\GaiottoXA,\RazamatYamazaki}, S-duality of class ${\cal S}$ theories implies that the operator acting on either maximal puncture of the bifundamental gives the same result.
It further implies that the operators $\CO_{U,V}$ commute for all poles and are self-adjoint with respect to the pairing \pairing.
In the rest of this paper we prove commutativity and a sufficient criterion for self-adjointness directly from the explicit formula \mainformula,
without assuming S-duality; the remaining contour questions are discussed in section 7.4.

\newsec{Special Cases}

\subsec{\it The pole $a^*=t^{1\over2}(pq)^{1/N}$}

For $(U,V)=(1,1)$ there are two types of configurations.
For ${\bf u}={\bf v}={\bf e}_I$ the holonomy is unchanged, and for ${\bf u}={\bf e}_I$, ${\bf v}={\bf e}_J$ ($I\ne J$) one has ${\bf n}'={\bf n}+{\bf e}_I-{\bf e}_J$. Hence
\eqn\Ooneone{
  \eqalign{
    (\CO_{1,1}\CI)_{{\bf n}}({\bf b})&=\sum_{I}F_I\;\CI_{{\bf n}}\big(b_I\to(pq)^{\frac1N-1}b_I,\;b_{i\neq I}\to(pq)^{\frac1N}b_i\big)\cr
    &\quad+\sum_{I\neq J}G_{IJ}\;\CI_{{\bf n}+{\bf e}_I-{\bf e}_J}\big(b_I\to p^{\frac1N-1}q^{\frac1N}b_I,\cr
    &\qquad\qquad\qquad\qquad b_J\to p^{\frac1N}q^{\frac1N-1}b_J,\;b_{i\neq I,J}\to(pq)^{\frac1N}b_i\big)\,.
  }}
With $n_{iI}:=n_i-n_I$, Theorem 3.1 gives
\eqn\FIGIJ{
  \eqalign{
    F_I&=c_F\prod_{i\neq I}
    \frac{\Th\big(q^{\bbr{n_{iI}}}\frac{t}{pq}\frac{b_I}{b_i};q^r\big)\,\Th\big(p^{\bbr{n_{iI}}}\frac{pq}{t}\frac{b_i}{b_I};p^r\big)}
    {\Th\big(q^{\bbr{-n_{iI}}}\frac{b_i}{b_I};q^r\big)\,\Th\big(p^{\bbr{-n_{iI}}}\frac{b_I}{b_i};p^r\big)}\,,\qquad
    c_F:=\frac{\Th(t;q^r)\,\Th(t^{-1};p^r)}{\Th((pq)^{-1};q^r)\,\Th(pq;p^r)}\,,\cr
    G_{IJ}&=\left(\frac{pq}{t}\right)^{Z}\prod_{i=1}^N
    \frac{\Th\big(q^{\bbr{n'_i-n'_I}}\,t\,b'_I/b'_i;\,q^r\big)}{\Th\big(q^{\bbr{n_I-n'_i}}\,b'_i/({\bar X}b_I);\,q^r\big)}\,
    \frac{\Th\big(p^{\bbr{n'_J-n'_i}}\,t\,b'_J/b'_i;\,p^r\big)}{\Th\big(p^{\bbr{n'_i-n_J}}\,b'_i/({\bar X}b_J);\,p^r\big)}\,,
  }}
where in $G_{IJ}$ one has ${\bar X}=(pq)^{1/N}$, ${\bf b}'={\bar X}\,(b_1,\dots,b_I/p,\dots,b_J/q,\dots,b_N)$ and ${\bf n}'={\bf n}+{\bf e}_I-{\bf e}_J$.
The operator acts on nearest neighbors of the holonomy lattice.
For $N=2$ this agrees with the operator (4.8)--(4.9) of \RazamatYamazaki, up to an overall constant and conventions for labeling the terms,
and with the operator $S^{(0,0,1)}$ of \MOY, which coincides with \Ooneone\ including the normalization (Proposition 5.8).

\subsec{\it The poles $a^*=t^{1\over2}q^{r/N}$ and $t^{1\over2}p^{r/N}$}

For $(U,V)=(0,r)$ the configurations are all compositions ${\bf v}$ of $r$ into $N$ non-negative parts, with ${\bf u}=0$, ${\bf n}'={\bf n}-{\bf v}$ and $b'_i=q^{r/N-v_i}b_i$:
\eqn\Ozeror{
  \eqalign{
    (\CO_{0,r}\CI)_{{\bf n}}({\bf b})=\sum_{|{\bf v}|=r}&\left(\frac{pq}{t}\right)^{Z}\prod_{i,j=1}^N\prod_{\beta=0}^{v_j-1}
    \frac{\Th\big(p^{\bbr{n'_j-n'_i+\beta}}q^{\beta}\,t\,b'_j/b'_i;\,p^r\big)}{\Th\big(p^{\bbr{n'_i-n_j+\beta}}q^{\beta-v_i}\,b_i/b_j;\,p^r\big)}\cr
    &\times\CI_{{\bf n}-{\bf v}}(q^{\frac rN-v_1}b_1,\dots,q^{\frac rN-v_N}b_N)\,.
  }}
This operator has $\binom{r+N-1}{N-1}$ terms. For generic $p$ it connects ${\bf n}$ to many points of the holonomy lattice.
The operator $\CO_{r,0}$ is obtained by $p\leftrightarrow q$ and ${\bf u}\leftrightarrow {\bf v}$, together with the reversal of all holonomies, ${\bf n}\to-{\bf n}$ and ${\bf n}'\to-{\bf n}'$.
This follows from the invariance of $\Gamma_k(x)$ under $p\leftrightarrow q$, $k\to-k$ (see appendix A); without the reversal of the holonomies the coefficients do not agree.
For $N=2$, with $y:=b_1/b_2$ and $n_{12}:=n_1-n_2$, the $r+1$ terms are labeled by $k=v_1=0,\dots,r$:
\eqn\Ozerorone{
  \eqalign{
    (\CO_{0,r}\CI)_{{\bf n}}({\bf b})&=\sum_{k=0}^{r}C_k\;\CI_{(n_1-k,\;n_2+k-r)}\big(q^{\frac r2-k}b_1,\;q^{k-\frac r2}b_2\big)\,,\cr
    C_k&=\left(\frac{pq}{t}\right)^{\frac{f(k)+f(n_{12}-k)-f(n_{12}-2k)}{r}}\cr
    &\qquad\times\prod_{\beta=0}^{k-1}\frac{\Th(p^{\bbr{\beta}}q^{\beta}t;p^r)\,\Th(p^{\bbr{n_{12}-2k+\beta}}q^{\beta+r-2k}\,t\,y;p^r)}
    {\Th(p^{\bbr{\beta-k}}q^{\beta-k};p^r)\,\Th(p^{\bbr{k-n_{12}+\beta}}q^{\beta+k-r}\,y^{-1};p^r)}\cr
    &\qquad\times\prod_{\beta=0}^{r-k-1}\frac{\Th(p^{\bbr{\beta}}q^{\beta}t;p^r)\,\Th(p^{\bbr{2k-n_{12}+\beta}}q^{\beta+2k-r}\,t\,y^{-1};p^r)}
    {\Th(p^{\bbr{k+\beta}}q^{\beta+k-r};p^r)\,\Th(p^{\bbr{n_{12}-k+\beta}}q^{\beta-k}\,y;p^r)}\,.
  }}
For $N=2$ the poles $(U,V)=(1,1)$, $(0,r)$ and $(r,0)$ are the poles $\alpha^2=tpq$, $tq^r$ and $tp^r$ of \MOY, with vanishing $U(1)$ holonomy.
The corresponding defects are denoted there by $S^{(0,0,1)}$, $S^{(0,1,0)}$ and $S^{(1,0,0)}$.
Their operators coincide with $\CO_{1,1}$, $\CO_{0,r}$ and $\CO_{r,0}$, including the normalization (Proposition 5.8 below).

\subsec{\it Macdonald limit}

We take the Macdonald limit $p\to0$ with $q$, $t$ and the flavor fugacities fixed and generic.
Following \RazamatYamazaki, we restrict to the poles with $U=0$, i.e.\ $a^*=t^{1\over2}q^{V/N}$ with $V\equiv0$ mod $r$, which remain at nonzero $a$ in this limit.
Let $I_V^0({\bf n}):=(pq/t)^{-\frac1r\sum_{i<j}f(n_i-n_j)}$ be the zero-point factor of \IV, and rescale the index of every theory as
$\widehat\CI_{\bf n}:=I_V^0({\bf n})^{1/2}\,\CI_{\bf n}$.
This is equivalent to the rescaling (5.1) of \RazamatYamazaki, where the hypermultiplet index is multiplied by $I_V^0({\bf m})^{1/2}I_V^0({\bf n})^{1/2}$ and the zero-point factor of the vector multiplet is dropped.
Under the rescaling the coefficient $C_{{\bf u},{\bf v}}$ is multiplied by $(I_V^0({\bf n})/I_V^0({\bf n}'))^{1/2}$.
Its total power of $p$ from the zero-point energies is the quantity studied in appendix A of \RazamatYamazaki,
with ${\bf m}=-{\bf n}'$.
This quantity is non-negative and vanishes if and only if ${\bf n}'$ is a permutation of ${\bf n}$.
For $U=0$ one has ${\bf u}=0$ and ${\bar X}/X_i=q^{V/N-v_i}$, so the arguments of the index do not move with $p$.
All theta functions in \mainformula\ are then of the form $\Th(cp^a;p^r)$ with $c$ independent of $p$ and $0\le a\le r-1$; their limits are $1-c$ for $a=0$ and $1$ for $0<a<r$.
For generic fugacities the denominator limits are nonzero, so only configurations with ${\bf n}'\in{\frak S}_N\cdot{\bf n}$ survive in the rescaled operator.
For $(U,V)=(0,r)$ these are the terms denoted $F_I$ and $G_{(I_1\cdots I_k)}$ in (5.4) of \RazamatYamazaki\ (not to be confused with $F_I$ and $G_{IJ}$ of \FIGIJ), where these operators were related to the Cherednik operators of the double affine Hecke algebra \cherednik.
For example, for $N=2$ only $k=0$, $k=r$ and $k=\bbr{n_{12}}$ survive in \Ozerorone. They reproduce the three terms of (5.8) of \RazamatYamazaki.
For $U>0$ both the theta functions and the shifted arguments have additional $p$-dependence, so this argument does not assert a termwise limit for those operators.

\subsec{\it $r=1$}

For $r=1$ all $\bbr{\cdot}$ vanish and the holonomy lattice is trivial. Then
$$
  \CO_{0,1}\propto\sum_I\prod_{i\ne I}\frac{\Th(\frac{pq}{t}\,b_i/b_I;p)}{\Th(b_i/b_I;p)}\;
  T_{I}\,,\qquad T_I:\ b_I\to q^{\frac1N-1}b_I,\ b_{i\neq I}\to q^{\frac1N}b_i\,.
$$
This is the elliptic RS operator in the variables $b_i^{-1}$ with coupling $pq/t$, as in \GaiottoXA.
The general $(U,V)$ give the higher operators of \GaiottoXA.

\newsec{Factorization}

The coefficients \mainformula\ of $\CO_{U,V}$ depend on the holonomies only through their residues mod $r$,
which enter through the functions $\Theta_{u,v}(k;x)$ of the shift lemma \shift.
Because of this reduction the operators are non-local on the holonomy lattice, and they do not look like difference operators of a familiar type.
In this section we show that this is an artifact of the normalization of the index.
After conjugation by an explicit function $\CG$ of $({\bf b},{\bf n})$, Gaussian in the holonomies, every $\CO_{U,V}$ becomes the product of two commuting operators of elliptic RS type:
one in the variables $\xi_i=b_ip^{n_i}$ with nome $p^r$, and one in the variables $\eta_i=b_iq^{-n_i}$ with nome $q^r$ (Theorem 5.6).

The argument has three steps.
First, we lift the holonomies to ${\bf n}\in\bbZ^N$ and replace the lens gamma function $\Gamma_k$ by a lifted version $\tGamma_k$, in which $k$ is not reduced mod $r$ (section 5.1).
The two differ by an explicit monomial (Lemma 5.1), and $\tGamma_k$ satisfies an exact shift lemma in which no residues appear (Lemma 5.2).
In the exact shift lemma the $p$-shifts produce theta functions with nome $q^r$ of $xq^{-k}$, and the $q$-shifts produce theta functions with nome $p^r$ of $xp^{k}$.
This is the origin of the sector variables $\eta$ and $\xi$ and of the sector operators (section 5.2).

Second, the monomials of Lemma 5.1 combine, multiplet by multiplet, into Gaussians of the holonomies that are local to each puncture (section 5.3). They are absorbed into the gauge function $\CG$.

Finally, we repeat the residue computation of section 3 for the lifted integrand (section 5.4).
This gives a lifted residue formula, whose coefficients are those of \mainformula\ with $\Theta$ replaced by $\tE$ and which separate into the two sectors (Proposition 5.5).
Comparing it with Theorem 3.1 through the Gaussians of the second step gives the factorization theorem (Theorem 5.6).

\subsec{\it Lifted gamma functions}

Throughout this section we lift holonomies to ${\bf n}\in\bbZ^N$, i.e.\ we pull back functions on $(\bbZ_r)^N$ along $\bbZ^N\to(\bbZ_r)^N$.
Define the {\it lifted} gamma function for $k\in\bbZ$ by
\eqn\tGdef{
  \tGamma_k(x):=\Gamma(xp^{k};pq,p^r)\,\Gamma(xq^{r-k};pq,q^r)\,,
}
so that $\Gamma_k=\tGamma_{\bbr{k}}$ by \Gammak.
This is the rarefied elliptic gamma function $\gamma^{(r)}(x,k;p,q)$ of \SpiridonovRarefied, eq.~(12), in which the discrete variable is not reduced mod $r$.
Lemmas 5.1 and 5.2 below are restatements, in our notation, of its quasi-periodicity in $k$ and of its elementary difference equations,
eqs.~(21) and (23) of \SpiridonovRarefied. We include short proofs for completeness.

The first lemma gives the price of the lift: $\tGamma_k$ and $\Gamma_k$ differ by a monomial in $x$ whose exponent is Gaussian in $k$.

\proclaim Lemma 5.1. Write $k=\bbr{k}+rs$ and $k_a:=\bbr{k}+ra$. For $s\ge0$,
$$
  \frac{\tGamma_k(x)}{\Gamma_k(x)}=\prod_{a=0}^{s-1}(-1)^{k_a}x^{-k_a}\left(\frac qp\right)^{k_a^2/2}(pq)^{k_a/2}\,,
$$
and for $s<0$ the right-hand side is replaced by the inverse of the product over $a=s,\dots,-1$.
Define the signed sums $S_n(k):=\sum_{a=0}^{s-1}k_a^n$ for $s\ge0$ and $S_n(k):=-\sum_{a=s}^{-1}k_a^n$ for $s<0$. Then, for all $s\in\bbZ$,
\eqn\Ssums{
  \eqalign{
    S_1(k)&=s\bbr{k}+\frac{rs(s-1)}2=\frac{f(k)-k(r-k)}{2r}\,,\cr
    S_2(k)&=s\bbr{k}^2+r\bbr{k}s(s-1)+\frac{r^2s(s-1)(2s-1)}6\,.
  }}

\proof Use the first relation in \Gammaqp\ on both factors of \tGdef. For $s\ge0$ this gives
$$
  \frac{\tGamma_k(x)}{\Gamma_k(x)}=\prod_{a=0}^{s-1}\frac{\Th(xp^{k_a};pq)}{\Th(xq^{-k_a};pq)}\,.
$$
Then apply $\Th(y(pq)^m;pq)=(-1)^my^{-m}(pq)^{-m(m-1)/2}\Th(y;pq)$ with $y=xq^{-k_a}$ and $m=k_a$.
The case $s<0$ is analogous.
For $s\ge0$ the first two expressions in \Ssums\ are the sums of an arithmetic progression and of its squares.
For $s<0$ one has $\sum_{a=s}^{-1}a=-s(s-1)/2$ and $\sum_{a=s}^{-1}a^2=-s(s-1)(2s-1)/6$, so the same polynomials in $s$ are obtained.
Finally, $f(k)-k(r-k)=-rs(r-2\bbr{k})+r^2s^2$ follows by substituting $k=\bbr{k}+rs$.
\qed

The second lemma is the gain: for $\tGamma_k$ the shift lemma holds without any reduction mod $r$.

\proclaim Lemma 5.2 (Exact lifted shift). Let
$$
  \tE_{u,v}(k;x):=\prod_{\alpha=0}^{u-1}\Th(x(pq)^\alpha q^{r-k};q^r)\prod_{\beta=0}^{v-1}\Th(x(pq)^\beta p^{k};p^r)\,.
$$
Then $\tGamma_{k-(u-v)}(p^uq^vx)=\tGamma_k(x)\,\tE_{u,v}(k;x)$ for all $k\in\bbZ$. Moreover, with $X:=p^uq^v$,
$$
  \lim_{x\to X^{-1}}(1-xX)\,\tGamma_{u-v}(x)=\big(\kappa\,\tE_{u,v}(u-v;X^{-1})\big)^{-1}\,.
$$

\proof Applying the second relation in \Gammaqp\ to one factor of \tGdef\ gives
$\tGamma_{k-1}(px)=\Th(xq^{r-k};q^r)\,\tGamma_k(x)$ and $\tGamma_{k+1}(qx)=\Th(xp^{k};p^r)\,\tGamma_k(x)$.
Iterating the first relation $u$ times and then the second $v$ times, starting from $(k-u,p^ux)$, gives the shift formula.
The residue follows as in Lemma 2.1, since $\tGamma_0=\Gamma_0$.
\qed

\noindent Compare $\tE_{u,v}$ with $\Theta_{u,v}$ in \shift: the residues $\bbr{\,\cdot\,}$ have disappeared.
Moreover, $\tE_{u,v}(k;x)$ has a simple structure.
The $u$ factors, which come from the $p$-shifts, are theta functions with nome $q^r$ and depend on $(x,k)$ only through $xq^{-k}$;
the $v$ factors, which come from the $q$-shifts, are theta functions with nome $p^r$ and depend on $(x,k)$ only through $xp^{k}$.
In the residue formula, $x$ is a ratio of fugacities and $k$ the corresponding difference of holonomies.
This suggests the change of variables of the next subsection.

\subsec{\it Sector variables and sector operators}

We introduce the {\it sector variables}
\eqn\sector{
  \xi_i:=b_i\,p^{n_i}\,,\qquad \eta_i:=b_i\,q^{-n_i}\,,\qquad R:=pq\,.
}
For the separate sector operators we first enlarge the space to $({\bf b},{\bf n})\in(\bbC^*)^N\times\bbZ^N$, without the determinant or total-holonomy constraint.
The map $({\bf b},{\bf n})\mapsto(\xi,\eta)$ is injective, since $0<|R|<1$; its image is cut out by $\xi_i/\eta_i\in R^{\bbZ}$.
The physical lifted space is the subset with $\prod_ib_i=1$ and $\sum_in_i\equiv0$ mod $r$; on it $\prod_i\xi_i=p^{\sum_in_i}$.
For a ratio $x=b_j/b_i$ with $k=n_j-n_i$ one has $xp^k=\xi_j/\xi_i$ and $xq^{-k}=\eta_j/\eta_i$, so the two kinds of theta functions in $\tE_{u,v}$ depend on $\xi$ and on $\eta$ respectively.
The shifts of the residue formula also separate.
A configuration $({\bf u},{\bf v})$ sends $({\bf b},{\bf n})$ to $({\bf b}',{\bf n}')$, and in the sector variables this reads
\eqn\sectorshift{
  \xi_i\to {\bar X}\,R^{-v_i}\xi_i\,,\qquad \eta_i\to {\bar X}\,R^{-u_i}\eta_i\,,
}
since $b'_ip^{n'_i}={\bar X}b_ip^{n_i}R^{-v_i}$ and $b'_iq^{-n'_i}={\bar X}b_iq^{-n_i}R^{-u_i}$.
Thus, apart from the common rescaling by $\bar X$, the $q$-type shifts (${\bf v}$) move only $\xi$, and the $p$-type shifts (${\bf u}$) move only $\eta$.

We define the following operators of elliptic RS type, one for each sector:
\eqn\sectorops{
  \eqalign{
    \CP^{\xi}_V&:=\sum_{|{\bf v}|=V}c_{\bf v}(\xi)\,T^{\xi}_{\bf v}\,,\qquad
    c_{\bf v}(\xi):=\prod_{i,j=1}^N\prod_{\beta=0}^{v_j-1}
    \frac{\Th(t\,R^{\beta-v_j+v_i}\xi_j/\xi_i;\,p^r)}{\Th(R^{\beta-v_i}\xi_i/\xi_j;\,p^r)}\,,\cr
    \CP^{\eta}_U&:=\sum_{|{\bf u}|=U}c_{\bf u}(\eta)\,T^{\eta}_{\bf u}\,,\qquad
    c_{\bf u}(\eta):=\prod_{i,j=1}^N\prod_{\alpha=0}^{u_j-1}
    \frac{\Th(q^r t\,R^{\alpha-u_j+u_i}\eta_j/\eta_i;\,q^r)}{\Th(q^rR^{\alpha-u_i}\eta_i/\eta_j;\,q^r)}\,,
  }}
where $T^\xi_{\bf v}:\xi_i\to R^{-v_i}\xi_i$ and $T^\eta_{\bf u}:\eta_i\to R^{-u_i}\eta_i$.
On the enlarged space these shifts are respectively $({\bf b},{\bf n})\mapsto((q^{-v_i}b_i)_i,{\bf n}-{\bf v})$ and $({\bf b},{\bf n})\mapsto((p^{-u_i}b_i)_i,{\bf n}+{\bf u})$.
They preserve the lattice condition $\xi_i/\eta_i\in R^{\bbZ}$, but separately need not preserve the physical constraints.
Thus $\CP^\xi$ has nome $p^r$ and step $R$, and $\CP^\eta$ has nome $q^r$ and step $R$.
The two are of the same form: using $\Th(q^ry;q^r)=\Th(y^{-1};q^r)$, one finds that $\CP^\eta_U$ is obtained from $\CP^\xi_U$ by the substitution $\xi_i\to\eta_i^{-1}$ and $(p^r,R,t)\to(q^r,R^{-1},t^{-1})$.
For $V=1$ the operator $\CP^\xi_1$ is proportional to the first elliptic RS operator with these parameters; for $V\ge2$ it is a polynomial in the elliptic RS operators, as we show in section 6.
These are precisely the operators that the residue computation of section 3 produces when $\Gamma_k$ is replaced by $\tGamma_k$:
we will see in Proposition 5.5 that the coefficient obtained from \mainformula\ by replacing $\Theta$ with $\tE$ is $c_{\bf v}(\xi)\,c_{\bf u}(\eta)$,
with all dependence on ${\bf v}$ through $\xi$ and all dependence on ${\bf u}$ through $\eta$.

\subsec{\it Gaussian gauge factors}

It remains to control the monomials of Lemma 5.1, which relate the lens integrand to the lifted one.
We define the Gaussian gauge function and the constant
\eqn\Gdef{
  \eqalign{
    \CG({\bf b},{\bf n})&:=\prod_{i<j}\left(\frac{b_i}{b_j}\right)^{n_i-n_j}\left(\frac{R}{t}\right)^{\frac{(n_i-n_j)^2}{2r}}\left(\frac pq\right)^{\frac{(n_i-n_j)^2}{2}}\,,\cr
    K_{U,V}&:=(-1)^{N|{\bf w}|}R^{\frac{U^2-V^2}{2}-\frac{|{\bf w}|^2}{2r}}\,t^{\frac{|{\bf w}|^2}{2r}+\frac{N|{\bf w}|}{2}}\,,
  }}
where $|{\bf w}|:=\sum_iw_i=U-V$.
The next two lemmas show that, for each multiplet, the monomials of Lemma 5.1 combine into such Gaussians.

\proclaim Lemma 5.3 (Hypermultiplet pair). For $k\in\bbZ$,
$$
  \left(\frac{R}{t}\right)^{\frac{f(k)}{2r}}\Gamma_{k}(t^{\frac12}y)\,\Gamma_{-k}(t^{\frac12}y^{-1})
  =(-1)^{k}\,y^{-k}\left(\frac{t}{R}\right)^{\frac{k^2}{2r}}\left(\frac qp\right)^{\frac{k^2}{2}}\,
  \tGamma_{k}(t^{\frac12}y)\,\tGamma_{-k}(t^{\frac12}y^{-1})\,.
$$

\proof Apply Lemma 5.1 to both factors.
By \Ssums\ one has $S_1(k)-S_1(-k)=-k$, $S_1(k)+S_1(-k)=(f(k)+k^2)/r$ and $S_2(k)+S_2(-k)=-k^2$,
where for $\bbr{k}\neq0$ one uses $-k=(r-\bbr{k})+r(-s-1)$. One finds
$$
  \Gamma_{k}\Gamma_{-k}=\tGamma_{k}\tGamma_{-k}\;(-1)^{(f+k^2)/r}\,y^{-k}\left(\frac qp\right)^{\frac{k^2}2}\left(\frac t{R}\right)^{\frac{f+k^2}{2r}}\,.
$$
Moreover $(f+k^2)/r=\bbr{k}+2s\bbr{k}+rs^2\equiv k$ mod 2.
The factor $(R/t)^{f/2r}$ from the zero-point energy cancels.
\qed

\noindent Lemma 5.3 is the counterpart, for a hypermultiplet pair, of the observation that lens elliptic hypergeometric sum/integrals can be written either with the non-periodic function $\gamma^{(r)}$ or with a gamma function made $r$-periodic by a Gaussian normalization factor; see \KelsYamazakiLens, eqs.~(2.3)--(2.6), and \SpiridonovRarefied, eqs.~(62)--(63).
Note that the zero-point factor $(R/t)^{f(k)/2r}$ of the index is exactly what is needed to make the right-hand side a Gaussian in $k$ rather than in $\bbr{k}$.

\proclaim Lemma 5.4 (Vector multiplet). $I_V^{({\bf m})}({\bf z})=\CG({\bf z},{\bf m})^2\;\widetilde I_V^{({\bf m})}({\bf z})$,
where $\widetilde I_V$ is obtained from \IV\ by dropping the zero-point factor and replacing every $\Gamma_k$ by $\tGamma_k$.

\proof Apply Lemma 5.1 to the four factors $\Gamma_{k}(y)$, $\Gamma_k(ty)$, $\Gamma_{-k}(y^{-1})$ and $\Gamma_{-k}(ty^{-1})$, with
$y=z_i/z_j$ and $k=m_i-m_j$, in the same way as in Lemma 5.3. \qed

\subsec{\it The lifted residue formula and the factorization theorem}

Let
$$
  \Phi:=I_H^{({\bf m},{\bf n})}(a,{\bf b},{\bf z})\,I_V^{({\bf m})}({\bf z})\,\CI_{-{\bf m}}({\bf z}^{-1},\dots)
$$
be the integrand of \glue, and let $\widetilde\Phi$ be the {\it lifted} integrand, obtained from $\Phi$ by replacing every $\Gamma_k$ by $\tGamma_k$ and dropping the zero-point factors.
By Lemma 5.1 the two differ by a monomial, so they have the same poles, and in particular the same pinching points \zpoles.
For a pinching configuration $({\bf u},{\bf v})$ and a lift ${\bf n}\in\bbZ^N$ we define the lifted counterpart of \mainformula,
\eqn\liftedC{
  \widetilde C_{{\bf u},{\bf v}}({\bf b};{\bf n}):=\prod_{i,j=1}^{N}
  \frac{\tE_{u_j,v_j}\big(n'_j-n'_i;\;t\,b'_j/b'_i\big)}{\tE_{u_j,v_j}\big(n'_i-n_j;\;b'_i/({\bar X}\,b_j)\big)}\,.
}
Unlike $C_{{\bf u},{\bf v}}$, this depends on the lift ${\bf n}$ and not only on its residues mod $r$.

\proclaim Proposition 5.5 (Lifted residue formula). The residue of $\widetilde\Phi$ at the pinching point of the configuration $({\bf u},{\bf v})$, taken as in Theorem 3.1, is
$\frac{\Gamma_0(t)}{N\kappa}\,\widetilde C_{{\bf u},{\bf v}}({\bf b};{\bf n})\,\CI_{{\bf n}'}({\bf b}',\dots)$. Moreover,
\eqn\liftedsep{
  \widetilde C_{{\bf u},{\bf v}}({\bf b};{\bf n})=c_{\bf v}(\xi)\,c_{\bf u}(\eta)\,,
}
where $\xi,\eta$ are built from $({\bf b},{\bf n})$ by \sector.

\proof The residue is computed as in the proof of Theorem 3.1, with Lemma 5.2 in place of Lemma 2.1; since $\widetilde\Phi$ has no zero-point factors, there is no factor $(R/t)^Z$.
For \liftedsep, write out \liftedC\ with the definition of $\tE$ in Lemma 5.2.
For the $v$-factors, which have nome $p^r$, the arguments are $t(b'_j/b'_i)p^{n'_j-n'_i}R^\beta$ and $(b'_i/({\bar X}b_j))p^{n'_i-n_j}R^\beta$.
By \sectorshift\ these equal $tR^{\beta-v_j+v_i}\xi_j/\xi_i$ and $R^{\beta-v_i}\xi_i/\xi_j$, which are the arguments in $c_{\bf v}(\xi)$.
The $u$-factors, which have nome $q^r$, give $c_{\bf u}(\eta)$ in the same way.
\qed

\noindent Thus the lifted residue formula already has the factorized form: its coefficients separate into a $\xi$-part and an $\eta$-part, without any reduction mod $r$.
It remains to compare it with Theorem 3.1.

\proclaim Theorem 5.6 (Factorization). For every pole $(U,V)$,
\eqn\factorization{
  \CO_{U,V}=K_{U,V}\;\CG^{-1}\circ \CD_{\bar X}\,\CP^{\eta}_U\,\CP^{\xi}_V\circ \CG\,,
}
where $\CD_{\bar X}:{\bf b}\to {\bar X}{\bf b}$ acts on the enlarged space of section 5.2 and $\CG$ acts by multiplication.
Only the complete product on the right-hand side is restricted to the physical lifted space.
Equivalently, for every configuration and every lift,
\eqn\CtildeC{
  C_{{\bf u},{\bf v}}({\bf b};{\bf n})=K_{U,V}\,\frac{\CG({\bf b}',{\bf n}')}{\CG({\bf b},{\bf n})}\,\widetilde C_{{\bf u},{\bf v}}({\bf b};{\bf n})
  =K_{U,V}\,\frac{\CG({\bf b}',{\bf n}')}{\CG({\bf b},{\bf n})}\,c_{\bf v}(\xi)\,c_{\bf u}(\eta)\,.
}

\proof Since $\CP^\xi$ and $\CP^\eta$ act on independent variables, the coefficient of $\CD_{\bar X}T^\eta_{\bf u}T^\xi_{\bf v}$ on the right-hand side of \factorization\
is $c_{\bf v}(\xi)\,c_{\bf u}(\eta)$, with $\xi,\eta$ built from $({\bf b},{\bf n})$ by \sector, and by \sectorshift\ the shift $\CD_{\bar X}T^\eta_{\bf u}T^\xi_{\bf v}$ is that of the configuration $({\bf u},{\bf v})$.
It therefore suffices to prove \CtildeC, and by \liftedsep\ only its first equality.
Rather than comparing $\Theta$ and $\tE$ factor by factor, we compare the integrands $\Phi$ and $\widetilde\Phi$.

\parhead{Locality.}Apply Lemma 5.3 to each hypermultiplet pair with $k_{ij}=m_i+n_j$ and $y=z_ib_ja$, and take the product over $i,j$.
Use $\prod_iz_i=\prod_jb_j=1$ and $\sum_{i,j}k^2_{ij}=N\sum_im_i^2+N\sum_jn_j^2+2|{\bf m}||{\bf n}|$, where $|{\bf m}|:=\sum_im_i$. Together with Lemma 5.4 this gives
\eqn\locality{
  \eqalign{
    \Phi&=\widetilde\Phi\cdot\frac{\CG({\bf z},{\bf m})}{\CG({\bf b},{\bf n})}\;\zeta(|{\bf m}|)\,\zeta(|{\bf n}|)\,
    \left(\frac{t}{R}\right)^{|{\bf m}||{\bf n}|/r}\left(\frac qp\right)^{|{\bf m}||{\bf n}|}a^{-N(|{\bf m}|+|{\bf n}|)}\,,\cr
    \zeta(k)&:=(-1)^{Nk}\left(\frac{R}{t}\right)^{-\frac{k^2}{2r}}\left(\frac pq\right)^{-\frac{k^2}{2}}\,.
  }}
Here we used $\prod_iz_i^{-Nm_i}=\prod_{i<j}(z_i/z_j)^{-(m_i-m_j)}$, which combines with $\CG({\bf z},{\bf m})^2$ from Lemma 5.4 into a single power of $\CG({\bf z},{\bf m})$.
The prefactor in \locality\ is {\it local}: it is a product of one factor for each puncture, times a coupling that involves only the total holonomies.

\parhead{Residues.}The prefactor in \locality\ is a monomial, holomorphic and non-vanishing.
Hence the residues of $\Phi$ are the prefactor at the pinching point times the residues of $\widetilde\Phi$.
By Theorem 3.1 and Proposition 5.5, $C_{{\bf u},{\bf v}}$ is therefore $\widetilde C_{{\bf u},{\bf v}}$ times the prefactor of \locality\ at the pinching point.

\parhead{Prefactor.}At the pinching point, $({\bf z},{\bf m})=({{\bf b}'}^{-1},-{\bf n}')$, and $\CG({{\bf b}'}^{-1},-{\bf n}')=\CG({\bf b}',{\bf n}')$.
Using $|{\bf n}'|=|{\bf n}|+|{\bf w}|$ and $a^*=t^{1\over2}p^{U/N}q^{V/N}$, the remaining factors combine to
$$
  \eqalign{
    &\zeta(|{\bf n}'|)\zeta(|{\bf n}|)\left(\frac{R}{t}\right)^{|{\bf n}'||{\bf n}|/r}\left(\frac pq\right)^{|{\bf n}'||{\bf n}|}(a^*)^{N|{\bf w}|}\cr
    &\qquad=(-1)^{N|{\bf w}|}\left(\frac{R}{t}\right)^{-\frac{|{\bf w}|^2}{2r}}\left(\frac pq\right)^{-\frac{|{\bf w}|^2}{2}}t^{\frac{N|{\bf w}|}{2}}p^{U|{\bf w}|}q^{V|{\bf w}|}=K_{U,V}\,,
  }
$$
which is independent of $|{\bf n}|$. This proves the theorem.
\qed

\noindent All coefficients in \CtildeC\ depend only on ratios of the fugacities, so this coefficient identity also holds on the enlarged space by a common rescaling to $\prod_ib_i=1$.
For each complete shift, $\prod_i b'_i=\prod_i b_i$ because ${\bar X}^N=p^Uq^V$, and $\sum_i n'_i=\sum_i n_i+U-V$.
Thus the complete product preserves the physical lifted space whenever $U\equiv V$ mod $r$.
Its action there is independent of how a function is extended off that space.
Since $\CG$ is not $r$-periodic in ${\bf n}$, the individual conjugated sector operators need not preserve $r$-periodic functions.
The complete product, however, has the coefficients $C_{{\bf u},{\bf v}}$ of Theorem 3.1, which depend on the holonomies only modulo $r$.
Hence \factorization\ descends to the physical holonomy lattice.

\proclaim Corollary 5.7. In the lifted sense described above, the lens system is gauge-equivalent to two decoupled systems of elliptic RS type: one with nome $p^r$ and step $R$ in the variables $\xi$,
and one with nome $q^r$ and step $R$ in the variables $\eta$. They are glued along the lattice $\xi_i/\eta_i\in R^{\bbZ}$.
In particular, apart from the uniform rescaling by ${\bar X}$, the $q$-type operators $\CO_{0,V}$ act only on $\xi$ and the $p$-type operators $\CO_{U,0}$ act only on $\eta$.

\parhead{Example.}Let $N=2$ and $(U,V)=(1,1)$, and write $y:=b_1/b_2$ and $n_{12}:=n_1-n_2$. Then $K_{1,1}=1$, ${\bar X}=R^{1/2}$,
$\CG=y^{n_{12}}(R/t)^{n_{12}^2/2r}(p/q)^{n_{12}^2/2}$, and
$$
  \eqalign{
    \CP^\xi_1&=\frac{\Th(t;p^r)}{\Th(R^{-1};p^r)}\left[\frac{\Th(tR^{-1}{\hat\xi};p^r)}{\Th({\hat\xi}^{-1};p^r)}\,T^\xi_{{\bf e}_1}+\frac{\Th(tR^{-1}{\hat\xi}^{-1};p^r)}{\Th({\hat\xi};p^r)}\,T^\xi_{{\bf e}_2}\right]\,,\cr
    \CP^\eta_1&=\frac{\Th(t^{-1};q^r)}{\Th(R;q^r)}\left[\frac{\Th(R{\hat\eta}^{-1}/t;q^r)}{\Th({\hat\eta};q^r)}\,T^\eta_{{\bf e}_1}+\frac{\Th(R{\hat\eta}/t;q^r)}{\Th({\hat\eta}^{-1};q^r)}\,T^\eta_{{\bf e}_2}\right]\,,
  }
$$
with ${\hat\xi}:=\xi_1/\xi_2=yp^{n_{12}}$ and ${\hat\eta}:=\eta_1/\eta_2=yq^{-n_{12}}$.
The four products $T^\eta_{{\bf e}_b}T^\xi_{{\bf e}_a}$ give the four configurations $({\bf u},{\bf v})=({\bf e}_b,{\bf e}_a)$ of \Ooneone.
For instance, $T^\eta_{{\bf e}_1}T^\xi_{{\bf e}_2}$ maps $(b_1,b_2,n_1,n_2)$ to $(b_1/p,b_2/q,n_1+1,n_2-1)$, which after $\CD_{\bar X}$ is the shift of the term $G_{12}$.

\proclaim Proposition 5.8. Let $N=2$, ${\bf b}=(b,b^{-1})$ and ${\bf n}=(n,-n)$, where $n\in\bbZ_r$ is the holonomy denoted by $m$ in \MOY.
Then $\CO_{1,1}$, $\CO_{0,r}$ and $\CO_{r,0}$ coincide, including the normalization, with the operators
$S^{(0,0,1)}$, $S^{(0,1,0)}$ and $S^{(1,0,0)}$ of \MOY\ (eqs.~(3.13), (3.19) and (3.14) there).

\proof The proof uses Theorem 5.6, which expresses every coefficient through the sector functions $c_{\bf v}(\xi)$ and $c_{\bf u}(\eta)$.
Since Theorem 3.1 does not depend on the lift of ${\bf n}$, the lift may be chosen separately for each term.
With a suitable choice, $\xi_1/\xi_2$ and $\eta_1/\eta_2$ become the variables used in \MOY, and the theta functions of the two sides agree up to monomials,
by $\Th(x^{-1};P)=-x^{-1}\Th(x;P)$. These monomials combine with $K_{U,V}$ and the ratio of Gaussians to the prefactors of \MOY.
The details are given in appendix A.
\qed

\noindent A related factorization appears for $N=2$ in \MOY, where $S^{(0,0,1)}=(R/t)^{1/r}X_qX_p$.
Here $X_p$ and $X_q$ are the elementary defects at the poles $\alpha^2=tp$ and $tq$ with nonzero $U(1)$ holonomy;
$X_p$ shifts the fugacity by powers of $p^{1/2}$ and has nome $q^r$, while $X_q$ shifts it by powers of $q^{1/2}$ and has nome $p^r$,
and both are traces of L-operators of Sklyanin algebras with shift $R$.
This parallels Theorem 5.6 for $(U,V)=(1,1)$, with $\CP^\eta_1$ and $\CP^\xi_1$ in the roles of $X_p$ and $X_q$,
and the doubled holonomy label $2m\in\bbZ_{2r}$ of \MOY\ plays a role similar to our lift ${\bf n}\in\bbZ^N$.
For $r\ge2$ the sector operators $\CP^\xi_V$ and $\CP^\eta_U$ alone are not residues of the index at vanishing $U(1)$ holonomy,
and the results of \MOY\ suggest that they are realized by defects with nonzero $U(1)$ holonomy.
We have not attempted to make this precise for general $N$.

\newsec{Commutativity}

In this section we prove that all the operators $\CO_{U,V}$ commute with each other (Theorem 6.4).
By the factorization theorem, this reduces to the commutativity of the sector operators $\CP^\xi_V$ among themselves, and of the $\CP^\eta_U$ among themselves (section 6.1).
Each sector operator is a sum over compositions ${\bf v}$ of $V$ of products of theta functions with a fixed nome and a fixed step.
Written additively, the operators with $V=1$ are the first of Ruijsenaars' commuting operators $D_1,\dots,D_N$ (section 6.2).
The operators with general $V$ are not among the $D_s$, but they satisfy a Wronski-type relation that expresses $\CP'_V$ through the $D_s$ and the $\CP'_{V'}$ with $V'<V$ (section 6.3).
This relation follows from an identity of Noumi and Sano \NoumiSano.
By induction, every $\CP'_V$ is a polynomial in the commuting operators $D_s$, and the theorem follows.

\subsec{\it Reduction to the sectors}

\proclaim Proposition 6.1. $[\CO_{U,V},\CO_{U',V'}]=0$ for all poles, provided that the operators $\CP^\xi_V$ of \sectorops\ commute among themselves for generic values of the nome, the step and $t$.

\proof Conjugation by $\CG$ and multiplication by the constant $K_{U,V}$ do not affect commutators, so by Theorem 5.6 it suffices to show that the operators $\CD_{\bar X}\CP^\eta_U\CP^\xi_V$ commute for different $(U,V)$.
$\CD_{\bar X}$ commutes with $\CP^\xi$ and $\CP^\eta$, since all coefficients depend only on ratios of the $b_i$.
The coefficients of $\CP^\xi$ depend only on $\xi$, and $T^\xi$ leaves $\eta$ fixed; the same holds with $\xi$ and $\eta$ exchanged.
Hence $\CP^\xi_V$ and $\CP^\eta_U$ commute on the enlarged space of section 5.2, whose image is invariant under the separate sector shifts.
What remains is the commutativity within each sector.
As noted below \sectorops, $\CP^\eta_U$ is obtained from $\CP^\xi_U$ by the substitution $\xi_i\to\eta_i^{-1}$ and $(p^r,R,t)\to(q^r,R^{-1},t^{-1})$, so it suffices to consider $\CP^\xi$ with generic parameters.
The complete products preserve the physical lifted space and descend to the holonomy lattice by Theorem 5.6, so their commutativity on the enlarged space implies the required commutativity after restriction.
\qed

\subsec{\it Additive form and Ruijsenaars' operators}

To prove commutativity within a sector, we write the sector operators additively.
Let $P$ be the nome and $e^{2\pi i\delta}$ the step, and define $\gamma$ by $Pe^{2\pi i\delta}/t=e^{2\pi i\gamma}$;
for $\CP^\xi$ these are $P=p^r$ and $e^{2\pi i\delta}=R$, and for $\CP^\eta$ they are $P=q^r$, $e^{2\pi i\delta}=R^{-1}$ and $t\to t^{-1}$.
Define $x_i$ by $\xi_i^{-1}=e^{2\pi ix_i}$, and use the odd theta function
$$
  [x]:=e^{-\pi ix}\,\Th(e^{2\pi ix};P)\,,\qquad [x]_m:=\prod_{\beta=0}^{m-1}[x+\beta\delta]\,,
$$
which satisfies $[-x]=-[x]$ and the Riemann relation.
Using $\Th(Py;P)=\Th(y^{-1};P)$ to rewrite the numerators of $c_{\bf v}(\xi)$ in \sectorops, one finds that, up to the ${\bf v}$-independent constant $e^{\pi i\gamma NV}$, the operator $\CP^\xi_V$ equals
\eqn\Pprime{
  \CP'_V:=\sum_{|{\bf v}|=V}c'_{\bf v}({\bf x})\,T^{{\bf v}\delta}_{\bf x}\,,\qquad
  c'_{\bf v}({\bf x}):=\prod_{i,j=1}^N\frac{[x_j-x_i+\gamma-v_i\delta]_{v_j}}{[x_j-x_i-v_i\delta]_{v_j}}\,,\qquad
  T^{{\bf v}\delta}_{\bf x}:x_i\to x_i+v_i\delta\,.
}
For $V=1$ only $j=k$ contributes to $c'_{{\bf e}_k}$, and one finds $\CP'_1=\frac{[\gamma-\delta]}{[-\delta]}D_1$, where
\eqn\Ruij{
  D_s:=\sum_{|I|=s}\prod_{i\in I,\,j\notin I}\frac{[x_i-x_j+\gamma]}{[x_i-x_j]}\;T^{\delta}_{I}\,,\qquad s=1,\dots,N\,,
}
with $T^\delta_I:x_i\to x_i+\delta$ for $i\in I$, are Ruijsenaars' operators \RuijsenaarsCMP. They commute with each other \RuijsenaarsCMP.
For $V\ge2$, however, $\CP'_V$ is not one of the $D_s$: it shifts a single variable by several steps, which no $D_s$ does.

\subsec{\it The Wronski relation}

\proclaim Proposition 6.2 (Wronski relation). For $V\ge1$,
\eqn\Wronski{
  \sum_{s=0}^{\min(V,N)}(-1)^s\,[s\gamma-V\delta]\;D_s\,\CP'_{V-s}=0\,,\qquad D_0:=1\,.
}

\proof We compare the coefficients of $T^{\bflambda\delta}_{\bf x}$ with $|\bflambda|=V$ on the left-hand side.
Write $\bfepsilon_I:=\sum_{i\in I}{\bf e}_i$. The product $D_s\CP'_{V-s}$ contributes the terms with $\bflambda={\bf v}+\bfepsilon_I$, and its coefficient of $T^{\bflambda\delta}_{\bf x}$ is
$$
  \sum_{|I|=s}\ \prod_{i\in I,\,j\notin I}\frac{[x_i-x_j+\gamma]}{[x_i-x_j]}\;c'_{\bflambda-\bfepsilon_I}({\bf x}+\bfepsilon_I\delta)\,.
$$
Two facts simplify this expression.
\item{(a)} From \Pprime\ one reads off the shift property
$$
  c'_{\bflambda-\bfepsilon_I}({\bf x}+\bfepsilon_I\delta)=c'_\bflambda({\bf x})\prod_{j\in I}\prod_{i=1}^N\frac{[x_j-x_i-\lambda_i\delta]}{[x_j-x_i+\gamma-\lambda_i\delta]}\,.
$$
In particular the right-hand side vanishes if $I$ contains an index $k$ with $\lambda_k=0$, because of the factor $[x_k-x_k]=0$. Hence only $I\subseteq{\rm supp}\,\bflambda$ contribute, and the common factor $c'_\bflambda({\bf x})$ can be divided out.
\item{(b)} For $k\notin I$ with $\lambda_k=0$, the factors $[x_j-x_k]/[x_j-x_k+\gamma]$, $j\in I$, of (a) cancel the factors $[x_j-x_k+\gamma]/[x_j-x_k]$ of $D_s$.
\par\noindent
Hence the coefficient of $T^{\bflambda\delta}_{\bf x}$ reduces to an identity in the variables $x_i$ with $\lambda_i>0$.
Set $\alpha_i:=x_i$ and $\beta_i:=x_i+\lambda_i\delta$ for these $i$, so that $\sum_i(\beta_i-\alpha_i)=V\delta$. The identity reads
$$
  \sum_{I}(-1)^{|I|}\,[|I|\gamma-V\delta]\prod_{i\in I,\,j\notin I}\frac{[\alpha_i-\alpha_j+\gamma]}{[\alpha_i-\alpha_j]}\prod_{i\in I,\,k}\frac{[\alpha_i-\beta_k]}{[\alpha_i-\beta_k+\gamma]}=0\,,
$$
where $I$ runs over all subsets of ${\rm supp}\,\bflambda$ and $j,k$ over ${\rm supp}\,\bflambda$.
Under $\alpha_i=-\varphi_i$, $\beta_i=-\psi_i$ this is Lemma 2.3 of \NoumiSano\ (with their variables $z_i,w_i$ renamed $\varphi_i,\psi_i$) with their parameter $a=\gamma$, multiplied by the constant $[\sum_i(\psi_i-\varphi_i)]=[-V\delta]$.
That lemma is a meromorphic identity in arbitrary $\varphi_i,\psi_i$, proved by induction using ellipticity.
\qed

\proclaim Proposition 6.3. For generic parameters, $\CP'_V\in\bbC[D_1,\dots,D_N]$ for all $V$.

\proof For generic parameters $[-V\delta]\ne0$ for every positive integer $V$; in each lens sector this follows from the non-resonance assumption of section 2.
The $s=0$ term of \Wronski\ can therefore be solved for:
$\CP'_V=-[-V\delta]^{-1}\sum_{s\ge1}(-1)^s[s\gamma-V\delta]\,D_s\,\CP'_{V-s}$. The claim follows by induction on $V$, starting from $\CP'_0=1$. \qed

\proclaim Theorem 6.4 (Commutativity). $[\CO_{U,V},\CO_{U',V'}]=0$ for all poles $(U,V)$ and $(U',V')$.

\proof By Proposition 6.3 and Ruijsenaars' theorem \RuijsenaarsCMP, the $\CP'_V$ of each sector commute with each other for generic parameters. Proposition 6.1 then gives the claim.
The resulting coefficient identities extend meromorphically wherever the operators are finite. \qed

\noindent Proposition 6.3 also shows that each sector is governed by the $N$ commuting operators $D_s$, and that the $\CP'_V$ generate, at most, the same commutative algebra.
For $N=2$ the Wronski relation is a three-term recursion, which expresses $\CP'_V$ as a polynomial of degree $V$ in $D_1$ with coefficients involving $D_2$;
in particular, on the kernel of $D_1$ one has $\CP'_V=0$ for odd $V$ and $\CP'_{2k}\propto D_2^k$.
This recursion is the analog of the fusion rule of \MOY, which expresses the $(1,0,0)$ defect as a polynomial of degree $r$ in the elementary operator $X_p$.

\newsec{Self-Adjointness Criterion}

In this section we prove a sufficient criterion for the operators $\CO_{U,V}$ to be self-adjoint with respect to the bilinear pairing \pairing, under explicit conditions on $(p,q)$ and on the test functions.
The idea is the following. Each term of $\CO_{U,V}$ shifts the argument $({\bf b},{\bf n})$ to $({\bf b}',{\bf n}')$.
Undoing this shift by a change of variables in $\langle f,\CO_{U,V}g\rangle$ moves the integration torus.
The criterion rests on two facts: a symmetry of the kernel of each term, which we call detailed balance and which is the exchange symmetry of the two maximal punctures of the bifundamental (section 7.1),
and the absence of poles of the kernel between the original and the shifted torus (section 7.2).
The theorem and its corollaries are in section 7.3.
Section 7.4 discusses what happens when the second condition fails, including an example with $(U,V)=(2,2)$ for which it fails in every parameter chamber.

\subsec{\it Detailed balance}

Throughout this section ${\frak c}=({\bf u},{\bf v})$ denotes a pinching configuration of $(U,V)$, and $x=({\bf b},{\bf n})$ a point of the fugacity space $\{\prod_ib_i=1\}$ times the holonomy lattice.
A function $f_{\bf n}({\bf b})$ is written $f(x)$, and we set $\Delta(x):=\Delta_{\bf n}({\bf b})$.
As in section 3, ${\bf n}':={\bf n}+{\bf w}$ denotes the shifted holonomy, and we write $C_{\frak c}(x):=C_{{\bf u},{\bf v}}({\bf b};{\bf n})$; this is exactly the coefficient \mainformula\ of Theorem 3.1.

We use three maps. The first is the shift appearing in the residue formula,
\eqn\maps{
  T_{\frak c}:({\bf b},{\bf n})\mapsto\Big(\Big(\frac{{\bar X}\,b_i}{X_i}\Big)_{i=1}^N,\;{\bf n}'\Big)\,,
}
with ${\bar X}=p^{U/N}q^{V/N}$ and $X_i=p^{u_i}q^{v_i}$ as in \wdef, so that $T_{\frak c}x=({\bf b}',{\bf n}')$ is the argument of $\CI$ in \resformula.
The second is the inversion $\iota:({\bf b},{\bf n})\mapsto({\bf b}^{-1},-{\bf n})$, which appears in the pairing \pairing.
The third is the composition $\sigma_{\frak c}:=\iota\,T_{\frak c}$, which sends $({\bf b},{\bf n})$ to $\big((X_i/({\bar X}b_i))_i,\,-{\bf n}'\big)$.
One checks directly that $T_{\frak c}\,\iota\,T_{\frak c}=\iota$; equivalently, $\sigma_{\frak c}$ is an involution.
Since $I_V$ is invariant under ${\bf z}\to{\bf z}^{-1}$, ${\bf m}\to-{\bf m}$, we have $\Delta\circ\iota=\Delta$.
In this notation the pairing and the operator read
$$
  \langle f,g\rangle=\sum_{\bf n}\oint_{\bbT}[d{\bf b}]\;\Delta(x)\,f(x)\,g(\iota x)\,,\qquad
  (\CO_{U,V}f)(x)=\sum_{\frak c}C_{\frak c}(x)\,f(T_{\frak c}x)\,,
$$
and hence
\eqn\twosides{
  \eqalign{
    \langle\CO_{U,V}f,g\rangle&=\sum_{\frak c}\sum_{\bf n}\oint_{\bbT}[d{\bf b}]\;\Delta(x)\,C_{\frak c}(x)\,f(T_{\frak c}x)\,g(\iota x)\,,\cr
    \langle f,\CO_{U,V}g\rangle&=\sum_{\frak c}\sum_{\bf n}\oint_{\bbT}[d{\bf b}]\;\Delta(x)\,C_{\frak c}(\iota x)\,f(x)\,g(T_{\frak c}\iota x)\,.
  }}
Let us introduce the kernel $\CK_{\frak c}$ of the ${\frak c}$-term as
\eqn\kernel{
  \CK_{\frak c}(x):=\Delta(x)\,C_{\frak c}(x)\,,
}
and rewrite the formulas as
\eqn\twosidestwo{
  \eqalign{
    \langle\CO_{U,V}f,g\rangle&=\sum_{\frak c}\sum_{\bf n}\oint_{\bbT}[d{\bf b}]\;\CK_{\frak c}(x)\,f(T_{\frak c}x)\,g(\iota x)\,,\cr
    \langle f,\CO_{U,V}g\rangle&=\sum_{\frak c}\sum_{\bf n}\oint_{\bbT}[d{\bf b}]\;\CK_{\frak c}(\iota x)\,f(x)\,g(T_{\frak c}\iota x)\,,
  }}
where in the second line we have used $\Delta\circ\iota=\Delta$.
Substituting $x=T_{\frak c}y$ in the second line and using $T_{\frak c}\iota T_{\frak c}=\iota$ and $\iota T_{\frak c}=\sigma_{\frak c}$, the ${\frak c}$-term of $\langle f,\CO_{U,V}g\rangle$ becomes
\eqn\shifted{
  \sum_{\bf n}\oint_{T_{\frak c}^{-1}\bbT}[d{\bf b}]\;\CK_{\frak c}(\sigma_{\frak c}y)\,f(T_{\frak c}y)\,g(\iota y)\,,
}
where $T_{\frak c}^{-1}\bbT:=\{|b_i|=|X_i/{\bar X}|\}$ is the shifted torus. Here we used that the measure $[d{\bf b}]$ is invariant and that ${\bf n}\mapsto{\bf n}'$ is a bijection of the holonomy lattice.
The integrand of \shifted\ differs from that of the ${\frak c}$-term in the first line of \twosidestwo\ only by $\CK_{\frak c}(\sigma_{\frak c}y)$ in place of $\CK_{\frak c}(y)$.
Hence the two ${\frak c}$-terms agree provided that the kernel is invariant under the involution $\sigma_{\frak c}$,
\eqn\detailed{
  \CK_{\frak c}(\sigma_{\frak c}x)=\CK_{\frak c}(x)\,,
}
and that the contour can be moved from $T_{\frak c}^{-1}\bbT$ back to $\bbT$.
We prove \detailed\ now and study the poles of the kernel in the next subsection.

To describe the kernel $\CK_{\frak c}$, let
\eqn\Bdef{
  B(x,\hat x):=\prod_{i\ne j}\left(\frac{R}{t}\right)^{\frac{f(n_i+\hat n_j)}{2r}}\,\Gamma_{n_i+\hat n_j}\big(t{\bar X}\,b_i\hat b_j\big)\,\Gamma_{-n_i-\hat n_j}\Big(\frac{1}{{\bar X}\,b_i\hat b_j}\Big)
}
for $x=({\bf b},{\bf n})$ and $\hat x=(\hat{\bf b},\hat{\bf n})$.
This is the bifundamental index \IH\ at the pole $a=a^*$, with its two maximal punctures at $x$ and $\hat x$ and with the diagonal factors $i=j$ removed.
Relabeling $i\leftrightarrow j$ shows that $B(x,\hat x)=B(\hat x,x)$: this is the exchange symmetry of the two punctures.

\proclaim Lemma 7.1 (Detailed balance). The coefficient of Theorem 3.1 can be written as
\eqn\Cgamma{
  C_{\frak c}(x)=\omega_{\frak c}\;\Delta(\sigma_{\frak c}x)\,B(x,\sigma_{\frak c}x)\,,\qquad
  \omega_{\frak c}:=\Big(\frac{\Gamma_0(t)}{\kappa}\Big)^{N-1}\prod_{i=1}^N\Big(\frac{R}{t}\Big)^{\frac{f(w_i)}{2r}}\frac{\Theta_{u_i,v_i}(0;t)}{\Theta_{u_i,v_i}(w_i;X_i^{-1})}\,.
}
Consequently $\CK_{\frak c}(x)=\omega_{\frak c}\,\Delta(x)\,\Delta(\sigma_{\frak c}x)\,B(x,\sigma_{\frak c}x)$, and the detailed balance \detailed\ holds.

\proof We go back to the residue computation in the proof of Theorem 3.1, before Lemma 2.1 is applied to the off-diagonal factors.
In the present notation the pinching point is ${\bf z}^*={{\bf b}'}^{-1}$ with gauge holonomy ${\bf m}=-{\bf n}'$.
By \resformula, $\frac{\Gamma_0(t)}{N\kappa}C_{\frak c}(x)$ is the residue of the integrand of \glue\ with $\CI$ removed, and this residue is the product of the following factors.
\item{(1)} For each $i$: the diagonal hypermultiplet factor $\Gamma_0(t)\Theta_{u_i,v_i}(0;t)$, the residue factor $1/(\kappa \Theta_{u_i,v_i}(w_i;X_i^{-1}))$, and the zero-point factor $(R/t)^{f(w_i)/2r}$.
\item{(2)} For each $i\ne j$: the hypermultiplet factors $\Gamma_{n_j-n'_i}(tX_ib_j/b_i)\,\Gamma_{n'_i-n_j}(b_i/(X_ib_j))$ with their zero-point factors $(R/t)^{f(n_j-n'_i)/2r}$.
Since $\sigma_{\frak c}x=({{\bf b}'}^{-1},-{\bf n}')$, $t{\bar X}b_i/b'_j=tX_jb_i/b_j$ and $b'_j/({\bar X}b_i)=b_j/(X_jb_i)$, these factors are precisely $B(x,\sigma_{\frak c}x)$, after relabeling $i\leftrightarrow j$.
\item{(3)} The vector multiplet index $I_V^{(-{\bf n}')}({{\bf b}'}^{-1})=I_V^{({\bf n}')}({\bf b}')=\Delta(T_{\frak c}x)=\Delta(\sigma_{\frak c}x)$.
\item{(4)} The factor $1/N$ from the residue in $a$.
\par\noindent
We divide by $\Gamma_0(t)/(N\kappa)$.
Then the factors (1) and (4) give $\omega_{\frak c}$, while (2) and (3) give $\Delta(\sigma_{\frak c}x)\,B(x,\sigma_{\frak c}x)$.
This proves \Cgamma.
Multiplying by $\Delta(x)$ gives the formula for $\CK_{\frak c}$.
Since $\sigma_{\frak c}^2=1$, the replacement $x\to\sigma_{\frak c}x$ exchanges the two factors $\Delta(x)$ and $\Delta(\sigma_{\frak c}x)$, and maps $B(x,\sigma_{\frak c}x)$ to $B(\sigma_{\frak c}x,x)=B(x,\sigma_{\frak c}x)$ by the symmetry of $B$.
Hence $\CK_{\frak c}(\sigma_{\frak c}x)=\CK_{\frak c}(x)$, which is \detailed.
\qed

\noindent The formula \Cgamma\ is the gamma-function form of \mainformula: applying Lemma 2.1 to the ratio $B(x,\sigma_{\frak c}x)\,\Delta(\sigma_{\frak c}x)$ factor by factor, as in the proof of Theorem 3.1, turns it back into the ratio of theta products $\Theta_{u,v}$ in \mainformula.
Together with Theorem 5.6 we thus have three expressions for the same coefficient $C_{{\bf u},{\bf v}}$: the ratio of theta products \mainformula, the gamma-function form \Cgamma, and the product $K_{U,V}\frac{\CG({\bf b}',{\bf n}')}{\CG({\bf b},{\bf n})}c_{\bf v}(\xi)\,c_{\bf u}(\eta)$ of sector functions.
The first is the most explicit, the third exhibits the factorization, and the second makes the exchange symmetry of the bifundamental, and hence detailed balance, manifest.

\subsec{\it Poles of the kernel}

By Lemma 7.1, $\CK_{\frak c}(x)$ is, up to a constant, a product over unordered pairs $\{i,j\}$ of functions of $y_{ij}:=b_i/b_j$ alone, since each of $\Delta(x)$, $\Delta(\sigma_{\frak c}x)$ and $B(x,\sigma_{\frak c}x)$ is such a product.
Writing out the gamma functions, the $\{i,j\}$ factor is $F_{ij}(y_{ij})\,F_{ji}(y_{ij}^{-1})$ with
\eqn\Fij{
  F_{ij}(y):=\frac{\Gamma_{n_i-n'_j}(tyX_j)}{\Gamma_{n_i-n_j}(ty)\,\Gamma_{n'_i-n'_j}(tyX_j/X_i)}\cdot
  \frac{\Gamma_{n'_i-n_j}(y/X_i)}{\Gamma_{n_i-n_j}(y)\,\Gamma_{n'_i-n'_j}(yX_j/X_i)}\,,
}
where the numerators come from $B$, the denominators from $\Delta(x)$ and $\Delta(\sigma_{\frak c}x)$, and we used $b'_i/b'_j=y_{ij}X_j/X_i$.

Each $\Gamma_k(z)$ is singular only at the points $z=p^aq^b$ with $b-a\equiv k$ mod $r$, where by \Gammak\ it has order
\eqn\orderGamma{
  o(a,b):=[a\le0]+[b\le0]-1\,,
}
poles being counted positively: a simple pole if $a,b\le0$, a simple zero if $a,b\ge1$, and no singularity otherwise.
One checks that all six gamma functions in \Fij\ are singular only at points $y=p^\alpha q^\beta$ or $y=p^\alpha q^\beta/t$ with $\alpha-\beta\equiv n_j-n_i$ mod $r$.
The singularities of the kernel are thus of two kinds: those depending on $t$, which turn out to lie far from the region swept by the contour, and those at the lattice points $p^\alpha q^\beta$, whose orders we now compute.
We write $u_{ij}:=u_i-u_j$, $v_{ij}:=v_i-v_j$ and $X_{ij}:=X_i/X_j=p^{u_{ij}}q^{v_{ij}}$, so that $|X_{ij}|=|p|^{u_{ij}}|q|^{v_{ij}}$.

\proclaim Lemma 7.2 (Singularities of the kernel). Assume $|R|<|t|<1$ and generic $t$.
\item{(a)} The $t$-dependent singularities of the $\{i,j\}$ factor lie at $|y_{ij}|\le|R/t|\min(1,|X_{ij}|)$ or at $|y_{ij}|\ge|t/R|\max(1,|X_{ij}|)$.
They are therefore strictly outside the closed annulus between $|y_{ij}|=1$ and $|y_{ij}|=|X_{ij}|$.
\item{(b)} The $t$-independent singularities lie at $y_{ij}=p^\alpha q^\beta$ with $\alpha-\beta\equiv n_j-n_i$ mod $r$, and the order there is
\eqn\orderformula{
  \ord_{{\frak c},ij}(\alpha,\beta)=-[\alpha=0]-[\beta=0]+[\alpha\in I_{\bf u}]+[\beta\in I_{\bf v}]\,,
}
where $I_{\bf u}:=\{\alpha\in\bbZ:\ -u_j\le\alpha\le u_i,\ \alpha\ne u_{ij}\}$ and $I_{\bf v}:=\{\beta\in\bbZ:\ -v_j\le\beta\le v_i,\ \beta\ne v_{ij}\}$.

\proof (a) The $t$-dependent part of $F_{ij}(y)$ is the first fraction in \Fij. At $y=p^\alpha q^\beta/t$, with $o$ as in \orderGamma, its order is
$$
  o(\alpha+u_j,\beta+v_j)-o(\alpha,\beta)-o(\alpha-u_{ij},\beta-v_{ij})=1+g(\alpha)+h(\beta)\,,
$$
where $g(\alpha):=[\alpha\le-u_j]-[\alpha\le0]-[\alpha\le u_{ij}]$ and $h$ is defined in the same way with ${\bf v}$ in place of ${\bf u}$.
Since $-u_j\le0$ and $-u_j\le u_{ij}$, one has $g\le0$, with $g(\alpha)=0$ if and only if $\alpha>\max(0,u_{ij})$; similarly for $h$.
Hence a pole requires $\alpha\ge\max(0,u_{ij})+1$ and $\beta\ge\max(0,v_{ij})+1$, and then
$|y|\le|R/t|\,|p|^{\max(0,u_{ij})}|q|^{\max(0,v_{ij})}\le|R/t|\min(1,|X_{ij}|)$.
The factor $F_{ji}(y^{-1})$ gives the inverted region. The last statement follows from $|R/t|<1$.
\par\noindent
(b) Collecting the orders of the $t$-independent factors of $F_{ij}(y)$ and $F_{ji}(y^{-1})$ at $y=p^\alpha q^\beta$ gives
$$
  \ord=\big[o(\alpha-u_i,\beta-v_i)+o(-\alpha-u_j,-\beta-v_j)\big]-\bar o(\alpha,\beta)-\bar o(\alpha-u_{ij},\beta-v_{ij})\,,
$$
where $\bar o(a,b):=o(a,b)+o(-a,-b)=[a=0]+[b=0]$.
The bracket equals $[\alpha\in[-u_j,u_i]]+[\beta\in[-v_j,v_i]]$, because $[a\le u_i]+[a\ge-u_j]-1=[a\in[-u_j,u_i]]$.
Since $u_{ij}\in[-u_j,u_i]$, one has $[\alpha\in[-u_j,u_i]]-[\alpha=u_{ij}]=[\alpha\in I_{\bf u}]$, and similarly for $\beta$. This gives \orderformula.
\qed

\noindent For the trivial configuration ${\bf u}={\bf v}=0$ the kernel $\CK_{\frak c}$ is $\Delta$ itself, since $C_{\frak c}=1$, and \orderformula\ reduces to $-[\alpha=0]-[\beta=0]$:
the double zero at $y_{ij}=1$ (for $n_i\equiv n_j$) and the simple zeros on the axes, which come from the vector multiplet.
For nontrivial configurations the sets $I_{\bf u}$ and $I_{\bf v}$ can produce genuine poles.

\subsec{\it The theorem}

Let $\ell_p:=-\log|p|$ and $\ell_q:=-\log|q|$, both positive. For a configuration ${\frak c}$ and $0\le\tau\le1$, let
$$
  \bbT^\tau_{\frak c}:=\{{\bf b}:\ |b_i|=|X_i/{\bar X}|^\tau\}\,,
$$
so that $\bbT^0_{\frak c}=\bbT$ and $\bbT^1_{\frak c}=T_{\frak c}^{-1}\bbT$.

\proclaim Theorem 7.3 (Self-adjointness criterion). Assume $|R|<|t|<1$ with $t$ generic.
Let $f$ and $g$ be holomorphic in ${\bf b}$ in a neighborhood of the tori $\{|b_i|=|{\bar X}/X_i|^\tau\}$, $0\le\tau\le1$, for all configurations ${\frak c}$ of $(U,V)$. Then
$$
  \langle\CO_{U,V}f,g\rangle=\langle f,\CO_{U,V}g\rangle\,,
$$
provided that the following condition holds:
\item{($\star$)} for every configuration ${\frak c}$, every ${\bf n}$ and every ordered pair $(i,j)$, there is no lattice point $(\alpha,\beta)\in\bbZ^2$ with $\alpha-\beta\equiv n_j-n_i$ mod $r$, $\ord_{{\frak c},ij}(\alpha,\beta)>0$, and $\alpha\ell_p+\beta\ell_q$ in the closed interval between $0$ and $u_{ij}\ell_p+v_{ij}\ell_q$.

\noindent Condition ($\star$) depends only on $(p,q)$. It says that no pole of the kernel lies between the unit torus and the shifted torus.

\proof Fix ${\frak c}$. By \shifted\ and \detailed, the ${\frak c}$-term of $\langle f,\CO_{U,V}g\rangle$ equals
$$
  \sum_{\bf n}\oint_{\bbT^1_{\frak c}}[d{\bf b}]\;\CK_{\frak c}(y)\,f(T_{\frak c}y)\,g(\iota y)\,,
$$
which differs from the ${\frak c}$-term of $\langle\CO_{U,V}f,g\rangle$ in \twosidestwo\ only by the integration torus, $\bbT^1_{\frak c}$ instead of $\bbT^0_{\frak c}$.
We show that the integrand is holomorphic in a neighborhood of every torus $\bbT^\tau_{\frak c}$, $0\le\tau\le1$. Cauchy's theorem then allows us to move the contour from $\bbT^1_{\frak c}$ to $\bbT^0_{\frak c}$, which proves the claim.
For $y\in\bbT^\tau_{\frak c}$ the fugacities of $T_{\frak c}y$ and of $\iota y$ have moduli $|{\bar X}/X_i|^{1-\tau}$ and $|{\bar X}/X_i|^{\tau}$, respectively, so $f(T_{\frak c}y)$ and $g(\iota y)$ are holomorphic by assumption.
The kernel $\CK_{\frak c}(y)$ depends on $y$ through the ratios $y_{ij}=b_i/b_j$ of Lemma 7.2, and for $y\in\bbT^\tau_{\frak c}$ one has $|y_{ij}|=|X_{ij}|^{\tau}$.
As $\tau$ runs over $[0,1]$, $|y_{ij}|$ sweeps the closed interval between $1$ and $|X_{ij}|$.
By Lemma 7.2(a) no $t$-dependent singularity lies in this range.
A $t$-independent pole $y_{ij}=p^\alpha q^\beta$ lies in it if and only if $\alpha\ell_p+\beta\ell_q=-\log|y_{ij}|$ lies between $0$ and $-\log|X_{ij}|=u_{ij}\ell_p+v_{ij}\ell_q$, which is excluded by ($\star$).
\qed

\noindent For generic $(p,q)$ the only lattice points with $\alpha\ell_p+\beta\ell_q$ at an end point of the interval are $(0,0)$ and $(u_{ij},v_{ij})$, where the order \orderformula\ is $\le0$ because $(u_{ij},v_{ij})\notin I_{\bf u}\times I_{\bf v}$.
Thus the kernel is regular on the two end tori, both sides of the identity are well defined, and ($\star$) is a condition on the open interval.

\proclaim Corollary 7.4. Under the holomorphy assumptions on $f,g$ in Theorem 7.3:
\item{(a)} $\CO_{1,1}$ is self-adjoint for all $|R|<|t|<1$. For $r=1$ the same holds for $\CO_{1,0}$ and $\CO_{0,1}$, in agreement with \GaiottoXA.
\item{(b)} $\CO_{0,V}$ is self-adjoint if $|p|<|q|^{V-1}$, and $\CO_{U,0}$ is self-adjoint if $|q|<|p|^{U-1}$.

\proof (a) For $(U,V)=(1,1)$ every configuration has ${\bf u}={\bf e}_a$ and ${\bf v}={\bf e}_b$, so $I_{\bf u},I_{\bf v}\subseteq\{0\}$ for all pairs, and \orderformula\ is never positive. Hence ($\star$) holds for all $(p,q)$.
The same argument applies to $(1,0)$ and $(0,1)$, which are poles only for $r=1$.
\par\noindent
(b) For ${\bf u}=0$ one has $I_{\bf u}=\emptyset$, so a positive order requires $\alpha\neq0$, $\beta\neq0$ and $\beta\in I_{\bf v}$.
If such a point lies in the interval of ($\star$), then $\alpha\ell_p=(\tau v_{ij}-\beta)\ell_q$ for some $\tau\in[0,1]$.
Since $\tau v_{ij}$ lies between $0$ and $v_{ij}$, we have $|\tau v_{ij}-\beta|\le\max(|\beta|,|v_{ij}-\beta|)$.
Both $\beta$ and $v_{ij}-\beta$ are nonzero integers in $[-v_j,v_i]$, and $v_i+v_j\le V$; their absolute values can reach $V$ only if one of $v_i,v_j$ vanishes and $\beta\in\{0,v_{ij}\}$, which is excluded.
Hence $0<|\alpha|\ell_p\le(V-1)\ell_q$, which is impossible if $|p|<|q|^{V-1}$. The case ${\bf v}=0$ follows by $p\leftrightarrow q$.
\qed

\noindent The bounds in Corollary 7.4(b) are sharp for the termwise pole-avoidance criterion in the following example.
For $N=2$ and $\CO_{0,r}$ with $r\ge2$, the configuration ${\bf v}=(r,0)$ with ${\bf n}=0$ has $I_{\bf v}=\{0,1,\dots,r-1\}$ for the pair $(1,2)$, so $(\alpha,\beta)=(1,1)$ is a pole of the kernel by \orderformula.
It lies in the interval of ($\star$), which is $[0,r\ell_q]$, as soon as $\ell_p+\ell_q\le r\ell_q$, i.e.\ $|p|\ge|q|^{r-1}$, which is exactly where the condition of Corollary 7.4(b) fails.
When a pole lies strictly between the tori, a contour deformation can produce residue contributions for that term; at a boundary a prescription for a pole on the contour may be needed.
This does not by itself prove failure of self-adjointness of the full operator: one must also control possible cancellations among the configurations and holonomy sectors.
In particular, for $r\ge2$ the sufficient conditions of Corollary 7.4(b) for $\CO_{0,r}$ and $\CO_{r,0}$ exclude each other, but this alone does not rule out simultaneous self-adjointness on the unit torus.

\subsec{\it Chambers and the contour prescription}

Fix $(U,V)$ and test functions $f,g$ as in Theorem 7.3, and regard both sides of the identity as functions of $(p,q,t)$.
The possible walls for the termwise contour deformation are
$\alpha\ell_p+\beta\ell_q=0$ and $\alpha\ell_p+\beta\ell_q=u_{ij}\ell_p+v_{ij}\ell_q$,
where a pole reaches an end torus.
Within a chamber where the integrands are regular on the contours and the test functions remain holomorphic, the matrix elements are analytic in the parameters.
Crossing a wall with fixed contours can add residue contributions; their net contribution to the full operator requires summing over configurations and holonomies.

Suppose that ($\star$) holds for $\CO_{U,V}$ in some chamber.
Corollary 7.4 supplies such chambers for $(1,1)$, $(0,V)$ and $(U,0)$ whenever these are allowed poles.
The two matrix elements then define equal analytic germs by Theorem 7.3.
Their analytic continuations along the same parameter path remain equal wherever those continuations exist, provided that the test functions also admit the required continuation.
Locally, this can be implemented by continuing the integration cycles term by term without letting poles cross them.
The continued matrix elements can differ from integrals over the undeformed unit torus in the destination chamber.
This construction is operator-dependent and does not by itself define a single bilinear pairing for all operators, or establish path independence of the continuation.
For $r\ge2$, for example, $\CO_{0,r}$ and $\CO_{r,0}$ start from different sufficient chambers.

\parhead{A mixed operator with no starting chamber.}Already for $N=2$ the criterion ($\star$) cannot hold for $\CO_{2,2}$, which is an allowed pole for every $r$.
Choose
$$
  {\bf u}=(2,0)\,,\qquad {\bf v}=(0,2)\,,\qquad {\bf n}=(-1,1)\quad\hbox{mod }r\,.
$$
For the pair $(1,2)$, $I_{\bf u}=\{0,1\}$ and $I_{\bf v}=\{-1,0\}$.
The point $(\alpha,\beta)=(1,-1)$ satisfies $\alpha-\beta=2\equiv n_2-n_1$ mod $r$ and has order $2$ by \orderformula.
Moreover,
$$
  \alpha\ell_p+\beta\ell_q=\ell_p-\ell_q
    ={1\over2}(u_{12}\ell_p+v_{12}\ell_q)\,.
$$
Thus the pole $y_{12}=p/q$ lies on the intermediate torus $\bbT^{1/2}_{\frak c}$ when $|p|\ne|q|$, and on both end tori when $|p|=|q|$.
There is no chamber from which the preceding continuation argument can start for this operator.
This is an obstruction to the termwise criterion, not a counterexample to self-adjointness of the summed operator.
Indeed, the same configuration exists for $r=1$, where the operators reduce to the ordinary-index defect operators (section 4.4), whose self-adjointness is expected from S-duality \GaiottoXA.
The polynomial expressions of section 6 concern the separate sector operators; by themselves they do not settle self-adjointness of the full operator, which also requires compatible function domains and integration cycles.
Proving self-adjointness for such mixed operators requires an additional argument, for instance one controlling cancellations of residues, or a contour prescription established independently of ($\star$).

\newsec{Discussion}

We have determined the surface defect operators of the ${\cal N}=2$ lens index for all poles at generic fugacities, proved their commutativity, and established a sufficient criterion for self-adjointness with respect to the bilinear vector multiplet pairing.
The criterion applies to $\CO_{1,1}$ and to the pure $p$- and $q$-type operators in suitable chambers; for general mixed operators, self-adjointness remains to be established.
The main tool was the factorization theorem, according to which the lens system, although non-local on the holonomy lattice,
is gauge-equivalent to two decoupled systems of elliptic RS type glued along a lattice.
We close with some open questions.

\item{$\bullet$} {\it Eigenfunctions.} The factorization suggests looking for joint eigenfunctions of the form $\CG^{-1}\phi(\xi)\chi(\eta)$,
where $\phi$ and $\chi$ are eigenfunctions of elliptic RS operators with nomes $p^r$ and $q^r$,
subject to the twisted periodicity under ${\bf n}\to{\bf n}+r {\bf e}_i$ required by the lens space.
We hope to return to this gluing condition, and to the resulting diagonal form of the lens index, in future work.

\item{$\bullet$} {\it Contour prescription.} An intrinsic contour prescription for general mixed operators, including $\CO_{2,2}$, is still missing.
For the operators covered by section 7.4, it also remains to determine whether the continued matrix elements arise from a common pairing for the whole commuting family.
Such a prescription should be compared with the analytic properties of class ${\cal S}$ lens indices, which are in general not holomorphic in the flavor fugacities.

\item{$\bullet$} {\it Other punctures and representations.} The residues at the poles \astar\ correspond to symmetric representations of $A_{N-1}$.
The factorization should extend to the operators associated with more general representations, proposed in appendix C of \RazamatYamazaki.

\item{$\bullet$} {\it Factorization and holomorphic blocks.} The two sectors of Theorem 5.6 have nomes $p^r$ and $q^r$.
These are the modular parameters of the two solid tori $D^2\times T^2$ into which $L(r,1)\times \S^1$ can be decomposed,
and from which partition functions on such manifolds can be rebuilt as sums of products of holomorphic blocks \NieriPasquetti.
It would be natural to identify the $\xi$ and $\eta$ sectors with the two solid tori, and the Gaussian $\CG$ with the gluing.

\item{$\bullet$} {\it $U(1)$ holonomy and Sklyanin algebras.} For $N=2$, the companion paper \MOY\ includes the holonomy of the $U(1)$ symmetry and finds elementary defects
that generate the other operators and are traces of L-operators of a lens Sklyanin algebra.
For general $N$, the residue formula of section 3 can presumably be extended to nonzero $U(1)$ holonomy.
The factorization theorem suggests that the resulting elementary operators are the sector operators $\CP^\xi_1$ and $\CP^\eta_1$,
which would then be related to higher rank generalizations of the Sklyanin algebra (see e.g.\ \refs{\CherednikSklyanin,\Hasegawa}).

\bigskip

\noindent{\bf Acknowledgments:}
The author would like to thank Shlomo S. Razamat for collaboration on \RazamatYamazaki, on which this work builds.
He would also like to thank Kazunobu Maruyoshi and Toshihiro Ota for collaboration on \MOY.
This work was supported in part by World Premier International Research Center Initiative (WPI), MEXT, Japan; by the Center for Data-Driven Discovery, Kavli IPMU (WPI); by the JSPS KAKENHI Grant No.\ 23K25865; by JST, Japan (CREST Grant No.\ JPMJCR26XA, Moonshot R\&D Grant No.\ JPMJMS256E); and by the IBM-UTokyo-sponsored research.
The author was assisted by Anthropic Claude and OpenAI ChatGPT in the computations and in writing and editing the manuscript.

\appendix{A}{Proof of Proposition 5.8}

Throughout this appendix $N=2$, ${\bf b}=(b,b^{-1})$ and ${\bf n}=(n,-n)$.
We show that each term of $\CO_{1,1}$, $\CO_{0,r}$ and $\CO_{r,0}$ coincides with the corresponding term of the operators $S^{(0,0,1)}$, $S^{(0,1,0)}$ and $S^{(1,0,0)}$ of \MOY, eqs.~(3.13), (3.19) and (3.14) there.
For each term three things are compared: the shift of $(b,n)$, the theta functions, and the prefactor.
Our starting point is Theorem 5.6, which writes every coefficient as
\eqn\appC{
  C_{{\bf u},{\bf v}}=K_{U,V}\,\frac{\CG({\bf b}',{\bf n}')}{\CG({\bf b},{\bf n})}\,c_{\bf v}(\xi)\,c_{\bf u}(\eta)\,.
}
The sector functions $c_{\bf v}(\xi)$ and $c_{\bf u}(\eta)$ depend on the lift of ${\bf n}$ through $\xi$ and $\eta$, whereas $C_{{\bf u},{\bf v}}$ does not (Theorem 3.1).
We exploit this freedom: for each term we choose the lift so that the sector variables become the variables used in \MOY.
The theta functions of the two sides then agree up to monomials, and the monomials, together with $K_{U,V}$ and the ratio of Gaussians in \appC, reproduce the prefactors of \MOY.

\parhead{Conventions.}As before $R=pq$, and we write $\Th(x)$ for $\Th(x;p^r)$ or $\Th(x;q^r)$ as indicated.
Only the two identities $\Th(x;P)=\Th(P/x;P)$ and $\Th(x^{-1};P)=-x^{-1}\Th(x;P)$ are needed.
For a lift ${\bf n}$ we write $n_{12}:=n_1-n_2$ and $n'_{12}:=n'_1-n'_2$; the quantities entering \appC\ are
$$
  \eqalign{
    &{\hat\xi}:=\xi_1/\xi_2=b^2p^{n_{12}}\,,\qquad {\hat\eta}:=\eta_1/\eta_2=b^2q^{-n_{12}}\,,\cr
    &\frac{\CG({\bf b}',{\bf n}')}{\CG({\bf b},{\bf n})}=\frac{(b'_1/b'_2)^{n'_{12}}}{b^{2n_{12}}}\left(\frac Rt\right)^{\frac{(n'_{12})^2-n_{12}^2}{2r}}\left(\frac pq\right)^{\frac{(n'_{12})^2-n_{12}^2}2}\,,
  }
$$
A lift with $n_1\equiv n$, $n_2\equiv-n$ and a prescribed value of $n_{12}$ exists whenever $n_{12}\equiv2n$ mod $r$,
for instance ${\bf n}=(n,n-n_{12})$, since only $\sum_in_i\equiv0$ is required.

\parhead{The operator $\CO_{0,r}$.}
Here ${\bf u}=0$, so $c_{\bf u}(\eta)=1$, and the configurations are ${\bf v}=(k,r-k)$ with $k=0,\dots,r$.
\parsub{Shift.}The term with ${\bf v}=(k,r-k)$ shifts $(b,n)\to(q^{r/2-k}b,n-k)$, as does the term with $M=k$ of (3.19) of \MOY. We write $K_k$ and $\widetilde c_k$ for the functions $K_M$ and $\widetilde c_M$ of \MOY\ at $M=k$.
\parsub{Lift.}Choose the lift with $n_{12}=-\bbr{-2n}$. Then ${\hat\xi}$ is the variable $\widetilde W_m(z)$ of \MOY\ with $m=n$ and $z=b$. All theta functions below have nome $p^r$.
\parsub{Theta functions.}Spelling out \sectorops, the four pairs $(i,j)$ give
$$
  c_{\bf v}(\xi)=\prod_{\beta=0}^{k-1}\frac{\Th(tR^\beta)}{\Th(R^{\beta-k})}\prod_{\beta=0}^{r-k-1}\frac{\Th(tR^\beta)}{\Th(R^{\beta-r+k})}
  \prod_{\beta=0}^{r-k-1}\frac{\Th(tR^{\beta-r+2k}{\hat\xi}^{-1})}{\Th(R^{\beta-k}{\hat\xi})}
  \prod_{\beta=0}^{k-1}\frac{\Th(tR^{\beta+r-2k}{\hat\xi})}{\Th(R^{\beta-r+k}{\hat\xi}^{-1})}\,.
$$
We convert each product into the form used in \MOY, keeping track of the monomials produced by $\Th(x^{-1})=-x^{-1}\Th(x)$.
In the first two products, $\Th(R^{-j})=-R^{-j}\Th(R^j)$ turns the denominators into those of $K_k(p^r)$ of \MOY, at the cost of $(-1)^rR^{[k(k+1)+(r-k)(r-k+1)]/2}$.
In the third product, $\Th(tR^{-i}{\hat\xi}^{-1})=-tR^{-i}{\hat\xi}^{-1}\Th(R^i{\hat\xi}/t)$ with $i=r-2k-\beta\in[1-k,r-2k]$ produces the factors $\Th({\hat\xi}R^i/t)$ of $\widetilde c_k({\hat\xi})$, at the cost of $(-1)^{r-k}t^{r-k}{\hat\xi}^{k-r}R^{-(r-k)(r-3k+1)/2}$.
In the fourth product the numerators are the factors $\Th(t{\hat\xi}R^i)$, $i\in[r-2k,r-k-1]$, of $\widetilde c_k({\hat\xi})$, and
$1/\Th(R^{-j}{\hat\xi}^{-1})=-R^j{\hat\xi}/\Th(R^j{\hat\xi})$ with $j\in[r-2k+1,r-k]$ gives its denominators $\Th({\hat\xi}R^{i+1})$, at the cost of $(-1)^k{\hat\xi}^kR^{k(2r-3k+1)/2}$.
The remaining denominators $\Th(R^{j}{\hat\xi})$, $j\in[-k,r-2k-1]$, are those of $\widetilde c_k({\hat\xi})$, after canceling its numerator $\Th({\hat\xi}R^{r-2k})$.
Altogether
$$
  c_{\bf v}(\xi)=K_k(p^r)\,\widetilde c_k({\hat\xi})\;t^{r-k}{\hat\xi}^{2k-r}R^{k(2r-2k+1)}\,.
$$
\parsub{Prefactor.}It remains to show that the monomial above, multiplied by $K_{0,r}$ and the ratio of Gaussians, is the prefactor of \MOY.
Here $K_{0,r}=R^{-r(r+1)/2}t^{-r/2}$, $b'_1/b'_2=q^{r-2k}b^2$ and $n'_{12}=n_{12}+\ell$ with $\ell:=r-2k$.
Collecting the exponents, the powers of $b$ cancel, $2\ell+2(2k-r)=0$, and the powers of $t$, $p$ and $q$ combine to
$$
  K_{0,r}\,\frac{\CG({\bf b}',{\bf n}')}{\CG({\bf b},{\bf n})}\,t^{r-k}{\hat\xi}^{2k-r}R^{k(2r-2k+1)}=\left(\frac Rt\right)^{\frac{(k-n_{12})(2k-r)}{r}}\,.
$$
For example, the exponent of $p$ is $-\frac{r(r+1)}2+\frac{\ell(2n_{12}+\ell)}{2r}+\frac{\ell(2n_{12}+\ell)}2-\ell n_{12}+k(2r-2k+1)$, which equals $-\ell(k-n_{12})/r$ after using $\ell+2k=r$.
Since $-n_{12}=\bbr{-2n}$, this is the prefactor of the term with $M=k$ of (3.19) of \MOY.

\parhead{The operator $\CO_{1,1}$.}
Here $K_{1,1}=1$, and the four configurations are ${\bf u}={\bf e}_I$, ${\bf v}={\bf e}_J$ with $I,J\in\{1,2\}$.
\parsub{Theta functions.}From \sectorops,
$$
  c_{{\bf e}_J}(\xi)=\frac{\Th(t;p^r)}{\Th(R^{-1};p^r)}\,\frac{\Th(tR^{-1}{\hat\xi}^{\mp1};p^r)}{\Th({\hat\xi}^{\pm1};p^r)}\,,\qquad
  c_{{\bf e}_I}(\eta)=\frac{\Th(t^{-1};q^r)}{\Th(R;q^r)}\,\frac{\Th(R{\hat\eta}^{\pm1}/t;q^r)}{\Th({\hat\eta}^{\mp1};q^r)}\,,
$$
with the upper signs for $J=2$ (resp.\ $I=2$) and the lower signs for $J=1$ (resp.\ $I=1$).
The constant factors agree with the prefactor of (3.13) of \MOY, since $\Th(t;p^r)\Th(t^{-1};q^r)=\Th(t^{-1};p^r)\Th(t;q^r)$ by $\Th(t^{-1})=-t^{-1}\Th(t)$.
\parsub{Lift.}We choose the lift with $n_{12}=\bbr{2n}$ for ${\bf v}={\bf e}_2$ and $n_{12}=-\bbr{-2n}$ for ${\bf v}={\bf e}_1$.
With this choice the remaining theta functions of $c_{{\bf e}_J}(\xi)c_{{\bf e}_I}(\eta)$ agree with those of (3.13) of \MOY\ up to a monomial, which we call $\rho$.
\parsub{Prefactor.}For each term we list the shift, the data entering the ratio of Gaussians, the monomial $\rho$, and the product of $\rho$ with the ratio of Gaussians.
\item{$\bullet$} ${\bf u}={\bf v}={\bf e}_2$: shift $(R^{1/2}b,n)$, $b'_1/b'_2=Rb^2$, $n'_{12}=n_{12}$. Then $\rho=\big(-t/(R{\hat\xi})\big)\big(-R{\hat\eta}/t\big)=R^{-n_{12}}$, and the Gaussian ratio is $R^{n_{12}}$. The product is $1$.
\item{$\bullet$} ${\bf u}={\bf v}={\bf e}_1$: shift $(R^{-1/2}b,n)$, $b'_1/b'_2=b^2/R$, $n'_{12}=n_{12}$. Then $\rho=(t{\hat\xi}/R)\big(R/(t{\hat\eta})\big)=R^{n_{12}}$, and the Gaussian ratio is $R^{-n_{12}}$. The product is $1$.
\item{$\bullet$} ${\bf u}={\bf e}_1$, ${\bf v}={\bf e}_2$: shift $((q/p)^{1/2}b,n+1)$, $b'_1/b'_2=(q/p)b^2$, $n'_{12}=n_{12}+2$. Then $\rho=t/(R{\hat\xi}{\hat\eta})$, and the Gaussian ratio is $b^4(p/q)^{n_{12}}(R/t)^{2(n_{12}+1)/r}$.
Since ${\hat\xi}{\hat\eta}=b^4(p/q)^{n_{12}}$, the product is $(R/t)^{(2+2n_{12}-r)/r}$.
\item{$\bullet$} ${\bf u}={\bf e}_2$, ${\bf v}={\bf e}_1$: shift $((p/q)^{1/2}b,n-1)$, $b'_1/b'_2=(p/q)b^2$, $n'_{12}=n_{12}-2$. Then $\rho=t{\hat\xi}{\hat\eta}/R$, and the Gaussian ratio is $b^{-4}(p/q)^{-n_{12}}(R/t)^{2(1-n_{12})/r}$.
The product is $(R/t)^{(2-2n_{12}-r)/r}$.
\par\noindent
In each case the result is $1$ or $(R/t)^{(2-r+2\bbr{2sn})/r}$ with $s=\pm1$ the holonomy shift, as in (3.13) of \MOY.

\parhead{The operator $\CO_{r,0}$.}
Rather than repeating the computation, we obtain $\CO_{r,0}$ from $\CO_{0,r}$ by a symmetry.
For $\bbr{k}\neq0$ one has $\bbr{-k}=r-\bbr{k}$, so $\Gamma_k(x)$ in \Gammak\ is invariant under $p\leftrightarrow q$, $k\to-k$.
For $\bbr{k}=0$ the same holds because $\Gamma(xp^r;pq,p^r)/\Gamma(x;pq,p^r)=\Th(x;pq)=\Gamma(xq^r;pq,q^r)/\Gamma(x;pq,q^r)$.
Since $f$ is even and the pole \astar\ is symmetric, the residue computation gives $\CO_{r,0}$ from $\CO_{0,r}$ by $p\leftrightarrow q$, ${\bf n}\to-{\bf n}$ (and hence ${\bf n}'\to-{\bf n}'$).
This is the relation between (3.14) and (3.19) of \MOY, which completes the proof.

\listrefs

\bye